\documentclass[a4paper,11pt]{article}
\pdfoutput=1 % if your are submitting a pdflatex (i.e. if you have images in pdf, png or jpg format)
\usepackage{jheppub}	% jheppub includes hyperref, color, natbib, amsmath, amssymb, epsfig, graphicx
\usepackage[utf8]{inputenc}
\usepackage[english]{babel}
\usepackage{makeidx}
\usepackage{amsfonts}
\usepackage{enumerate}
\usepackage{mathrsfs}
\usepackage{tensor}
\usepackage{float}
\usepackage{xcolor,tikz,pgfplots}

\usepackage[autostyle]{csquotes}

\usepackage{subcaption}
\usepackage{braket}
\usepackage{caption}
\usepackage{sidecap}

\usepackage{youngtab}

\usepackage{braket}

\usetikzlibrary{matrix,calc,positioning,decorations.markings,decorations.pathmorphing,decorations.pathreplacing}%tikzmark,
\usetikzlibrary{arrows,cd,shapes}
\usepackage{bbding}
\tikzset{%Define standard arrow tip
    >=stealth',
    punkt/.style={
           rectangle,
           rounded corners,
           draw=black, very thick,
           text width=7.4em,
           minimum height=2em,
           text centered},
    punkt2/.style={
           rectangle,
           rounded corners,
           draw=black!20!red, very thick,
           text width=7em,
           minimum height=2em,
           text centered},
    punktL/.style={
           rectangle,
           rounded corners,
           draw=black!20!red, very thick,
           text width=8.8em,
           minimum height=2em,
           text centered},
    pil/.style={
           ->,
           thick,
           shorten <=2pt,
           shorten >=2pt,},
    pil2/.style={
           <->,
           thick,
           shorten <=2pt,
           shorten >=2pt,}}

\usepackage{float}
\usepackage{booktabs}

\usepackage{bm}

\usepackage{feynmp}
\newcommand\encircle[1]{%
  \tikz[baseline=(X.base)] 
    \node (X) [draw, shape=circle, inner sep=0] {\strut #1};}
\usepackage{amsmath}

\numberwithin{equation}{section}

\title{The Reissner-Nordstr\"om-Tangherlini solution from scattering amplitudes of charged scalars}
\author[a]{Simone D'Onofrio,}
\author[b]{Federica Fragomeno,}
\author[a]{Claudio Gambino,}
\author[c]{Fabio Riccioni}

\affiliation[a]{Dipartimento di Fisica, Universit\`a di Roma ``La Sapienza'', \\
	Piazzale Aldo Moro 2, 00185 Roma, Italy}
\affiliation[b]{Department of Physics and Astronomy, York University,\\
4700 Keele Street, Toronto, Ontario M3J 1P3, Canada}
\affiliation[c]{I.N.F.N.  Sezione  di  Roma, Dipartimento di Fisica,  Universit\`a di Roma ``La Sapienza'', \\
	Piazzale Aldo Moro 2, 00185 Roma, Italy}
\emailAdd{donofrio.1749089@studenti.uniroma1.it}
\emailAdd{fedefrag@yorku.ca}
\emailAdd{gambino.1793203@studenti.uniroma1.it}
\emailAdd{Fabio.Riccioni@roma1.infn.it}

\abstract{The metric and the electromagnetic potential generated by a static, spherically symmetric charged massive object in any dimension are given by the Reissner-Nordstr\"om-Tangherlini solution. We derive the expansion of this solution up to third post-Minkowskian order by computing the classical contribution of  scattering amplitudes describing the emission of either a graviton or a photon from a massive charged scalar field up to two loops. In four and five dimensions these amplitudes develop ultraviolet divergences that are cancelled by higher-derivative counterterms in a way that generalises what was recently shown to happen in the chargeless case. This renormalisation procedure produces logarithmic terms that match exactly those produced in the post-Minkowskian expansion of the classical solution in de Donder gauge. }


\preprint{PREPRINT}
 \clearpage

\begin{document}
\maketitle

\section{Introduction}
We universally think of general relativity as the low energy effective field theory of a quantum theory of gravity. More precisely, we think of the Einstein-Hilbert action as the first term of a higher-derivative expansion where infinitely many operators are suppressed at low energy by inverse powers of the Planck mass.  Gravitational interactions are mediated by the exchange of spin-2 gravitons, and performing a perturbative expansion in  $\hslash$ one determines graviton vertices as well as vertices of gravitons interacting with matter \cite{Feynman:1963ax,DeWitt:1967yk,DeWitt:1967ub,DeWitt:1967uc,tHooft:1974toh}. Using the tools of effective field theory, one can then for instance compute quantum scattering amplitudes involving massive particles and determine $\hslash$ corrections to the gravitational potential \cite{Donoghue:1993eb,Donoghue:1994dn}.

The most remarkable outcome of this analysis is that by performing a loop expansion one obtains at any loop order not only quantum corrections, but also terms that are zeroth order in $\hslash$, {\it i.e.} entirely classical \cite{Iwasaki:1971vb,Donoghue:1994dn,Bjerrum-Bohr:2002fj,Donoghue:1996mt,Holstein:2004dn,Bjerrum-Bohr:2018xdl,Kosower:2018adc,Cheung:2018wkq}. In particular, while 
the Schwarzschild metric at first order in the post-Minkowskian expansion arises from the scattering amplitude of a massive scalar emitting a graviton at tree level, the analysis of  \cite{Bjerrum-Bohr:2002fj} shows that the next order arises from a 1-loop amplitude containing a three-graviton vertex. 
Similarly, one reproduces  at second post-Minkowskian order the Kerr metric by computing loop amplitudes involving spinors \cite{Bjerrum-Bohr:2002fj}, as well as the terms proportional to the electric charge of the Reissner-Nordstr\"om and Kerr-Newman metrics by considering the 1-loop scattering of charged particles with photons running in the loop \cite{Donoghue:2001qc}.

More recently, a systematic procedure to extract the classical contribution of loop amplitudes of massive scalars interacting with gravitons in any dimension was given in \cite{Bjerrum-Bohr:2018xdl}, which also shows how such computations coincide with the original work of \cite{Duff:1973zz} at second post-Minkowskian order. Applying these techniques, the Schwarzschild-Tangherlini metric \cite{Tangherlini:1963bw} at fourth order in the post-Minkowskian expansion was shown to arise from  gravitational scattering amplitudes of massive scalars up to three loops \cite{Mougiakakos:2020laz}.\footnote{At tree-level and one loop the agreement  between the Schwarzschild-Tangherlini metric and the amplitude results in any dimension was earlier shown in \cite{KoemansCollado:2018hss,Jakobsen:2020ksu}.}
The computations of the amplitudes in \cite{Mougiakakos:2020laz} are performed in de Donder gauge, and in order to compare the results with the classical metric one has to write down the latter in the same gauge, in which logarithmic terms appear  in the post-Minkowskian expansion starting from second order in five dimensions and from third order in four dimensions. Correspondingly, the amplitudes develop ultraviolet  divergences \cite{Jakobsen:2020ksu,Mougiakakos:2020laz} 
which are renormalised by the inclusion of specific higher-derivative couplings \cite{Mougiakakos:2020laz},\footnote{Higher-derivative couplings had already been introduced in \cite{Goldberger:2004jt} in the context of the world-line formalism.} and as a result one obtains the remormalised energy-momentum tensor which gives exactly the logarithmic terms of the Schwarzschild-Tangherlini metric in de Donder gauge.\footnote{There are additional renormalised terms in  the energy-momentum tensor that do not affect the metric \cite{Mougiakakos:2020laz}. We will not discuss this any further in this paper.}

In this paper we want to repeat the analysis of \cite{Mougiakakos:2020laz} for the case of charged scalars in any dimension. We will show that the long-distance metric derived from the amplitude computation matches the Reissner-Nordstr\"om-Tangherlini metric \cite{Tangherlini:1963bw}. We will make use of the prescription of \cite{Bjerrum-Bohr:2018xdl} to isolate the classical contribution from each Feynman diagram. 
We will perform the analysis up to two loops, and we will show that the terms of the metric proportional to the electric charge are exactly reproduced up to third post-Minkowskian order by the diagrams in which photons circulate in the loop. In de Donder gauge there are also logarithmic terms proportional to the charge, and these are exactly reproduced by renormalising the divergent terms in the amplitude adding higher-derivative couplings. Remarkably, these couplings are precisely the same as in the chargeless case, but the coefficient in front of them is modified by the addition of a term proportional to a given power of the charge.

The techniques to extract the classical contribution of loop amplitudes can also be applied to diagrams in  which the scalar field emits a  photon. We compute such contributions up to two loops, and we derive from the resulting current the first three terms in the post-Minkowskian expansion of the electromagnetic potential. We compare the result with the post-Minkowskian expansion of the  potential in the Reissner-Nordstr\"om-Tangherlini solution, finding again perfect agreement. In five dimensions the amplitudes develop an ultraviolet divergence, which is renormalised by the same counterterm that renormalises the metric. We compute the logarithmic terms that arise and we show that again they match exactly the logarithms in the post-Minkowskian expansion of the 
 Reissner-Nordstr\"om-Tangherlini potential.

The paper is organised as follows. In section \ref{sec:dedonder} we 
  determine  the post-Minkowskian expansion of the Reissner-Nordstr\"om-Tangherlini solution in de Donder gauge.   In section \ref{sec:amplitudes}
 we show  how to relate the post-Minkowskian expansion of the  metric and the gauge potential to the loop expansion of the amplitude for the emission of either a graviton or a photon from a massive charged scalar. We use these results to compute in section \ref{sec:metric} and  section \ref{sec:gaugepotential} the metric  and the gauge potential respectively from scattering amplitudes up to two loops.
 %With respect to the chargeless case, new terms arise from the diagrams in which photons run in the loops.
 In section \ref{sec:divergences}
 we discuss how non-minimal couplings are included to cancel the ultraviolet divergences, producing logarithmic terms in the metric.
 The final outcome is that  also  the terms containing the electric charge in the post-Minkowskian expansion of the metric in de Donder gauge, as well as the electromagnetic potential,  are exactly reproduced by scattering amplitude computations. Finally,
section \ref{Sec:Discussion} contains a discussion and our conclusions. The paper also contains three appendices. In appendix \ref{App:FeynmanRules} we give all the  expressions for the propagators and vertices that are used in the paper. In appendix \ref{App:masterintegrals} we derive for completeness the Fourier transforms  that are used  to determine the metric and the potential from the amplitudes given in momentum space. Finally in appendix \ref{App:LoopRed} we list all the results for the loop integrals that are needed to evaluate all the amplitudes in sections \ref{sec:metric} and \ref{sec:gaugepotential}.

\section{The Reissner-Nordstr\"om-Tangherlini solution in de Donder gauge}\label{sec:dedonder}

The Reissner–Nordstr\"om solution gives the metric and the electromagnetic potential of a spherically symmetric charged mass distribution in four dimensions. Its generalisation to arbitrary dimension $D=d+1$ was given by Tangherlini in \cite{Tangherlini:1963bw}, and the resulting metric has the expression
\begin{equation}\label{RNTmetric}
    ds^2 = \bigg( 1-\frac{\mu_m}{r^{d-2}}+\frac{\mu_Q}{r^{2(d-2)}} \bigg)dt^2 - \frac{dr^2}{1-\frac{\mu_m}{r^{d-2}}+\frac{\mu_Q}{r^{2(d-2)}}} - r^2 d\Omega_{d-1}^2 \ ,
\end{equation}
where $\mu_m$ and $\mu_Q$ are related to the mass $m$ and charge $Q$ of the black hole by
\begin{equation} \label{mus}
    \mu_m = \frac{16 \pi G_N m }{(d-1)\Omega_{d-1}} \quad \text{and} \quad \mu_Q=\frac{8 \pi G_N Q^2}{(d-2)(d-1)\Omega_{d-1}^2} \ ,
\end{equation}
while the electromagnetic potential is
\begin{equation}\label{Classical_EM_Potential}
    A_\mu(r)=\delta_\mu^0 \frac{1}{(d-2)\Omega_{d-1}}\frac{Q}{r^{d-2}}\ ,
\end{equation}
where $\Omega_{d-1}=\frac{2 \pi^{d/2}}{\Gamma(d/2)}$ is the area of the $(d-1)$-sphere.

In this section we want to perform the post-Minkowskian expansion of the metric and the electromagnetic potential in de Donder gauge. As far as  the metric is concerned, the analysis is the same as the one performed in \cite{Mougiakakos:2020laz} for the case of the Schwarzschild-Tangherlini solution, and we refer to that paper for further details. Given in general a metric of the form
\begin{equation}\label{generalmetric}
    ds^2 = C(r)dt^2 - \frac{dr^2}{C(r)} - r^2 d\Omega_{d-1}^2 \ ,
\end{equation}
its expression in cartesian coordinates becomes
\begin{equation}\label{metricCr}
     ds^2 = C(r)dt^2-d\vec{x}^2- \frac{1-C(r)}{C(r)} \frac{(\vec{x}\cdot d\vec{x})^2}{r^2} \ .
\end{equation} 
We want to determine the change of coordinates such that the transformed metric satisfies the de Donder gauge condition
\begin{equation}\label{deDonderWEAK}
    \eta^{\mu\nu} (g_{\beta\nu,\mu} + g_{\mu\beta,\nu} -g_{\mu\nu,\beta})=0 \ .
\end{equation}
Given that the metric in \eqref{metricCr} only depends on $r$, we look for transformations that rescale the spatial coordinates by an $r$-dependent (positive) function, namely
\begin{equation}\label{Coordiante_Transformation_fx}
(t, \vec{x}) \rightarrow (t,f(r)\vec{x}) \ \ {\rm with}\ \  r = \sqrt{x^i x^i} \ .
 \end{equation}
The metric in the new coordinates becomes
\begin{equation}\label{metricCart}
    ds^2 = h_0(r)dt^2-h_1(r)d\vec{x}^2- h_2(r) \frac{(\vec{x}\cdot d\vec{x})^2}{r^2} \ ,
\end{equation}
where $h_0 (r)$, $h_1 (r)$ and $h_2 (r)$ are determined in terms of $C(r)$ and $f(r)$ as 
\begin{equation}
\begin{aligned}\label{hiC}
   h_0(r) &= C(f(r)r) \\
 h_1(r) &= f(r)^2\\
 h_2(r) &= -f(r)^2+\frac{(f (r)+rf'(r))^2}{C(f(r)r)} \ . \\
\end{aligned}
\end{equation}
The de Donder condition \eqref{deDonderWEAK} then leads to the equation
\begin{equation}\label{gaugeR}
    r \frac{d}{dr} \Big( h_0(r) + (d-2) h_1(r) - h_2(r) \Big) = 2(d-1) h_2(r) \ ,
\end{equation}
that substituting the relations in \eqref{hiC} becomes a differential equation for $f(r)$ \cite{Mougiakakos:2020laz}.

We want to write down the metric \eqref{RNTmetric} in de Donder gauge, which means that we want to solve eq. \eqref{gaugeR}  for 
\begin{equation}
C(r) =   1-\frac{\mu_m}{r^{d-2}}+\frac{\mu_Q}{r^{2(d-2)}}  \ . \label{Cr}
\end{equation}
Following \cite{Mougiakakos:2020laz}, we define  
\begin{equation}
    {\rho}(r) = \frac{\Gamma(\frac{d}{2}-1) \pi^{1-d/2}}{r^{d-2}} \ , \label{defrho}
\end{equation}
so than  substituting eq. \eqref{Cr} in \eqref{hiC}, one gets
\begin{equation}\label{hiRN}
\begin{aligned}
h_0(r) &=  1 - 4m G_N \frac{d-2}{d-1}\frac{\rho(r)}{f(r)^{d-2}} \bigg( 1-\frac{\alpha}{2m} \frac{\rho (r)}{f(r)^{d-2}}\bigg) \\
 h_1(r) &= f(r)^2\\
 h_2(r) &=-f(r)^2+\frac{\left(f(r)+(2-d)\rho (r) \frac{df
(r)}{d\rho}\right)^2}{ 1 - 4m G_N \frac{d-2}{d-1}\frac{\rho(r)}{f(r)^{d-2}} \left( 1-\frac{\alpha}{2m} \frac{\rho (r)}{f(r)^{d-2}}\right)} \ ,
\end{aligned}
\end{equation}
where we have introduced for convenience the fine structure constant $\alpha= \frac{Q^2}{4\pi}$ in natural units.
We plug  these expressions into eq. \eqref{gaugeR}, which we rewrite as an equation dependent on the variable $\rho$ 
\cite{Mougiakakos:2020laz},
\begin{equation}
(2-d) \rho \frac{d}{d\rho} \Big( h_0(r) + (d-2) h_1(r) - h_2(r) \Big) = 2(d-1) h_2(r) \ . \label{eqforfinrho}
\end{equation}
and we want to solve this equation perturbatively for $f(r)$ as a power expansion in $\rho$,
\begin{equation}
    f(r) = 1 + \sum_{n=1}^{\infty} a_{n} \rho(r)^n \ , \label{frhoanyd}
\end{equation}
that is order by order in the post-Minkowskian expansion.\footnote{In this paper we define  the post-Minkowskian expansion as an expansion in $\rho (r)$,  which due to the presence of the charge does not coincide with an expansion in  $G_N$.}
We have done this numerically and although we have managed to determine all the coefficients of the metric up to eighth order in the post-Minkowskian expansion, here we give the expression for $f(r)$, and therefore for $h_0(r)$, $h_1(r)$ and $h_2(r)$, up to $\rho^3$, which is  the order needed for comparison with the 2-loop calculation in the next sections. The result is
    \begin{equation} \label{franydsolved}
    \begin{aligned}
    f(r) &=1+\frac{2 m G_N }{d-1}\rho (r)-\frac{  \alpha G_N \left(d^2-4 d+3\right)+4 \left(d^2-4 d+5\right) m^2 G_N^2}{(d-4) (d-1)^2}\rho (r)^2\\&+\frac{2 m G_N}{3 (d-1)^3 \left(d^2-7 d+12\right)} \Big(\alpha  G_N\left(6 d^4-53 d^3+167 d^2-223 d+103\right) \\ &+2 \left(7 d^4-57 d^3+172 d^2-232 d+122\right) m^2 G_N^2\Big)\rho (r)^3 +O\left(\rho (r)^4\right) \ .
    \end{aligned}
\end{equation}
Plugging this into \eqref{hiRN} and expanding up to order $\rho^3$ one finally obtains
\begin{equation}\label{hid}
\begin{aligned}
h_0& (r)  =  1-\frac{4 (d-2) }{d-1}m G_N {\rho } + \left(\frac{2 (d-2)}{d-1} \alpha G_N + \frac{8 (d-2)^2  }{(d-1)^2} m^2 G_N^2\right) {{\rho }^2}
\\ &+\left(-\frac{4 (d-2)^2 (3 d-11)   }{(d-4) (d-1)^2} m \alpha G_N^2
+\frac{8 (7-3 d) (d-2)^3 }{(d-4) (d-1)^3} m^3  G_N^3\right){\rho }^3 +O( {\rho }^4)\\
h_1& (r) = 1+ \frac{4  }{d-1}m G_N {\rho }+\left(-\frac{2 (d-3) }{(d-4) (d-1)} \alpha G_N +
\frac{4 \left(-2 d^2+9 d-14\right) } {(d-4) (d-1)^2}m^2 G_N^2 \right) {\rho }^2
\\&+\left(\frac{8 (3 d^3-25 d^2+69 d-65)}{3 (d-4) (d-3) (d-1)^2} m\alpha G_N^2+\frac{8 \left(7 d^4-63 d^3+214 d^2-334 d+212\right)}{3 (d-4) (d-3) (d-1)^3}m^3 G_N^3 \right) \rho^3\\
& +O(\rho^4)\\
 h_2&(r) = \left( \frac{2 (d-2)^2 }{(d-4) (d-1)}\alpha G_N+\frac{4 (d-2)^2 (3 d-2)} {(d-4) (d-1)^2}m^2 G_N^2 \right)\rho^2\\& +\left(\frac{4 (d-2)^2  \left(-3 d^3+19   d^2-33  d+17 \right)}{(d-4) (d-3) (d-1)^3}m\alpha G_N^2  \right. \\& \left.+\frac{8 (d-2)^2 \left(-2 d^3+13 d^2-25 d+10\right) }{(d-4) (d-3) (d-1)^3}m^3 G_N^3 \right)\rho^3   +O(\rho^4) \ .
\end{aligned}
\end{equation}
As it is obvious on dimensional grounds, for each power $n$ of $\rho$ there are terms proportional to $(mG_N)^{n-j} (\alpha G_N)^{\frac{j}{2}}$ for each non-negative and even $j$ such that $n-j$ is non-negative.

The metric obtained using this procedure is not well defined in four and five dimensions, that is when $d=3$ and $d=4$, in which eqs. \eqref{hid} are singular \cite{Mougiakakos:2020laz}.  This means that the  ansatz in eq. \eqref{frhoanyd} has to be changed. Proceeding perturbatively, every time there is a divergent coefficient we have to add to the term at that order all the terms with a power of logarithm permitted by the order of the polynomial. Every new term comes with a new coefficient in the expansion. The outcome is that by implementing this new ansatz in eq. \eqref{eqforfinrho}
all the  coefficients will be fixed apart from the first divergent one. In the case $d=3$ the first divergent coefficient is at the third power of $\rho$, so the new ansatz for $f(r)$ is
\begin{equation}\label{ansatzd=3}
    f(r)_{d=3} = 1+ a_1 \rho+ a_2 \rho^2 + (a_{3,2} \log(\rho)^2 + a_{3,1}\log(\rho) + a_3) \rho^3 + O(\rho^4) \ .
\end{equation}
Plugging this into eq. \eqref{eqforfinrho} % and expressing the result in terms of the variable $\rho$ in \eqref{defrho} 
one solves for all the coefficients up to a constant. The result is 
\begin{equation}
\begin{aligned}
    f(r)_{d=3} & =1+m G_N  \rho+2  m^2G_N^2 \rho^2+\frac{2}{3} mG_N  \left(2 \alpha G_N \log \left(\frac{2 mG_N  \rho}{c}\right)\right.\\ & \left. - m^2G_N^2 \log \left(\frac{2 mG_N  \rho}{c}\right)\right) \rho^3+O\left(\rho^4\right) \ ,
\end{aligned}
\end{equation}
which shows that $a_{3,2}$ in \eqref{ansatzd=3} vanishes, while  $a_3$ is not determined. We have placed the undetermined constant inside the logarithm for convenience. The other constants in the argument of the logarithm have been inserted for dimensional reasons.

Using \eqref{hiRN} with $d=3$, the resulting metric is 
\begin{equation}\label{hi3}
\begin{aligned}
    h_0(r)_{d=3} &= 1-\frac{2 m G_N}{r} +\left({\alpha G_N }+2 m^2 G_N^2 \right) \frac{1}{r^2}+\left(-2 m \alpha G_N^2 + 2 m^3 G_N^3 \right) \frac{1}{r^3} +O\left(\frac{1}{r^4}\right) \\
    h_1(r)_{d=3} &=1+\frac{2 m G_N}{r}+\frac{5 m^2G_N^2 }{r^2}\\ & + \left(\frac{8}{3 } m \alpha G_N^2  \log \left(\frac{2 m G_N }{c r}\right) +\frac{4}{3} m^3 G_N^3 \left(3-\log \left(\frac{2 m G_N }{c r}\right)\right) \right)\frac{1}{r^3}+O\left(\frac{1}{r^4}\right) \\
    h_2(r)_{d=3} &= \left(  -\alpha  G_N -7 m^2 G_N^2\right)\frac{1}{r^2} +\left(\frac{2}{3} m\alpha G_N^2 \left(-12 \log \left(\frac{2 m G_N }{c r}\right)  -7  \right)\right.\\
    & \left.+\frac{2 }{3 } m^3 G_N^3\left(6 \log \left(\frac{2 m G_N }{c r}\right)-19\right)\right) \frac{1}{r^3} + O\left(\frac{1}{r^4}\right)  \ .
\end{aligned}
\end{equation}

In the case $d=4$ we have a divergence already at second order in $\rho$, and the ansatz for $f$ is 
\begin{equation}\label{Log_fx_d4}
    f(r)_{d=4} = 1+ a_1 \rho+ (a_{2,1} \log(\rho) + a_2) \rho^2 + (a_{3,2} \log(\rho)^2 + a_{3,1}\log(\rho) + a_3) \rho^3 + O(\rho^4) \ .
\end{equation}
Again, solving for $f$  gives 
\begin{equation} \label{frd=4}
\begin{aligned}
    f &(r)_{d=4} = 1+\frac{2}{3} m G_N \rho+\frac{1}{18}  G_N \left(-20 \log \left(\frac{8 m G_N \rho}{3 c}\right) G_N m^2-3 \alpha  \log \left(\frac{8 m G_N \rho}{3 c}\right)\right) \rho^2\\&+\frac{1}{81} m G_N^2 \left(180 \log \left(\frac{8 m G_N \rho}{3 c}\right) G_N m^2+32 G_N m^2+66 \alpha +27 \alpha  \log \left(\frac{8 m G_N \rho}{3 c}\right)\right) \rho^3\\&+O\left(\rho^4\right) \ ,
\end{aligned}
\end{equation}
and plugging the result in eq. \eqref{hiRN} one gets
\begin{equation}\label{hi4}
\begin{aligned}
   h_0& (r)_{d=4} = 1-\frac{8 m G_N}{3 \pi  r^2}
   +\left(\frac{4 \alpha G_N}{3 \pi ^2 } +\frac{32 m^2 G_N^2}{9 \pi ^2 }\right)\frac{1}{r^4}+\left(-\frac{8 m \alpha G_N^2}{9 \pi ^3 } \left(\log \left(\frac{8 m G_N}{3 c \pi  r^2}\right)  +4 \right)\right. \\& \left. +\frac{32 m^3 G_N^3}{27 \pi ^3 } \left(-5 \log \left(\frac{8 m G_N}{3 c \pi  r^2}\right)-3\right)\right) \frac{1}{r^6}+O\left(\frac{1}{r^8}\right) \\
   h_1 & (r)_{d=4} = 1+\frac{4 m G_N}{3 \pi  r^2}+ 
   \left(-\frac{\alpha G_N}{3 \pi ^2 } \log \left(\frac{8  m G_N}{3 c \pi  r^2}\right)+\frac{4 m^2 G_N^2}{9 \pi ^2} \left(1-5 \log \left(\frac{8 m G_N}{3 c \pi  r^2}\right)\right) \right) \frac{1}{r^4}\\&+ \left(\frac{4 m \alpha G_N^2 }{27 \pi ^3 }  \left(3 \log \left(\frac{8 m G_N}{3 c \pi  r^2}\right)  +11 \right)+\frac{16 m^3 G_N^3 }{81 \pi ^3}\left(15 \log \left(\frac{8 m G_N}{3 c \pi  r^2}\right)+4\right)\right)\frac{1}{r^6}+O\left(\frac{1}{r^8}\right) \\
   h_2& (r)_{d=4} =\left( \frac{2 \alpha G_N}{3 \pi ^2 } \left(2   \log \left(\frac{8 m G_N}{3 c \pi  r^2}\right)-1 \right) +\frac{40 m^2 G_N^2}{9 \pi ^2 } \left(2 \log \left(\frac{8 m G_N}{3 c \pi  r^2}\right)+1\right)\right) \frac{1}{r^4}\\ & +\left(-\frac{8 m \alpha G_N^2}{9 \pi ^3 } \left(\log \left(\frac{8 m G_N}{3 c \pi  r^2}\right)  +13  \right)+\frac{32 m^3 G_N^3 }{27 \pi ^3 } \left(-5 \log \left(\frac{8 m G_N }{3 c \pi  r^2}\right)-4\right)\right)\frac{1}{r^6} \\
   & +O\left( \frac{1}{r^8}\right) \ .
\end{aligned}
\end{equation}

Having determined the post-Minkowskian expansion of the metric in de Donder gauge, we now proceed to compute the same expansion for the potential. We thus have to perform the coordinate transformation \eqref{Coordiante_Transformation_fx} on the potential in eq. \eqref{Classical_EM_Potential}. The potential satisfies the Lorentz gauge condition, and in particular only its time component is non-vanishing and only depends on $r$. This implies that the only effect of the change of coordinates is to rescale $r$, so that the potential in de Donder gauge becomes
\begin{equation} \label{potentialdedondergauge}
    A^{\rm dD}_0 (r) =  \frac{1}{(d-2)\Omega_{d-1}}\frac{Q}{(f(r)r)^{d-2}}\ .
\end{equation}
Plugging eq. \eqref{franydsolved} into this equation and expanding up to cubic order in $\rho$ one gets
\begin{equation}\label{EM_Potential_deDonder}
\begin{aligned}
    A^{\rm dD}_0(r)& =\frac{Q}{4\pi} \Biggl( \rho-\frac{2(d-2) }{(d-1)}m G_N \rho ^2 \\
     & +\left(\frac{2(d-2)^2(3d-7)}{(d-4)(d-1)^2} m^2 G_N^2  + \frac{(d-3) (d-2)}{(d-4)(d-1)}\alpha G_N\right)\rho^3\Biggl)+O(\rho^4)\ .
\end{aligned}
\end{equation}
\noindent The electromagnetic potential in this gauge is not well defined at third post-Minkowskian order in five dimensions ({\it i.e.} $d=4$) because of the divergence in eq. \eqref{EM_Potential_deDonder} which is inherited from the divergence of $f(r)$. Plugging in \eqref{potentialdedondergauge} the function $ f(r)_{d=4} $ in eq. \eqref{frd=4} one gets 
\begin{equation}\label{EM_Potential_deDonderd=4}
\begin{aligned}
    A^{\rm dD}_0 & (r)_{d=4}=\frac{Q}{4\pi}\Biggl( \frac{1}{\pi}\frac{1}{r^2}-\frac{4}{3\pi^2}G_N m\frac{1}{r^4}\\
    & + \left(\frac{4G_N^2 m^2}{9\pi^3}\left(5\log\left(\frac{8G_Nm}{3\pi  c r^2}\right)+3\right) +\frac{\alpha G_N}{3 \pi ^3 }\log \left(\frac{8 G_N m}{3 \pi  c r^2}\right) \right)\frac{1}{r^6}\Biggl)+O\left(\frac{1}{r^8}\right)\ .
\end{aligned}
\end{equation}

In the rest of the paper we will show how all the expressions we have derived in this section for the metric and the eletromagnetic potential of the  Reissner-Nordstr\"om-Tangherlini solution in de Donder gauge up to third post-Minkwskian order can be obtained from amplitude computations up to two loops.

\section{Classical limit from scattering amplitudes}\label{sec:amplitudes}

In this section we will review the procedure discussed in \cite{Bjerrum-Bohr:2018xdl,Mougiakakos:2020laz} to extract classical contributions from loop amplitudes. The procedure will be applied to amplitudes describing the emission of either a graviton or a photon, with gravitons and photons circulating in the loop. In order to set up the conventions, we first write down the classical action  
\begin{equation}\label{Total_Action}
    S=\int d^{d+1}x \sqrt{-g}\left( -\frac{2}{\kappa^2} R -\frac{1}{4} F_{\mu\nu}F_{\alpha\beta}g^{\mu\alpha}g^{\nu\beta}+(D_\mu\phi)^* (D_\nu \phi)g^{\mu\nu}-m^2\phi^* \phi \right)\ ,
\end{equation}
where $\kappa^2=32\pi G_N$ and the covariant derivative is defined as $D_\mu \phi= (\partial_\mu +i Q A_\mu)\phi$. In our conventions the Ricci tensor is defined as $R_{\mu\nu} = R^\alpha{}_{\mu\alpha\nu}$ and the metric has signature $(+,-,...,-)$.

From (\ref{Total_Action}) we can derive the Einstein equations in generic dimension 
\begin{equation}\label{Einsteins_Equations}
    R_{\mu\nu} -\frac{1}{2}g_{\mu\nu} R= \frac{\kappa^2}{4}  T_{\mu\nu}  \ ,
\end{equation}
where $T_{\mu\nu}(x)$ is the stress-energy  tensor.
We want to derive the metric that solves  perturbatively  this equation plugging in as a source the classical  stress-energy tensor  that results from the quantum emission of a graviton by the  scalar field. 
As we will review later in this section, the perturbative loop expansion coincides with the post-Minkowskian expansion with respect to the variable $\rho$ defined in eq. \eqref{defrho}, and as explained in the previous section for each power $n$ of $\rho$ we expect for dimensional reasons terms proportional to $\alpha$ with power up to $\lfloor \frac{n}{2} \rfloor$. At tree level,
one gets a pointlike source on mass $m$ which generates the Newton potential, while the loop corrections generate precisely the contributions to the stress-energy tensor resulting from the self-interactions of the graviton and from the electromagnetic potential \cite{Donoghue:2001qc,Bjerrum-Bohr:2002fj}. 
Following the notation of the previous section, we expand the stress-energy tensor as
\begin{equation}
T_{\mu\nu}=\sum_{n=0}^{+\infty}\sum_{\substack{j=0 \\ j \ \text{even}}}^{n}T_{\mu\nu}^{(n, j)}\ ,
\end{equation}
and correspondingly we expand  the metric as
\begin{equation}\label{MetricSumDefinition}
    g_{\mu\nu} = \eta_{\mu\nu}+\kappa\, h_{\mu\nu} = \eta_{\mu\nu}+\kappa \sum_{n=1}^{+\infty}\sum_{\substack{j=0 \\ j \ \text{even}}}^{n}h_{\mu\nu}^{(n, j)}\ .
\end{equation}
We impose the de Donder gauge condition \eqref{deDonderWEAK}, which by linearity holds separately for each term $h_{\mu\nu}^{(n, j)}$ in the expansion:
\begin{equation}
    \partial_\lambda h^{\lambda}{}_{\nu}{}^{(n, j)}-\frac{1}{2}\partial_\nu h^{(n, j)}=0\ .
\end{equation}
In the same gauge we compute the amplitudes.
Substituting  the expansions above in \eqref{Einsteins_Equations}, one gets at any order the  equation
\begin{equation}\label{Box_MetricPerturbation}
    \Box h_{\mu\nu}^{(n, j)}(x) = -\frac{\kappa}{2} \left(T_{\mu\nu}^{(n-1, j)}(x) - \frac{1}{d-1}\eta_{\mu\nu}T^{(n-1, j)}(x)\right)\ ,
\end{equation}  
which says that the $n-1$-th post-Minkowskian term of the stress-energy tensor is the source of the $n$-th post-Minkowskian term of the metric. The interpretation of this phenomenon is due to the non-linear nature of gravity. Indeed, as already mentioned above, while the first post-Minkowskian order of the metric is reconstructed by the matter contribution of the stress-energy tensor, higher orders correspond to terms in which the gravitational field interacts with itself \cite{Bjerrum-Bohr:2002fj}. This explains also that the non linearity of gravity is translated to self-interaction terms of the gravitons, which are encoded in loop diagrams.  

In the very same way, it is possible to work out a similar expression for the electromagnetic potential. From the action in \eqref{Total_Action}, the field equations of the electromagnetic potential in Feynman gauge are
\begin{equation}
    \Box A_\mu(x) = j_\mu(x) \ ,
\end{equation}
where $j_\mu(x)$ is the electromagnetic current. 
Expanding the electromagnetic current and the potential in a post-Minkowskian series like
\begin{equation}
j_{\mu}=\sum_{n=0}^{+\infty}\sum_{\substack{j=0 \\ j \ \text{even}}}^{n}j_{\mu}^{(n, j)} \quad \text{and} \quad A_{\mu}=\sum_{n=1}^{+\infty}\sum_{\substack{j=0 \\ j \ \text{even}}}^{n}A_{\mu}^{(n, j)}\ ,
\end{equation}
the field equations at each order are
\begin{equation}\label{Eq_Motion_A}
    \Box A_\mu^{(n, j)}(x) = j_\mu^{(n-1, j)} (x) \ .
\end{equation}

We now review the techniques to extract the classical contributions of loop amplitudes and  apply them to diagrams with photons and gravitons in the loops, describing the emission of either a graviton 
or a photon.
% Now we are interested in extracting the classical limit of the Reissner-Nordstr\"om-Tangherlini solution, which means find those terms that are long range and that survive after the $\hbar\rightarrow 0$ limit. 
Following \cite{Mougiakakos:2020laz}, considering the matrix element of the gravitational source
\begin{equation}
     T_{\mu\nu}(q^2)\equiv\bra{p_2}T_{\mu\nu}(0)\ket{p_1}\ , 
\end{equation}
where $q=p_1-p_2$ is the transferred  momentum and  the non-covariant normalization of particle states is implied,\footnote{In particular we use the normalization in which $\braket{p_2|p_1}=(2\pi)^d\delta^{(d)}(\Vec{p_1}-\Vec{p_2})$.}
we know that the classical contribution of the stress-energy tensor in momentum space can be computed from processes of the kind
\begin{figure}[H]
    \begin{minipage}{.5\textwidth} 
\begin{equation*}
\hspace{2cm}
    \begin{fmffile}{TreeDiagramGravExt}
\begin{fmfgraph*}(100,120)
 \fmfleft{I1,I2}
 \fmfright{o}
 \fmf{fermion,label=$p_1$,l.s=left}{I1,i1}
  \fmf{plain}{i1,a,e,c,b,d,i2}
  \fmf{fermion,label=$p_2$,l.s=left}{i2,I2}
 \fmffreeze
 \fmf{photon,label=$\vdots$,l.s=right,label.dist=-0.18w}{i1,v}
 \fmf{dbl_wiggly}{v,i2}
  \fmf{dbl_wiggly,label=$\hspace{-1.2cm}\vdots$,l.s=right,label.dist=-0.26w}{b,v}
  \fmf{photon,label=,l.s=right,label.dist=0.05w}{v,e}
 \fmf{dbl_wiggly,tension=4,label=$q$,l.s=left}{v,o}
 \fmflabel{${\mu\nu}$}{o}
\fmfv{decor.shape=circle,decor.filled=empty,
decor.size=.4w,label=tree, label.dist=-0.10w}{v}
\fmfv{label=$\vspace{-0.8cm}j\Bigg\{$}{e}
\fmfv{label=$\vspace{-1.2cm}n-j\Bigg\{$}{i2}
\end{fmfgraph*}
\end{fmffile}
\end{equation*}
    \end{minipage}%
    \begin{minipage}{0.5\textwidth}
        \begin{equation}\label{SpecificLoop_GravitonEmission_Process}
            = -\frac{i\, \kappa}{2} \, \sqrt{4E_1 E_2} \, T_{\mu\nu}^{(l,j)}(q^2)\ , \hspace{5cm}
        \end{equation}
    \end{minipage}
\end{figure}
%in which the stress-energy tensor is expressed in a loop expansion as well, and where, here and after, we imply only the classical contributions of both sources and fields. 
\noindent
where $n-j$ gravitons and an even number $j$ of photons are attached to a massive line, and a tree internal structure ends up with a graviton emission. Therefore, the number of loops in the amplitude is $l=n-1$, which relates the loop order with the order in the post-Minkowskian expansion. As a consequence, considering the static limit, eq. \eqref{Box_MetricPerturbation} can be rewritten as 
\begin{equation}\label{h_EMT}
    h_{\mu\nu}^{(l+1,j)}(\vec{x}) = -\frac{\kappa}{2}\int \frac{d^d\vec{q}}{(2\pi)^d} \frac{e^{i\vec{q}\cdot \vec{x}}}{\vec{q}^2}\left(T^{(l,j)}_{\mu\nu}(\vec{q}^2) - \frac{1}{d-1}\eta_{\mu\nu}T^{(l,j)}(\vec{q}^2) \right) \ ,
\end{equation}
where $T_{\mu\nu}^{(l,j)} (\vec{q}^2)$ is the Fourier transform of $T_{\mu\nu}^{(l,j)} (\vec{x})$.

In order to recover the post-Minkowskian expansion of the metric, we can infer that the classical limit of $T_{\mu\nu}^{(l,j)} (\vec{q}^2)$ must be 
\begin{equation}\label{EMT_propto_MasterIntegral}
    T_{\mu\nu}^{(l,j)}(\vec{q}^2) \propto \int \prod_{i=1}^{l}\frac{d^d\vec{\ell_i}}{(2\pi)^d}\frac{\vec{q}^2}{\left(\prod_{i=1}^{l}\vec{\ell_i}^2\right) \left(\vec{q}-\vec{\ell_1}-...-\vec{\ell_l}\right)^2}=J_{(l)}(\vec{q}^2) \ ,
\end{equation}
where $J_{(l)}(\vec{q}^2)$ is the massless $l$-loop `sunset' master integral, as shown in appendix \ref{App:masterintegrals}.
From this last expression, exploiting the fact that the stress-energy  tensor is conserved and Lorentz covariant, we can express it in terms of  form factors $c_1^{(l,j)}(d)$ and $c_2^{(l,j)} (d)$ as 
\begin{equation}\label{EMT_FormFactors}
\begin{aligned}
    & T_{\mu\nu}^{(l)}(\vec{q}^2) = \sum_{\substack{j=0 \\ j \ \text{even}}}^{l+1} T_{\mu\nu}^{(l,j)} (\vec{q}^2)= \\
    &    \sum_{\substack{j=0 \\ j \ \text{even}}}^{l+1}m \pi^l \left(c_1^{(l, j)}(d)\delta_\mu^0\delta_\nu^0+c_2^{(l, j)}(d)\left(-\frac{q_\mu q_\nu}{\vec{q}^2}-\eta_{\mu\nu}\right)\right)(G_N m)^{l-j}(\alpha G_N)^\frac{j}{2}J_{(l)}(\Vec{q}^2)\ .
    \end{aligned}
\end{equation}
% where the non-covariant normalization is implied. 
%In order to recover the metric perturbation out of this calculation, starting from (\ref{Box_MetricPerturbation}) and considering the static limit, we can write
%\footnote{Note that the normalisation of the integral is a consequence of the normalisation in eq. \eqref{EMT_FormFactors}, which both differ from the one used in \cite{Donoghue:2001qc,Bjerrum-Bohr:2002fj,Mougiakakos:2020laz}.}
%\begin{equation}\label{h_EMT}
 %   h_{\mu\nu}^{(l+1)}(\vec{x}) = -\frac{\kappa}{2}\int \frac{d^d\vec{q}}{(2\pi)^d} \frac{e^{i\vec{q}\cdot \vec{x}}}{\vec{q}^2}\left(T^{(l)}_{\mu\nu}(\vec{q}^2) - \frac{1}{d-1}\eta_{\mu\nu}T^{(l)}(\vec{q}^2) \right) \ ,
%\end{equation}
Substituting this relation in eq. \eqref{h_EMT}, one gets order by order in $l$ and $j$
\begin{equation}\label{MetricPerturb_loop_Expansion}
    \begin{aligned}
    \kappa\, & h_{\mu\nu}^{(l+1, j)}(\vec{x}) =\\ &
    -16\, \pi^{l+1} \int \frac{d^d\vec{q}}{(2\pi)^d} \frac{e^{i\vec{q}\cdot \vec{x}}}{\vec{q}^2}\left( c_1^{(l, j)}(d)\left(\delta_\mu^0\delta_\nu^0-\frac{\eta_{\mu\nu}}{d-1}\right) + c_2^{(l, j)}(d)\left(-\frac{q_\mu q_\nu}{\vec{q}^2}+\frac{\eta_{\mu\nu}}{d-1}\right)\right) \\
    & \times (G_N m)^{l+1-j}(\alpha G_N)^\frac{j}{2}J_{(l)}(\Vec{q}^2)\ .
\end{aligned}
\end{equation}
Using the master integral identities \eqref{AppIdent1} and \eqref{AppIdent2} 
%  in appendix \ref{App:masterintegrals}, 
%which show how to compute such Fourier transforms, 
one then determines $h_{00}^{(l+1,j)}(r)$ and $h_{ij}^{(l+1,j)}(r)$, which are the only non-vanishing components of the metric. 
 From these, one extracts the functions $h^{(l+1,j)}_i(r)$ defined in (\ref{metricCart}), obtaining \cite{Mougiakakos:2020laz}
\begin{equation} \label{metric_pert}
    \begin{aligned}
h_{0}^{(l+1,j)}(r) =&-\frac{16}{d-1}\left((d-2) c_{1}^{(l,j)}(d)+c_{2}^{(l,j)}(d)\right)\left(\frac{\rho}{4}\right)^{l+1}(G_N m)^{l+1-j}(\alpha G_N)^\frac{j}{2} \\
h_{1}^{(l+1,j)}(r) =&\frac{16}{d-1}\left(c_{1}^{(l,j)}(d)-\left(1+\frac{d-1}{2-l(d-2)}\right) c_{2}^{(l,j)}(d)\right)\\&\times\left(\frac{\rho}{4}\right)^{l+1}(G_N m)^{l+1-j}(\alpha G_N)^\frac{j}{2} \\
h_{2}^{(l+1,j)}(r) =&16 \frac{(d-2)(l+1)}{2-l(d-2)} c_{2}^{(l,j)}(d)\left(\frac{\rho}{4}\right)^{l+1}(G_N m)^{l+1-j}(\alpha G_N)^\frac{j}{2} \ .
\end{aligned}
\end{equation}
 These expressions impose an explicit relation between the metric and the amplitude calculation of processes like (\ref{SpecificLoop_GravitonEmission_Process}), from which  the form factors are derived.

We now discuss how the form factors in \eqref{EMT_FormFactors} can be extracted from the amplitudes in \eqref{SpecificLoop_GravitonEmission_Process}.
Using the Feynman rules in appendix \ref{App:FeynmanRules}, the stress-energy tensor is given by
\begin{equation}
\begin{aligned}\label{EMT_Amplitude_Complete}
   & -\frac{i\, \kappa}{2}\, \sqrt{4E_1 E_2}\, T_{\mu\nu}^{(l,j)}(q^2)=\\ 
   & \int \prod_{i=1}^l\frac{d^{d+1}\ell_i}{(2\pi)^{d+1}}\frac{(i)^l\left(\prod_{i=1}^{l+1-j}i\left( \tau_{\phi^2 h}\right)_{\mu_i\nu_i}P^{\mu_i\nu_i, \lambda_i \sigma_i}\right)\left(\prod_{i=1}^{j}-i\left( \tau_{\phi^2 A}\right)^{\alpha_i}\right)}{\prod_{i=1}^l\left( \left(p_1-\sum_{k=1}^i\ell_k\right)^2-m^2+i\epsilon\right)}\\
   & \times \frac{\mathcal{M}_{\lambda_1\sigma_1, ..., \lambda_{l-j+1}\sigma_{l-j+1}, \alpha_1, ..., \alpha_j, \mu\nu}(p_1, p_2, \ell_1, ..., \ell_l)}{\prod_{i=1}^{l+1}\left(\ell_i^2+i\epsilon\right)}\ ,
\end{aligned}
\end{equation}
where $\mathcal{M}$ parametrises the tree structure of the amplitude, and in which the momentum conservation implies $p_1-p_2=q=\ell_1+...+\ell_{l+1}$. Following \cite{Bjerrum-Bohr:2018xdl,Mougiakakos:2020laz}, we can integrate out the temporal component of each internal momentum, fixing $\ell_i^0=0$ for any $i=1, ..., l+1$ in the argument of the loop integral, and considering a combinatorial factor $\frac{1}{(l+1)!}$ in front. 
Moreover, the vertices attached to the massive line, due to the static and long range limit,  
\begin{equation}\label{Static_LowEnergy_Limit}
    q^0=E_1-E_2=0 \qquad \text{and} \qquad |\vec{p_i}|\ll m \ , 
\end{equation}
take the form
\begin{equation} \label{static_vertex}
    \left(\tau_{\phi^2 h}\right)_{\mu\nu}\simeq -i\, \kappa \, m^2\delta_\mu^0\delta_\nu^0\quad \text{and} \quad  \left(\tau_{\phi^2 A}\right)_{\alpha}\simeq -2\, i\, Q\, m\, \delta_\alpha^0\ ,
\end{equation}
acting essentially like projectors on the temporal component. From these considerations, eq. (\ref{EMT_Amplitude_Complete}) becomes 
\begin{equation}
\begin{aligned} \label{Tmunufromamplitude}
    -\frac{i\, \kappa}{2}\, 2m\, T_{\mu\nu}^{(l,j)}(\vec{q}^2)&=\frac{1}{(l+1)!}\int\prod_{i=1}^l\frac{d^d\Vec{\ell_i}}{(2\pi)^d} (-1)^{j+l+1}2^{j-l}m^{l+2-j}\kappa^{l-j+1}Q^j\frac{\prod_{i=1}^{l-j+1}P^{00, \lambda_i\sigma_i}}{\prod_{i=1}^{l+1}\Vec{\ell_i}^2}\\
    &\times \mathcal{M}_{\lambda_1\sigma_1, ..., \lambda_{l-j+1}\sigma_{l-j+1}, 0, ..., 0, \mu\nu}(p_1, p_2, \ell_1, ..., \ell_l)\Bigl|_{\ell_i^0=0}\ ,
\end{aligned}
\end{equation}
where
\begin{equation}
    P_{00, \mu\nu}=\delta_\mu^0\delta_\nu^0-\frac{\eta_{\mu\nu}}{d-1}\ .
\end{equation}
Comparing this expression with eq. \eqref{EMT_propto_MasterIntegral}, one finally concludes that the classical limit is obtained extracting from \eqref{Tmunufromamplitude} all the pieces proportional to the master integral $J_{(l)} (\vec{q}^2 )$.

This analysis can be easily repeated in order to compute the classical limit of the electromagnetic potential. Following the same procedure as before, defining the electromagnetic current as 
\begin{equation}
    j_{\mu}(q^2)\equiv\bra{p_2}j_{\mu}(0)\ket{p_1}\ , 
\end{equation}
its classical contribution in momentum space can be computed from an $l$-loop expansion of photon emission scattering amplitudes,
\begin{figure}[H]
    \begin{minipage}{.5\textwidth} 
\begin{equation*}
\hspace{2cm}
    \begin{fmffile}{SpecificLoop_PhotonEmission_Process}
\begin{fmfgraph*}(100,120)
 \fmfleft{I1,I2}
 \fmfright{o}
 \fmf{fermion,label=$p_1$,l.s=left}{I1,i1}
  \fmf{plain}{i1,a,e,c,b,d,i2}
  \fmf{fermion,label=$p_2$,l.s=left}{i2,I2}
 \fmffreeze
 \fmf{photon,label=$\vdots$,l.s=right,label.dist=-0.18w}{i1,v}
 \fmf{dbl_wiggly}{v,i2}
  \fmf{dbl_wiggly,label=$\hspace{-1.2cm}\vdots$,l.s=right,label.dist=-0.26w}{b,v}
  \fmf{photon,label=,l.s=right,label.dist=0.05w}{v,e}
 \fmf{photon,tension=4,label=$q$,l.s=left}{v,o}
 \fmflabel{${\mu}$}{o}
\fmfv{decor.shape=circle,decor.filled=empty,
decor.size=.4w,label=tree, label.dist=-0.10w}{v}
\fmfv{label=$\vspace{-0.8cm}j+1\Bigg\{$}{e}
\fmfv{label=$\vspace{-1.2cm}n-j-1\Bigg\{$}{i2}
\end{fmfgraph*}
\end{fmffile}
\end{equation*}
    \end{minipage}%
    \begin{minipage}{0.5\textwidth}
        \begin{equation}\label{SpecificLoop_PhotonEmission_Process}
           = -i \, \sqrt{4E_1 E_2}\, j_\mu{}^{(l,j)}(q^2)\ , \hspace{5cm}
        \end{equation}
    \end{minipage}
\end{figure}
\noindent
in which the current is expanded in a loop series as well. Attached to the massive line, we have $n-j-1$ gravitons and an odd number $j+1$ of photons, and analogously to the previous case a tree internal structure ends up with a photon emission. In the static limit, eq. \eqref{Eq_Motion_A}  therefore becomes
\begin{equation}\label{ClassicalEM_potential_withCurrent}
    A_\mu^{(l+1,j)}(\vec{x}) = \int \frac{d^dq}{(2\pi)^d}\frac{e^{i\vec{q}\cdot \vec{x}}}{\vec{q}^2}j_\mu^{(l,j)}(\vec{q}^2) \ ,
\end{equation}
where again $j_\mu^{(l,j)}(\vec{q}^2)$ is the Fourier transform of $j_\mu^{(l,j)}(\vec{x})$.
Following the same argument discussed in the metric case, one concludes  that the classical limit of the electromagnetic current must be  proportional to the master integral $ J_{(l)}(\vec{q}^2)$, from which an expansion in terms of  form factors $c^{(l,j)} (d)$ leads to 
\begin{equation}\label{Electro_Current_FormFactor}
    j_\mu^{(l)}(\vec{q}^2)=\sum_{\substack{j=0 \\ j \ \text{even}}}^{l+1} j_\mu^{(l, j)}(\vec{q}^2) =\sum_{\substack{j=0 \\ j \ \text{even}}}^{l+1} Q \,  \delta_\mu^0 \, c^{(l, j)}(d)  (G_N m)^{l-j}(\alpha G_N)^\frac{j}{2}J_{(l)}(\Vec{q}^2)\ .
\end{equation}
Using  the Fourier transform of the master integral in eq. \eqref{AppIdent1}, one finally   obtains 
\begin{equation}
    A_\mu^{(l+1, j)}(r) = Q\, \delta_\mu^0c^{(l, j)}(d)  (G_N m)^{l-j}(\alpha G_N)^\frac{j}{2}\left(\frac{\rho}{4\pi}\right)^{l+1} \ .
\end{equation}
We observe that only the temporal component of the potential is non vanishing. 
%To summarise, 
The outcome of this analysis is that one obtains an explicit relation between the photon emission amplitudes in \eqref{SpecificLoop_PhotonEmission_Process} and the classical electromagnetic potential.

% we recover the post-Minkowskian expansion in terms of form factors, which can be computed from the electromagnetic current as in (\ref{SpecificLoop_PhotonEmission_Process}).

All manipulations performed to extract the classical contribution to the stress-energy tensor apply for the electromagnetic current as well. 
The electromagnetic current can be directly written as
\begin{equation}\label{Electro_Current_Amplitudes}
\begin{aligned}
    -i\,2m\, j_{\mu}^{(l,j)}(\vec{q}^2)=\frac{1}{(l+1)!}\int\prod_{i=1}^l\frac{d^d\Vec{\ell_i}}{(2\pi)^d} (-1)^{l+j}2^{j+1-l}m^{l+1-j}\kappa^{l-j}Q^{j+1}\frac{\prod_{i=1}^{l-j}P^{00, \lambda_i\sigma_i}}{\prod_{i=1}^{l+1}\Vec{\ell_i}^2}\\
    \times \mathcal{M}_{\lambda_1\sigma_1, ..., \lambda_{l-j+1}\sigma_{l-j+1}, 0, ..., 0, \mu}(p_1, p_2, \ell_1, ..., \ell_l)\Bigl|_{\ell_i^0=0}\ ,
\end{aligned}
\end{equation}
where again $\mathcal{M}$ parametrises the tree structure of the amplitude. The classical limit is obtained selecting out of \eqref{Electro_Current_Amplitudes} the contribution proportional to $J_{(l)}(\Vec{q}^2)$, from which the form factors can be read. 

% where again the covariant normalization of  states is implied. 

%and replacing (\ref{Electro_Current_FormFactor}) we finally obtain
%\begin{equation}\label{EM_Potential_NoFormFactors}
%    A_\mu^{(l+1, j)}(\vec{x}) = Q\, \delta_\mu^0\int \frac{d^dq}{(2\pi)^d}\frac{e^{i\vec{q}\cdot \vec{x}}}{\vec{q}^2}c^{(l, j)}(d)  (G_N m)^{l-j}(\alpha G_N)^\frac{j}{2}J_{(l)}(\Vec{q}^2) \ .
%\end{equation}

\section{Metric from scattering amplitudes}\label{sec:metric}

In the following, we compute the interaction of massive charged scalar fields with the gravitational field in an arbitrary number of dimensions $D=d+1$. By means of Feynman rules, we perform quantum computations up to 2-loop order and from the resulting stress-energy tensor we  recover the metric perturbation components in de Donder gauge computed in section \ref{sec:dedonder}. We show the appearance of divergences at 1 and 2 loops in $d=4$ and at 2 loops  in $d=3$, which will be treated in detail in section \ref{sec:divergences}.

\subsection*{Tree level}

\begin{figure}[h]
\centering
    \begin{minipage}{.3\textwidth} 
    \begin{equation*}
    \begin{fmffile}{vertexscal}
    \begin{fmfgraph*}(70,50)
    \fmfleft{i1,i2}
    \fmfright{o}
    \fmf{fermion,label=$p_1$}{i1,v}
    \fmf{fermion,label=$p_2$,l.s=right}{v,i2}
    \fmf{dbl_wiggly,label=$q$,l.s=left}{v,o}
    \fmflabel{${\mu\nu}$}{o}
    \fmfdot{v}
    \end{fmfgraph*}
    \end{fmffile}
    \end{equation*}
    \end{minipage}
    \caption{Tree-level diagram for graviton emission.}
    \label{tree_level}
\end{figure}
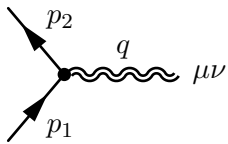

For completeness, we first consider the tree level amplitude, corresponding to $l=0,\ j=0$ in \eqref{SpecificLoop_GravitonEmission_Process}, given in figure \ref{tree_level}.
We consider on-shell particles, $p_1^2=p_2^{2}=m^2$, with transferred momentum  $q=p_1-p_2$. The stress-energy tensor arising at tree-level is
\begin{equation}
    \frac{-i\kappa}{2}\sqrt{4E_1 E_2}T_{\mu\nu}^{(0,0)}(q^2) = (\tau_{\phi^2h})_{\mu\nu}\ ,
    \label{T_0loop}
\end{equation} 
where $(\tau_{\phi^2h})$ is the 2 scalars - 1 graviton vertex in \eqref{2scalars1graviton}. Using \eqref{static_vertex} one gets
\begin{equation}
    T_{\mu\nu}^{(0,0)}(\vec{q}^2) = m \, \delta_\mu^0\delta_\nu^0\ .
\end{equation} 
Therefore, comparing this result with eq. \eqref{EMT_FormFactors}, the coefficients $c_1^{(0,0)}$ and $c_2^{(0,0)}$ are
\begin{equation}
\begin{aligned}
    c_1^{(0,0)} (d) &= 1 \\
    c_2^{(0,0)} (d) &= 0
\end{aligned}
\end{equation}
and from \eqref{metric_pert}  one gets the metric components 
\begin{equation}
    \begin{aligned}
        h_0^{(1,0)}(r) &= -4\frac{d-2}{d-1}G_N m\,\rho\\
        h_1^{(1,0)}(r) &= \frac{4}{d-1}G_N m\,\rho\\
        h_2^{(1,0)}(r) &= 0 \ ,
    \end{aligned}
    \label{h_i^1}
\end{equation}
where $\rho$ is defined in \eqref{defrho}. These results are in agreement with the post-Minkowskian expansion of the metric \eqref{hid} for the terms proportional to $ m G_N\rho$.

\subsection*{1-loop order}
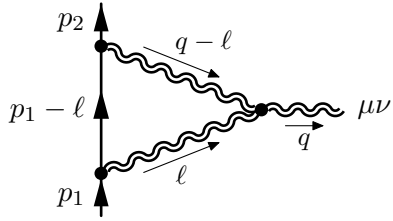
\begin{figure}[h]
    \centering
    \begin{minipage}{0.5\textwidth}
        \centering \begin{equation*}
\begin{fmffile}{sdfasfsdflllsdq}
\begin{fmfgraph*}(100,80)

\DeclareGraphicsRule{*}{mps}{*}{}
\newcommand{\marrow}[5]{%
    \fmfcmd{style_def marrow#1 
    expr p = drawarrow subpath (1/4, 3/4) of p shifted 6 #2 withpen pencircle scaled 0.4; 
    label.#3 (btex #4 etex, point 0.5 of p shifted 6 #2); 
    enddef;}
    \fmf{marrow#1, tension=0}{#5}}
\unitlength=1mm
 \fmfleft{i1,i2}
 \fmfright{o}
  \fmf{fermion,tension=3,label=$p_1$,l.s=left}{i1,v1}
 \fmf{fermion,tension=3,label=$p_2$,l.s=left}{v2,i2}
  \fmf{fermion,label=$p_1 - \ell$,l.s=left}{v1,v2}
  \fmffreeze
  \fmf{dbl_wiggly}{v1,v3}
 \fmf{dbl_wiggly,l.s=left}{v3,v2}
 \fmf{dbl_wiggly,tension=4}{v3,o}
 \fmflabel{${\mu\nu}$}{o}
 \marrow{a}{down}{bot}{$q$}{v3,o}
 \marrow{b}{up}{top}{$\ \ \ \ \ q-\ell $}{v2,v3}
 \marrow{c}{down}{bot}{$\ell$}{v1,v3}
\fmfdot{v1,v2,v3}
\end{fmfgraph*}
\end{fmffile}
\end{equation*}
    \end{minipage}
    \caption{1-loop diagram for graviton emission with 2 internal gravitons.}
    \label{1loop_grav}
\end{figure}

We now  compute the contribution to the stress-energy tensor that arises at 1-loop order. There are two diagrams that contribute, \textit{i.e.} $l=1, \ j=0$ and $l=1, \ j=2$. The first one is obtained evaluating the diagram that involves a 3-gravitons vertex, as in figure \ref{1loop_grav}.
The amplitude of this diagram is 
\begin{equation}
\begin{aligned}
     \frac{-i\kappa}{2}\sqrt{4E_1 E_2}&T^{(1,0)}_{\mu\nu}(q^2)  = \\
   & \int\frac{d^{d+1}\ell}{(2\pi)^{d+1}} \frac{-i P^{\alpha\beta,\lambda\kappa}P^{\gamma\delta,\rho\sigma}  \left(\tau_{\phi^2h}\right)_{\,\alpha\beta}\,\left(\tau_{\phi^2h}\right)_{\,\gamma\delta}\,\left(\tau_{h^2h}\right)_{\mu\nu,\rho\sigma,\lambda\kappa}(\ell,q)}{(\ell^2+i\epsilon)((\ell-q)^2+i\epsilon)((\ell-p_1)^2-m^2+i\epsilon)}
    \end{aligned}
\end{equation}
where $P^{\alpha\beta,\gamma\delta}$ is defined in \eqref{PropGraviton} and the 3 graviton vertex $(\tau_{h^2h})$ is defined in eq. \eqref{VertDonoghue}.
The resulting coefficients of the stress-energy tensor are \cite{Mougiakakos:2020laz}
\begin{equation}
    \begin{aligned}
       &  c_1^{(1,0)} (d) =  -2\frac{4d^2-15d+10}{(d-1)^2}\\
       &  c_2^{(1,0)} (d) = -\frac{2(d-2)(3d-2)}{(d-1)^2} 
    \end{aligned}
\end{equation}
and from \eqref{metric_pert} we get
\begin{equation}
    \begin{aligned}
        h_0^{(2,0)} (r) &= \frac{8(d-2)^2}{(d-1)^2}(G_N m)^2\rho^2 \\
        h_1^{(2,0)} (r) &= -\frac{4(2d^2 - 9d + 14)}{(d-4)(d-1)^2}(G_N m)^2\rho^2 \\
        h_2^{(2,0)} (r) &= \frac{4(d-2)^2(3d-2)}{(d-4)(d-1)^2}(G_N m)^2\rho^2 \ .
    \end{aligned}
    \label{h_ST^2}
\end{equation}
These components are in any dimension  in agreement with the post-Minkowskian expansion of the metric \eqref{hid} for the terms proportional to $ m^2G_N^2\rho^2$. In particular, $h_1^{(2,0)}$ and $  h_2^{(2,0)}$ are divergent in five dimensions.

\begin{figure}[!htb]
    \centering
    \begin{minipage}{0.5\textwidth}
        \centering \begin{equation*}
\begin{fmffile}{sdfasfsdfsdq}
\begin{fmfgraph*}(100,80)

\DeclareGraphicsRule{*}{mps}{*}{}
\newcommand{\marrow}[5]{%
    \fmfcmd{style_def marrow#1 
    expr p = drawarrow subpath (1/4, 3/4) of p shifted 6 #2 withpen pencircle scaled 0.4; 
    label.#3 (btex #4 etex, point 0.5 of p shifted 6 #2); 
    enddef;}
    \fmf{marrow#1, tension=0}{#5}}
\unitlength=1mm

 \fmfleft{i1,i2}
 \fmfright{o}
  \fmf{fermion,tension=3,label=$p_1$,l.s=left}{i1,v1}
 \fmf{fermion,tension=3,label=$p_2$,l.s=left}{v2,i2}
  \fmf{fermion,label=$p_1 - \ell$,l.s=left}{v1,v2}
    \fmffreeze
  \fmf{photon}{v3,v1}
 \fmf{photon,l.s=left}{v2,v3}
 \fmf{dbl_wiggly,tension=4}{v3,o}
 \fmflabel{${\mu\nu}$}{o}
 \marrow{a}{down}{bot}{$q$}{v3,o}
 \marrow{b}{up}{top}{$\ \ \ \ q-\ell$}{v2,v3}
 \marrow{c}{down}{bot}{$\ell$}{v1,v3}
 \fmfdot{v1,v2,v3}
\end{fmfgraph*}
\end{fmffile}
\end{equation*}
    \end{minipage}%
    \caption{1-loop diagram for graviton emission with 2 internal photons.}
    \label{1loop_photon}
\end{figure}
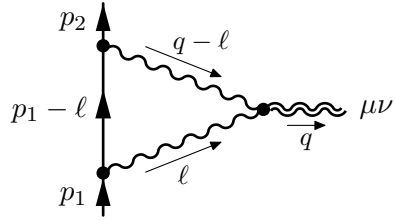

The other contribution at 1-loop order is given by the diagram that contains two photons drawn in figure \ref{1loop_photon}.
The resulting contribution to the stress-energy tensor is
\begin{equation}
\begin{aligned}
    \frac{-i\kappa}{2}\sqrt{4E_1 E_2}&T^{(1,2)}_{\mu\nu}(q^2) =\\ &  \int\frac{d^{d+1}\ell}{(2\pi)^{d+1}} \frac{-i\left(\tau_{\phi^2A}\right)_{\alpha}\left(\tau_{\phi^2A}\right)_{\beta}\left(\tau_{A^2h}\right){}_{\mu\nu}{}^{\alpha,\beta}(\ell,\ell-q)}{(\ell+i\epsilon)^2((\ell-q)^2+i\epsilon)((p_1-\ell)^2-m^2+i\epsilon)} \ , 
  \end{aligned}
\end{equation}
where $(\tau_{\phi^2A})$ is the 2 scalars - 1 photon vertex in \eqref{2scalars1photon} and $(\tau_{A^2h})$ is the 2 photons - 1 graviton vertex in \eqref{2photon1graviton}. Following the procedure in section \ref{sec:amplitudes} one gets
\begin{equation}
   T^{(1,2)}_{\mu\nu}(\Vec{q}^2) = i\frac{Q^2}{\kappa} \int \frac{d^{d}\ell}{(2\pi)^{d}} \frac{\left(\tau_{A^2h}\right)_{\mu\nu}{}^{0, 0}(\ell,\ell-q)\Big|_{\ell^0=0}}{\vec{\ell}^2 (\Vec{q}-\Vec{\ell})^2} \, , 
\end{equation}
where the numerator is explicitly
\begin{equation}
   \left(\tau_{A^2h}\right)_{\mu\nu}{}^{0, 0}(\ell,\ell-q)\Big|_{\ell^0=0} =i\kappa\left( \left(\delta_\mu^0\delta_\nu^0-\frac{\eta_{\mu\nu}}{2}\right) \ell\cdot (\ell-q)+\ell_\mu \ell_\nu -\frac{1}{2}(q_\mu \ell_\nu +q_\nu \ell_\mu) \right)\ .
\end{equation}
Then the resulting component of the stress-energy tensor, using the identities in appendix \ref{App:LoopRed}, are
\begin{equation}
        T^{(1,2)}_{00}(\vec{q}^2)= -\frac{Q^2}{4}J_{(1)}(\vec{q}^2)\ ,
    \label{T_0012_Ddim}
\end{equation}
and 
\begin{equation}
    \delta^{ij}T^{(1,2)}_{ij}(\Vec{q}^2)  =  \frac{Q^2}{4}(2-d)J_{(1)}(\vec{q}^2) \,. \label{traceT12_Ddim}
\end{equation}
% We list in the following the result, substituting the definition of the fine structure constant $\alpha=\frac{Q^2}{4}$ 
% \begin{equation}
%     \left\{
%    \begin{aligned}
%    T^{(1,2)\, 00} (q^2)=& -\alpha J_{(1)}(\vec{q}^2) \\ 
%    \delta_{ij}T^{(1,0)ij} (q^2) =&  \alpha(2-d)J_{(1)}(\vec{q}^2) \,.
%    \end{aligned}
%    \right.
% \end{equation}
% Now we can compare these stress-energy tensor components with the following general form that can be derived from \eqref{EMT_FormFactors} 
% \begin{equation}
 %   \left\{
 %   \begin{aligned}
 %       T^{(l,j)00} =& m (G_N m)^{l-j}(G_N\alpha)^{\frac{j}{2}} (c_1^{(l,j)} - c_2^{(l,j)}) J_l(q^2)\\
 %       \delta^{ij}T^{(l,j)}_{ij} =& m (G_N m)^{l-j}(G_N\alpha)^{\frac{j}{2}} (d-1) c_2^{(l,j)}J_l(q^2) \,.
 %       \label{EMT_00_ij}
 %  \end{aligned}
 %   \right.
% \end{equation}
From eqs. \eqref{T_0012_Ddim} and \eqref{traceT12_Ddim} one finally gets
\begin{equation}
    \begin{aligned}
    & c_1^{(1,2)}(d)  = \frac{-2d+3}{d-1} \\
    & c_2^{(1,2)} ( d) = \frac{2-d}{d-1}
    \end{aligned}
\end{equation}
and again using eq. \eqref{metric_pert} one obtains
\begin{equation}
    \begin{aligned}
        h_0^{(2,2)} (r)&= -\frac{(-2d+4)}{d-1}G_N\alpha\rho^2 \\
        h_1^{(2,2)} (r)&= \frac{-2d + 6}{(d-1)(d-4)}G_N\alpha\rho^2 \\
        h_2^{(2,2)} (r)&= 2\frac{(d-2)^2}{(d-1)(d-4)} G_N\alpha\rho^2 \ .
    \end{aligned}
    \label{h_RN}
\end{equation}
These metric perturbations  are  for any $d$  in agreement with the post-Minkowskian expansion of the metric \eqref{hid} for the terms proportional to $\alpha G_N\rho^2$, and $h_1^{(2,2)}$ and $h_2^{(2,2)}$ are  divergent in five dimensions, as expected.

\subsection*{2-loop order}
The third order post-Minkowskian contributions to the metric are given by the 2-loop diagrams of the type in \eqref{SpecificLoop_GravitonEmission_Process}.
Some of these diagrams give the same contribution
since in the classical limit the amplitude is invariant under the exchange of photon and graviton lines. Since all the vertices in appendix \ref{App:FeynmanRules} are defined with the right symmetry factor inside, we have to add to each diagram the corresponding multiplicity factor.
At 2-loop order ($n=3$) one gets the diagrams in  \eqref{SpecificLoop_GravitonEmission_Process} for $j=0$ and for $j=2$, which give correspondingly the metric contributions $h_{\mu\nu}^{(3,0)}$ and $h_{\mu\nu}^{(3,2)}$.

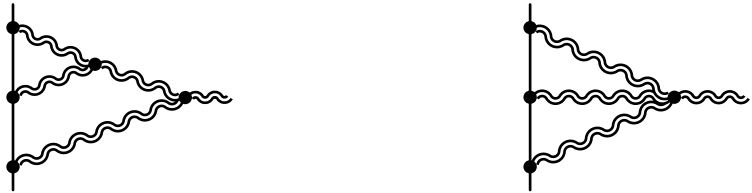
\begin{figure}[h]
    \centering
    \begin{minipage}{0.5\textwidth}
        \centering \begin{equation*}
\begin{fmffile}{2loopGrav2Vert}
\begin{fmfgraph*}(90,70)

 \fmfleft{i1,i2}
 \fmfright{o}
  \fmf{plain,tension=3}{i1,v1}
  \fmf{plain}{v1,a,v2}
 \fmf{plain,tension=3}{v2,i2}
 \fmffreeze
\fmf{dbl_wiggly,tension=0.05}{v1,v3}
\fmf{dbl_wiggly,tension=0.01}{a,b}
\fmf{dbl_wiggly,tension=0.1}{v2,b,v3}
\fmf{dbl_wiggly,tension=0.4}{v3,o}
\fmfdot{v1,v2,a,b,v3}
\end{fmfgraph*}
\end{fmffile}
\end{equation*}
    \end{minipage}%
 \centering
    \begin{minipage}{0.4\textwidth}
        \centering \begin{equation*}
\begin{fmffile}{2loopGraviton1Vert}
\begin{fmfgraph*}(90,70)

 \fmfleft{i1,i2}
 \fmfright{o}
  \fmf{plain,tension=3}{i1,v1}
  \fmf{plain}{v1,a,v2}
 \fmf{plain,tension=3}{v2,i2}
 \fmffreeze
\fmf{dbl_wiggly,tension=0.1}{v1,v3}
\fmf{dbl_wiggly,tension=0.1}{a,v3}
\fmf{dbl_wiggly,tension=0.1}{v2,v3}
\fmf{dbl_wiggly,tension=0.6}{v3,o}
\fmfdot{v1,v2,a,v3}
\end{fmfgraph*}
\end{fmffile}
\end{equation*}
    \end{minipage}%
    \caption{2-loop diagrams for graviton emission with $j=0$, {\it i.e.} only graviton internal lines.}
    \label{figure:2loopMassive}
\end{figure}

The first case ($j=0$) corresponds to the diagrams in which there are only graviton internal lines, which are given in figure \ref{figure:2loopMassive}.
The computation of these diagrams was carried out in \cite{Mougiakakos:2020laz} and we report here the result for the coefficients $c_1(d)$ and $c_2 (d)$:
\begin{equation}
\begin{aligned}
c_{1}^{(2,0)}(d)=&\frac{32}{3(d-4)(d-1)^{3}}\left(9 d^{4}-70 d^{3}+203 d^{2}-254 d+104\right)\\
c_{2}^{(2,0)}(d)=&\frac{64(d-2)}{3(d-4)(d-1)^{3}}\left(2 d^{3}-13 d^{2}+25 d-10\right) \ .
\end{aligned}
\end{equation}
From these, one computes the metric components
\begin{equation}
\begin{aligned}
h_0^{(3,0)}(r) =&\frac{8 (7-3 d) (d-2)^3 }{(d-4) (d-1)^3}m^3G_N^3 \rho^3\\
h_1^{(3,0)}(r) =& \frac{8 \left(7 d^4-63 d^3+214 d^2-334 d+212\right) }{3 (d-4) (d-3) (d-1)^3}m^3G_N^3 \rho^3\\
 h_2^{(3,0)}(r) =& \frac{8 (d-2)^2 \left(-2 d^3+13 d^2-25 d+10\right) }{(d-4) (d-3) (d-1)^3}m^3G_N^3 \rho^3 \ .
\end{aligned}
\end{equation}
These results are in agreement with the post-Minkowskian expansion of the metric \eqref{hid} for the terms proportional to $ m^3G_N^3\rho^3$. In particular, they are all divergent in five dimensions, while $h_1^{(3,0)}$ and $h_2^{(3,0)}$ are also divergent in four dimensions.

In the case  $j=2$ we have to consider diagrams with two photons and one graviton emitted from the scalar line. There are in total three diagrams and we will compute the stress-energy tensor of each diagram with the right multiplicity and then sum all of them.
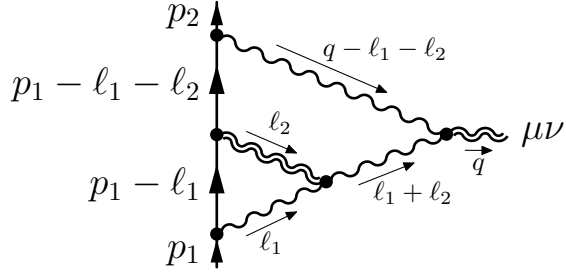
\begin{figure}[H]
    \raggedright
    \Large \encircle{a}
    \begin{minipage}{1\textwidth}
        \raggedright \begin{equation*}
\begin{fmffile}{saasaassa}
\begin{fmfgraph*}(120,100)
\DeclareGraphicsRule{*}{mps}{*}{}
\newcommand{\marrow}[5]{%
    \fmfcmd{style_def marrow#1 
    expr p = drawarrow subpath (1/4, 3/4) of p shifted 6 #2 withpen pencircle scaled 0.4; 
    label.#3 (btex #4 etex, point 0.5 of p shifted 6 #2); 
    enddef;}
    \fmf{marrow#1, tension=0}{#5}}
\unitlength=1mm
 \fmfleft{i1,i2}
 \fmfright{o}
  \fmf{fermion,label=$p_1$,l.s=left,tension=3}{i1,v2}
   \fmf{fermion,label=$p_1-\ell_1$,l.s=left}{v2,a}
  \fmf{fermion,label=$p_1-\ell_1-\ell_2$,l.s=left}{a,v1}
 \fmf{fermion,label=$p_2$,l.s=left,tension=3}{v1,i2}
 \fmffreeze
\fmf{photon,tension=0.05}{v1,v3}
\fmf{dbl_wiggly,tension=0.01}{a,b}
\fmf{photon,tension=0.1}{v2,b,v3}
\fmf{dbl_wiggly,tension=0.4}{v3,o}
\fmflabel{$\mu\nu$}{o}
\marrow{a}{down}{bot}{$q$}{v3,o}
\marrow{b}{down}{bot}{$\quad \quad \ell_1+\ell_2$}{b,v3}
\marrow{c}{up}{top}{$\qquad \qquad q-\ell_1-\ell_2$}{v1,v3}
\marrow{d}{up}{top}{$\ \ \ell_2$}{a,b}
\marrow{e}{down}{bot}{$ \ell_1$}{v2,b}
\fmfdot{a,b,v2,v1,v3}
\end{fmfgraph*}
\end{fmffile}
\end{equation*}
    \end{minipage}%
    \caption{2-loop diagram for  graviton emission with 1 internal graviton and 3 internal photons.}
    \label{figure:2loopChargedA}
\end{figure}
The first diagram is given in figure \ref{figure:2loopChargedA}, and its  contribution to the energy-momentum tensor is
\begin{equation}
    \begin{aligned}
    &-\frac{i\kappa}{2} \sqrt{4E_1 E_2} T^{(2,2)(\text{a})}_{\mu\nu}(q^2) =\\
    & \,6  \times\int \frac{d^{d+1}\ell_1}{(2\pi)^{d+1}}\frac{d^{d+1}\ell_2}{(2\pi)^{d+1}}\frac{\left(\tau_{\phi^2 A}\right)^\alpha
    \left(\tau_{\phi^2 A}\right)^\beta \left(\tau_{\phi^2 h}\right)^{\rho\sigma}}{\left((p_1-\ell_1)^2-m^2+i\epsilon\right)\left((p_1-\ell_1-\ell_2)^2-m^2+i\epsilon\right)} \\
    &\times\frac{P_{\rho\sigma,\eta\chi}
    \left(\tau_{A^2 h}\right)_{\mu\nu,\gamma,\alpha} (k,k')
    \left(\tau_{A^2 h}\right)^{\eta\chi}{}_{\beta}{}^\gamma (p,p')}{\left(\ell_1^2+i\epsilon\right)\left(\ell_2^2+i\epsilon\right)\left((\ell_1+\ell_2)^2+i\epsilon\right)\left((\ell_1+\ell_2-q)^2+i\epsilon\right)} 
    \ ,
    \end{aligned}
\end{equation}
where we introduced the new momenta 
\begin{equation}\label{TwoLoop_newmomenta}
\begin{aligned} 
    p&=\ell_1 &k&=\ell_1+\ell_2 \\
    p'&=\ell_1+\ell_2 &k'&=\ell_1+\ell_2-q \ .
\end{aligned}
\end{equation}
The factor multiplying the integral is the multiplicity of the diagram in figure \ref{figure:2loopChargedA}.
From the analysis  of section \ref{sec:amplitudes} one obtains
\begin{equation}
\begin{aligned}\label{T22aafterl0integral}
      T^{(2,2)(\text{a})}_{\mu\nu}(\vec{q}^2) =& mQ^2\int \frac{d^d\vec{\ell}_1}{(2\pi)^d}\frac{d^d\vec{\ell}_2}{(2\pi)^d} \frac{\left(\tau_{A^2h}\right)_{\mu\nu,\gamma,0}(k,k')P_{00,\eta\chi}\left(\tau_{A^2h}\right)^{\eta\chi}{}_{0}{}^\gamma(p,p')\bigl|_{\ell_i^0=0}}{\vec{\ell}_1^2\vec{\ell}_2^2(\vec{\ell}_1+\vec{\ell}_2-\vec{q})^2(\vec{\ell}_1+\vec{\ell}_2)^2} \ .
\end{aligned}
\end{equation}
Using the definition \eqref{2photon1graviton} of the vertex $\left(\tau_{A^2h}\right)$ the numerator becomes
\begin{equation}
\begin{aligned}
   \left(\tau_{A^2h}\right)_{\mu\nu,\gamma,0}(k,k')&P_{00,\eta\chi}\left(\tau_{A^2h}\right)^{\eta\chi}{}_{0}{}^\gamma(p,p')\bigl|_{\ell_i^0=0}=\\& -\kappa^2 \frac{d-2}{d-1}p\cdot p' \Biggl(k\cdot k' \left(\delta^0_\mu \delta^0_\nu - \frac{1}{2}\eta_{\mu\nu}\right) + \frac{1}{2}(k_\mu k'_\nu + k_\nu k'_\mu)\Biggl) \ .
\end{aligned}
\end{equation}
Substituting this in eq. \eqref{T22aafterl0integral} and using the identities in appendix \ref{App:LoopRed}, one obtains 
that the time-like component of the stress-energy tensor is
\begin{equation}
\begin{aligned}
     T^{(2,2)(\text{a})}_{00}(\vec{q}^2)
     =&\frac{1}{12}\frac{d-2}{d-1}mQ^2\kappa^2 J_{(2)}(\vec{q}^2) \ ,
\end{aligned}
\end{equation}
while for the trace of the space-like components one gets 
\begin{equation}
    \begin{aligned}
    \delta^{ij}T^{(2,2)(\text{a})}_{ij}(\vec{q}^2) =& \frac{1}{12}\frac{(d-2)^2}{d-1}mQ^2\kappa^2  J_{(2)}(\vec{q}^2) \ .
    \end{aligned}
\end{equation}

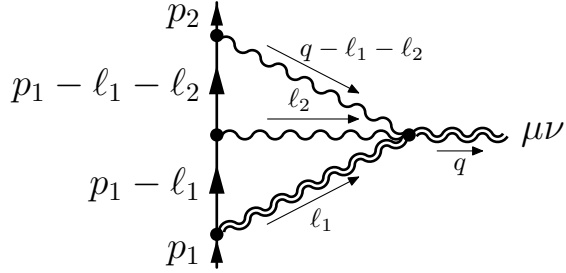
\begin{figure}[H]
    \raggedright
    \Large \encircle{b}
    \begin{minipage}{1\textwidth}
        \centering \begin{equation*}
\begin{fmffile}{22111ad}
\begin{fmfgraph*}(120,100)
\DeclareGraphicsRule{*}{mps}{*}{}
\newcommand{\marrow}[5]{%
    \fmfcmd{style_def marrow#1 
    expr p = drawarrow subpath (1/4, 3/4) of p shifted 6 #2 withpen pencircle scaled 0.4; 
    label.#3 (btex #4 etex, point 0.5 of p shifted 6 #2); 
    enddef;}
    \fmf{marrow#1, tension=0}{#5}}
\unitlength=1mm

 \fmfleft{i1,i2}
 \fmfright{o}
  \fmf{fermion,label=$p_1$,l.s=left,tension=3}{i1,v1}
   \fmf{fermion,label=$p_1-\ell_1$,l.s=left}{v1,a}
  \fmf{fermion,label=$p_1-\ell_1-\ell_2$,l.s=left}{a,v2}
 \fmf{fermion,label=$p_2$,l.s=left,tension=3}{v2,i2}
 \fmffreeze
\fmf{dbl_wiggly,tension=0.1}{v1,v3}
\fmf{photon,tension=0.1}{a,v3}
\fmf{photon,tension=0.1}{v2,v3}
\fmf{dbl_wiggly,tension=0.6}{v3,o}
\marrow{a}{down}{bot}{$q$}{v3,o}
\marrow{b}{down}{bot}{$\ \ \ell_1$}{v1,v3}
\marrow{c}{up}{top}{$\ell_2  \ \ \ $}{a,v3}
\marrow{d}{up}{top}{$\ \ \ \ \ \ \ \ \ \ \ q-\ell_1-\ell_2$}{v2,v3}
\fmflabel{$\mu\nu$}{o}
\fmfdot{v1,v2,v3,a}
\end{fmfgraph*}
\end{fmffile}
\end{equation*}
    \end{minipage}%
    \caption{2-loop diagram for graviton emission with 2 internal photons and 1 internal graviton.}    \label{fig:2loopChargedB}
\end{figure}
The second diagram that we consider is the one in figure \ref{fig:2loopChargedB}, which gives the contribution to the stress-energy tensor
\begin{equation}
    \begin{aligned}
    & -\frac{i\kappa}{2}\sqrt{4E_1 E_2} T^{(2,2)(\text{b})}_{\mu\nu}(q^2) =\\ &3\times  \int \frac{d^{d+1}\ell_1}{(2\pi)^{d+1}}\frac{d^{d+1}\ell_2}{(2\pi)^{d+1}}\frac{i\left(\tau_{\phi^2 A}\right)^\alpha
    \left(\tau_{\phi^2 A}\right)^\beta \left(\tau_{\phi^2 h}\right)^{\rho\sigma}}{\left((p_1-\ell_1)^2-m^2+i\epsilon\right)\left((p_1-\ell_1-\ell_2)^2-m^2+i\epsilon\right)} \\
    &\times\frac{P_{\rho\sigma,\gamma\delta}
    \left(\tau_{A^2 h^2}\right){}_{\mu\nu}{}^{\gamma\delta}{}_{\alpha,\beta} (p,p')}{\left(\ell_1^2+i\epsilon\right)\left(\ell_2^2+i\epsilon\right)\left((\ell_1+\ell_2-q)^2+i\epsilon\right)}\ ,
\end{aligned}\end{equation}
where we defined the momenta $p=\ell_2$ and $p' = \ell_1+\ell_2-q$ and where $(\tau_{A^2h^2})$ is the 2 photons - 2 gravitons vertex in \eqref{VertA2h2}.
Applying \eqref{Static_LowEnergy_Limit} and integrating the temporal component of the loop momenta we have
\begin{equation}
\begin{aligned} \label{T22bafterintegration}
     T^{(2,2)(\text{b})}_{\mu\nu}(\vec{q}^2) =&\,\frac{imQ^2}{2}\int \frac{d^d\vec{\ell}_1}{(2\pi)^d}\frac{d^d\vec{\ell}_2}{(2\pi)^d} \frac{P_{00,\gamma\delta} \left(\tau_{A^2 h^2}\right){}_{\mu\nu}{}^{\gamma\delta}{}_{0,0} (p,p')\bigl|_{\ell_i^0=0}}{\vec{\ell}_1^2\vec{\ell}_2^2(\vec{\ell}_1+\vec{\ell}_2-\vec{q})^2} \ .
\end{aligned}
\end{equation}
Using the vertex \eqref{VertA2h2} the numerator of the stress-energy tensor is
\begin{equation}
    \begin{aligned}
  &P_{00,\gamma\delta}\left(\tau_{A^2 h^2}\right){}_{\mu\nu}{}^{\gamma\delta}{}_{0,0} (p,p')\bigl|_{\ell_i^0=0} =\\&-\frac{i\kappa^2}{4}\left(\frac{2(d-3)}{d-1}\left(p_\mu p'_\nu + p_\nu p'_\mu-p\cdot p' \eta_{\mu\nu}\right) + \frac{2(3d-7)}{d-1}p\cdot p' \delta^0_\mu\delta^0_\nu\right) \ ,
\end{aligned}
\end{equation}
\noindent
and substituting back in eq. \eqref{T22bafterintegration}, the resulting components of the stress-energy tensor are 
\begin{equation}
\begin{aligned}
     T^{(2,2)(\text{b})}_{00}(\vec{q}^2)  =&-\frac{1}{12}\frac{d-2}{d-1} Q^2m\kappa^2J_{(2)}(\vec{q}^2) \ ,
\end{aligned}
\end{equation}
and
\begin{equation}
    \begin{aligned}
    \delta^{ij}T^{(2,2)(\text{b})}_{ij}(\vec{q}^2) =& -\frac{1}{24}\frac{(d-3)(d-2)}{d-1} mQ^2\kappa^2 J_{(2)}(\vec{q}^2) \ .
    \end{aligned}
\end{equation}

\begin{figure}[H]
    \raggedright\Large \encircle{c}
    \begin{minipage}{1\textwidth}
        \raggedright \begin{equation*}
\begin{fmffile}{2loopchargedgraphC}
\begin{fmfgraph*}(120,100)
\DeclareGraphicsRule{*}{mps}{*}{}
\newcommand{\marrow}[5]{%
    \fmfcmd{style_def marrow#1 
    expr p = drawarrow subpath (1/4, 3/4) of p shifted 6 #2 withpen pencircle scaled 0.4; 
    label.#3 (btex #4 etex, point 0.5 of p shifted 6 #2); 
    enddef;}
    \fmf{marrow#1, tension=0}{#5}}
\unitlength=1mm
 \fmfleft{i1,i2}
 \fmfright{o}
 \fmf{fermion,label=$p_1$,l.s=left,tension=3}{i1,v1}
   \fmf{fermion,label=$p_1-\ell_1$,l.s=left}{v1,a}
  \fmf{fermion,label=$p_1-\ell_1-\ell_2$,l.s=left}{a,v2}
 \fmf{fermion,label=$p_2$,l.s=left,tension=3}{v2,i2}
 \fmffreeze
\fmf{dbl_wiggly,tension=0.05}{v2,v3}
\fmf{photon,tension=0.01}{a,b}
\fmf{photon,tension=0.1}{v1,b}
\fmf{dbl_wiggly,tension=0.1}{b,v3}
\fmf{dbl_wiggly,tension=0.4}{v3,o}
\marrow{a}{down}{bot}{$q$}{v3,o}
\marrow{b}{down}{bot}{$\ \quad \ \ell_1+\ell_2$}{b,v3}
\marrow{c}{up}{top}{$\qquad \ \ \ \ \ q-\ell_1-\ell_2$}{v2,v3}
\marrow{d}{up}{top}{$\ \ \ell_2$}{a,b}
\marrow{e}{down}{bot}{$\ell_1$}{v1,b}
\fmfdot{a,b,v1,v2,v3}
\end{fmfgraph*}
\end{fmffile}
\end{equation*}
    \end{minipage}%
    \caption{2-loop diagram for graviton emission with 2 internal gravitons and 2 internal photons.} \label{fig:2loopChargedC}
\end{figure}
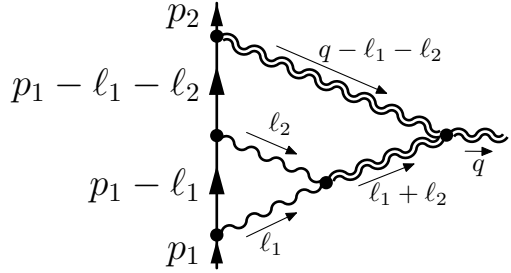
The last diagram for the emission of a graviton at 2-loop order is the one in figure \ref{fig:2loopChargedC}. The corresponding stress-energy tensor is 
\begin{equation}
    \begin{aligned}
    &-\frac{i\kappa}{2}\sqrt{4E_1 E_2} T^{(2,2)(\text{c})}_{\mu\nu}(q^2) =\\ 
    &3 \times  \int \frac{d^{d+1}\ell_1}{(2\pi)^{d+1}}\frac{d^{d+1}\ell_2}{(2\pi)^{d+1}}\frac{-\left(\tau_{\phi^2 A}\right)^\alpha
    \left(\tau_{\phi^2 A}\right)^\beta \left(\tau_{\phi^2 h}\right)^{\rho\sigma}}{\left((p_1-\ell_1)^2-m^2+i\epsilon\right)\left((p_1-\ell_1-\ell_2)^2-m^2+i\epsilon\right)} \\&\times
    \frac{P_{\rho\sigma}{}^{\eta\chi}P_{\theta\xi}{}^{\gamma\delta}
    \left(\tau_{A^2 h}\right)^{\theta\xi}{}_{\alpha,\beta} (p,p') \left(\tau_{h^2h}\right)_{\mu\nu,\gamma\delta,\eta\chi} (k,k')}{\left(\ell_1^2+i\epsilon\right)\left(\ell_2^2+i\epsilon\right)\left((\ell_1+\ell_2)^2+i\epsilon\right)\left((\ell_1+\ell_2-q)^2+i\epsilon\right)} 
    \ ,
    \end{aligned}
\end{equation}
where we have introduced the new momenta 
\begin{equation}\label{TwoLoop_newmomentaC}
\begin{aligned} 
    p&=\ell_1 &k&=\ell_1+\ell_2 \\
    p'&=-\ell_2 &k'&=q \ .
\end{aligned}
\end{equation}
Integrating over the time-like components of the loop momenta gives
\begin{equation}\label{Twoloop_grav_EMTnumB}
\begin{aligned}
     T^{(2,2)(\text{c})}_{\mu\nu}(\vec{q}^2) =&-\frac{mQ^2}{2}\int \frac{d^d\vec{\ell}_1}{(2\pi)^d}\frac{d^d\vec{\ell}_2}{(2\pi)^d} \frac{\left(\tau_{A^2h}\right)^{\alpha\beta}{}_{0,0}(p,p')P_{\alpha\beta}{}^{\gamma\delta}P_{00}{}^{\rho\sigma}\left(\tau_{h^2h}\right)_{\mu\nu,\gamma\delta,\rho\sigma} (k,k')\bigl|_{\ell_i^0=0}}{\vec{\ell}_1^2\vec{\ell}_2^2(\vec{\ell}_1+\vec{\ell}_2-\vec{q})^2(\vec{\ell}_1+\vec{\ell}_2)^2} \ .
\end{aligned}
\end{equation}
Introducing the vertex $\Tilde{\tau}_{h^2h}$ defined in \eqref{VertTildeDef} the numerator of the integrand can be written as
\begin{equation}
    \begin{aligned}
    \left(\tau_{A^2h}\right)^{\alpha\beta}{}_{0,0}(p,p')P_{\alpha\beta}{}^{\gamma\delta}P_{00}{}^{\rho\sigma}\left(\tau_{h^2h}\right)_{\mu\nu,\gamma\delta,\rho\sigma} (k,k') = \left(\tau_{A^2h}\right)^{\alpha\beta}{}_{0,0}(p,p')\left(\tilde{\tau}_{h^2h}\right)_{\mu\nu,\alpha\beta,00} (k,k') \ .
    \end{aligned}
\end{equation}
Using the vertexes \eqref{VertTilde} and \eqref{2photon1graviton} we obtain the expression of the numerator to be
\begin{equation}
    \begin{aligned}
     &\left(\tau_{A^2h}\right)^{\alpha\beta}{}_{0,0}(p,p')\left(\tilde{\tau}_{h^2h}\right)_{00,\alpha\beta,00} (k,k')\bigl|_{\ell_i^0=0}=\\ &\frac{\kappa^2}{4}\frac{1}{d-1}\left(p\cdot p' \left(-dk'^2+ \frac{(3d-3)(d-2)}{d-1}k^2+(d-2)(k-k')^2\right)+ (4d-8) (k'\cdot p) (k'\cdot p')\right)
    \end{aligned}
\end{equation}
for the $00$-component and 
\begin{equation}
    \begin{aligned}
     \delta^{ij}&\left(\tau_{A^2h}\right)^{\alpha\beta}{}_{0,0}(p,p')\left(\tilde{\tau}_{h^2h}\right)_{ij,\alpha\beta,00} (k,k')\bigl|_{\ell_i^0=0}=\\ &-\frac{\kappa^2}{4} \frac{d-2}{d-1} \left(p\cdot p'\left((d-2) k^2 +(d-4) (k-k')^2 + 3d k'^2\right) + 4 ( k'\cdot p) (k'\cdot p')\right)
    \end{aligned}
\end{equation}
for the trace of the spatial components.
Substituting in \eqref{Twoloop_grav_EMTnumB}, inserting the momenta \eqref{TwoLoop_newmomentaC} and using the master integral identities in appendix \ref{App:LoopRed} (in particular \eqref{LR_Loop2_l1q_l2q}) we have 
\begin{equation}
\begin{aligned}
     T^{(2,2)(\text{c})}_{00}(\vec{q}^2) =\frac{(3d^2-19d+32)}{24(d-4)(d-1)} mQ^2\kappa^2J_{(2)}(\vec{q}^2) \ ,
\end{aligned}
\end{equation}
and
\begin{equation}
\begin{aligned}
     \delta^{ij}T^{(2,2)(\text{c})}_{ij}(\vec{q}^2) =\frac{(d-2)(5d^2-27d+30)}{24(d-4)(d-1)}mQ^2\kappa^2J_{(2)}(\vec{q}^2) \ .
\end{aligned}
\end{equation}

Finally, summing all the contributions, we obtain the total energy-momentum tensor 
\begin{equation}
    T^{(2,2)}_{\mu\nu}(\vec{q}^2)=\sum_{i=\text{a}, \text{b}, \text{c}}T^{(2,2)(i)}_{\mu\nu}(\vec{q}^2)
\end{equation}
with
\begin{equation}\label{Twoloop_grav_EMTcomponents}
        \begin{aligned}
   T_{00}^{(2,2)}(\vec{q}^2) &= \frac{(3d^2-19d+32)}{24(d-4)(d-1)}mQ^2\kappa^2J_{(2)}(\vec{q}^2)
    \end{aligned}
\end{equation}
and
\begin{equation}
\delta^{ij}T^{(2,2)}_{ij} (\vec{q}^2)= \frac{ (d-2) (3 d^2-16 d+17)}{ 12(d-4) (d-1)}{mQ^2\kappa^2}J_{(2)}(\vec{q}^2) \ ,
\end{equation}
from which we can compute the coefficients $c^{(l,j)}_i(d)$, obtaining
\begin{equation}
        \begin{aligned}
   c_1^{(2,2)}(d) &= \frac{ 16\left(9 d^3-66 d^2+149 d-100\right)}{3(d-4) (d-1)^2} \\
        c_2^{(2,2)}(d) &=  \frac{32 (d-2) \left(3 d^2-16 d+17\right)}{3 (d-4) (d-1)^2} \ .
    \end{aligned}
\end{equation}
From these coefficients, using eq. \eqref{metric_pert} we finally obtain the metric components $h_i^{(3,2)}$ 
\begin{equation}\label{}
    \begin{aligned}
    h_0^{(3,2)} (r) =& -\frac{4(d-2)^2(3d-11)}{(d-4)(d-1)^2}m\alpha G_N^2{\rho}^3\\
    h_1^{(3,2)}(r)=&\frac{8(3d^3-25d^2+69d-65)}{3(d-4)(d-3)(d-1)^2} m\alpha G_N^2{\rho}^3\\
    h_2^{(3,2)}(r) = & -\frac{4(d-2)^2(3d^3-19d^2+33d-17)}{(d-4)(d-3)(d-1)^3}m\alpha G_N^2 {\rho}^3 \ .
    \end{aligned}
\end{equation}
We can compare this result with the part proportional to $ m\alpha G_N^2$ of \eqref{hid} and observe that they match exactly. In particular, all these terms diverge in $d=4$, and $ h_1^{(3,2)}$ and $  h_2^{(3,2)} $ also diverge in $d=3$.

As a summary, we have shown that  the loop computations give exactly the expression for the metric of the Reissner-Nordstr\"om-Tangherlini solution in de Donder gauge given in eq. \eqref{hid}. 
At third post-Minkowskian order, in $d=3$ only $h_1$ and $h_2$ diverge while in $d=4$ all the components diverge. To obtain the values of the metrics component in these dimensions we need to perform a renormalisation process which will be described in  section \ref{sec:divergences}. In the next section, we first proceed to compute the gauge potential from scattering amplitudes.

\section{Electromagnetic potential from scattering amplitudes}\label{sec:gaugepotential}
In this section we perform the  computation of the photon emission process in \eqref{SpecificLoop_PhotonEmission_Process} up to 2-loop order, and we recover the expression of the electromagnetic potential by directly evaluating eq. \eqref{ClassicalEM_potential_withCurrent} from  the electromagnetic current. We show the appearance of divergences at 2 loops in $d=4$, which will be treated in detail in section \ref{sec:divergences}. As already noticed in section \ref{sec:amplitudes}, only the time component of the gauge potential is non vanishing, so in the following it will be the only component taken into account. 

\subsection*{Tree level}
As for the metric, we start the analysis at tree level for completeness. In the notation of section \ref{sec:amplitudes}, the tree-level diagram corresponds to the term $l=0,\ j=0$ in the expansion \eqref{SpecificLoop_PhotonEmission_Process}.
\begin{figure}[h]
\newcommand{\marrow}[5]{%
    \fmfcmd{style_def marrow#1 
    expr p = drawarrow subpath (1/4, 3/4) of p shifted 6 #2 withpen pencircle scaled 0.4; 
    label.#3 (btex #4 etex, point 0.5 of p shifted 6 #2); 
    enddef;}
    \fmf{marrow#1, tension=0}{#5}}
\centering
    \begin{minipage}{.3\textwidth} 
    \begin{equation*}
    \begin{fmffile}{vertexphotonemissiontree}
    \begin{fmfgraph*}(70,50)
    \fmfleft{i1,i2}
    \fmfright{o}
    \fmf{fermion,label=$p_1$}{i1,v}
    \fmf{fermion,label=$p_2$,l.s=right}{v,i2}
    \fmf{photon,l.s=left}{v,o}
    \marrow{a}{down}{bot}{$q$}{v,o}
    \fmflabel{${\nu}$}{o}
    \fmfdot{v}
    \end{fmfgraph*}
    \end{fmffile}
    \end{equation*}
    \end{minipage}
    \caption{Tree-level diagram for photon emission.}
    \label{figure:TreeLevel_PhotonEmission}
\end{figure}
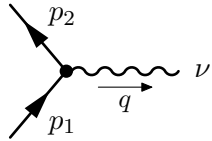
Evaluating the matrix element of the current due to the diagram in figure \ref{figure:TreeLevel_PhotonEmission}, and using the Feynman rules in section \ref{App:FeynmanRules}, one obtains 
\begin{equation}
    -i \, \sqrt{4E_1 E_2}\, j_\nu^{(0, 0)}(\vec{q}^2) = \left(\tau_{\phi^2 A}\right)_\nu\ .
\end{equation}
Considering the classical limit, the tree level electromagnetic current is simply the charge,
\begin{equation}
    j_0^{(0, 0)}(\vec{q}^2) = Q\ ,
\end{equation}
and replacing it inside eq. \eqref{ClassicalEM_potential_withCurrent} and using the Fourier transform idendities of the master integral in appendix \ref{App:masterintegrals}, we get
\begin{equation}
    A_0^{(1, 0)}(r) =\frac{Q}{4\pi}\rho\ ,
\end{equation}
which is exactly the first order of the expression in \eqref{EM_Potential_deDonder}.

\subsection*{1-loop}
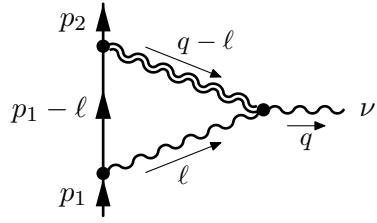
\begin{figure}[h]
    \centering
    \begin{minipage}{0.5\textwidth}
        \centering \begin{equation*}
\begin{fmffile}{oneloopphotonemission}
\begin{fmfgraph*}(100,80)
\DeclareGraphicsRule{*}{mps}{*}{}
\newcommand{\marrow}[5]{%
    \fmfcmd{style_def marrow#1 
    expr p = drawarrow subpath (1/4, 3/4) of p shifted 6 #2 withpen pencircle scaled 0.4; 
    label.#3 (btex #4 etex, point 0.5 of p shifted 6 #2); 
    enddef;}
    \fmf{marrow#1, tension=0}{#5}}
\unitlength=1mm
 \fmfleft{i1,i2}
 \fmfright{o}
  \fmf{fermion,tension=3,label=$p_1$,l.s=left}{i1,v1}
 \fmf{fermion,tension=3,label=$p_2$,l.s=left}{v2,i2}
  \fmf{fermion,label=$p_1 - \ell$,l.s=left}{v1,v2}
  \fmffreeze
  \fmf{photon}{v1,v3}
 \fmf{dbl_wiggly,l.s=left}{v3,v2}
 \fmf{photon,tension=4}{v3,o}
 \fmflabel{${\nu}$}{o}
 \marrow{a}{down}{bot}{$q$}{v3,o}
 \marrow{b}{up}{top}{$\ \ \ \ \ q-\ell $}{v2,v3}
 \marrow{c}{down}{bot}{$\ell$}{v1,v3}
 \fmfdot{v1,v2,v3}
\end{fmfgraph*}
\end{fmffile}
\end{equation*}
    \end{minipage}
    \caption{1-loop diagram for photon emission.}
    \label{figure:1Loop_PhotonEmission}
\end{figure}

At this loop order we can only have $j=0$. So the 1-loop process is given by the amplitude in figure \ref{figure:1Loop_PhotonEmission}.
\noindent
The electromagnetic current associated to this process reads
\begin{equation}
-i \, \sqrt{4E_1E_2}\, j_{\nu}^{(1, 0)}(q^2) = 2\times \int \frac{d^{d+1}\ell}{(2\pi)^{d+1}} \frac{i\, P_{\alpha\beta, \sigma\rho}\left(\tau_{\phi^{2} A}\right)^{\mu}\left(\tau_{\phi^{2} h}\right)^{\alpha\beta}\left(\tau_{A^{2} h}\right)^{\sigma\rho}{}_{\mu,\nu}\left(\ell, q\right)}{\left((p_1-\ell)^2-m^2+i\epsilon\right)\left(\ell^2+i\epsilon\right)\left((q-\ell)^2+i\epsilon\right)}\ ,
\end{equation}
where a factor 2 is implied in order to consider the multiplicity of the diagram. Performing the classical limit to obtain an expression like eq. \eqref{Electro_Current_Amplitudes}, one gets
\begin{equation}
j_{0}^{(1, 0)}(\vec{q}^2) = -\frac{i}{2} \kappa Q m \int \frac{d^d\ell}{(2\pi)^d} \frac{P_{00, \alpha \beta}\left(\tau_{A^{2} h}\right)^{\alpha \beta}{}_{0,0}\left(\ell, q\right)\big|_{\ell^0=0}}{\vec{\ell}^2(\vec{q}-\vec{\ell})^2}\ .
\end{equation}
Considering the tensorial contraction 
\begin{equation}
    P_{00, \alpha \beta}\left(\tau_{A^{2} h}\right)^{\alpha \beta}{}_{0,0}\left(\ell, q\right)\big|_{\ell^0=0}=-i\kappa\frac{d-2}{d-1}\Vec{\ell}\cdot\Vec{q}\ ,
\end{equation}
%the electromagnetic current takes the form
%\begin{equation}
%j_0^{(1, 0)}= -\frac{1}{2}m Q\kappa^2\frac{d-2}{d-1} \int \frac{d^d\ell}{(2\pi)^d} \frac{\vec{\ell}\cdot\vec{q}}{\vec{\ell}^2(\vec{q}-\vec{\ell})^2}\ .
%\end{equation}
we can express the current in term of the master integral through the identities in appendix \ref{App:LoopRed}. In particular using (\ref{LR_1Loop_1}) one end up with the compact expression
\begin{equation}
    j_0^{(1, 0)}(\vec{q}^2)= -\frac{1}{4}m Q\kappa^2\frac{d-2}{d-1}J_{(1)}(\Vec{q}^2)\ ,
\end{equation}
recovering eq. \eqref{Electro_Current_FormFactor}, from which it is possible to extract the form factor $c^{(1, 0)}(d)$.
Replacing the above expression inside the (\ref{ClassicalEM_potential_withCurrent}), exploiting the property \eqref{AppIdent1}, one obtains
\begin{equation}
A_0^{(2, 0)}(r) = - \frac{Q}{4\pi} \frac{2(d-2)}{d-1}m G_N \rho^2\ ,
\end{equation}
which shows, at this order, a perfect agreement with the expression \eqref{EM_Potential_deDonder}.

\subsection*{2-loop}
\begin{figure}[h]
    \raggedright
    \Large \encircle{a}
    \begin{minipage}{1\textwidth}
        \raggedright \begin{equation*}
\begin{fmffile}{PhotonEMission_2Loop_2Photon2Gravitons}
\begin{fmfgraph*}(120,100)
\DeclareGraphicsRule{*}{mps}{*}{}
\newcommand{\marrow}[5]{%
    \fmfcmd{style_def marrow#1 
    expr p = drawarrow subpath (1/4, 3/4) of p shifted 6 #2 withpen pencircle scaled 0.4; 
    label.#3 (btex #4 etex, point 0.5 of p shifted 6 #2); 
    enddef;}
    \fmf{marrow#1, tension=0}{#5}}
\unitlength=1mm
 \fmfleft{i1,i2}
 \fmfright{o}
 \fmf{fermion,label=$p_1$,l.s=left,tension=2}{i1,v1}
   \fmf{fermion,label=$p_1-\ell_1$,l.s=left}{v1,a}
  \fmf{fermion,label=$p_1-\ell_1-\ell_2$,l.s=left}{a,v2}
 \fmf{fermion,label=$p_2$,l.s=left,tension=2}{v2,i2}
 \fmffreeze
\fmf{dbl_wiggly,tension=0.05}{v2,v3}
\fmf{dbl_wiggly,tension=0.01}{a,b}
\fmf{photon,tension=0.1}{v1,b}
\fmf{photon,tension=0.1}{b,v3}
\fmf{photon,tension=0.4}{v3,o}
\marrow{a}{down}{bot}{$q$}{v3,o}
\marrow{b}{down}{bot}{$\ \ \ \ \ \ \ell_1+\ell_2$}{b,v3}
\marrow{c}{down}{bot}{$\ell_1$}{v1,b}
\marrow{d}{up}{top}{$\ell_2$}{a,b}
\marrow{e}{up}{top}{$\ \ \ \ \quad \ \ \ \ \ \ q-\ell_1-\ell_2$}{v2,v3}
\fmfdot{a,b,v1,v2,v3}
 \fmflabel{${\nu}$}{o}
\end{fmfgraph*}
\end{fmffile}
\end{equation*}
    \end{minipage}
    \caption{2-loop diagram for photon emission with 2 internal photons and 2 internal gravitons.}
    \label{fig:PhotonEMission_2Loop_2Photon2Gravitons}
\end{figure}
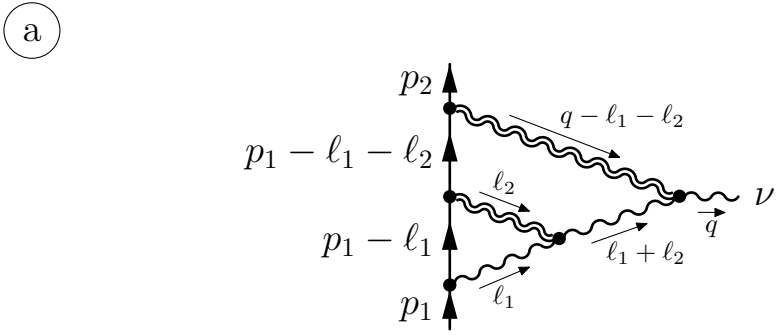

At 2 loops we can have either $j=0$ or $j=2$. We first discuss the case $j=0$, for which there are  three different diagrams, which differ by the internal tree structure. The first amplitude we consider is the one in figure \ref{fig:PhotonEMission_2Loop_2Photon2Gravitons}, which has 2 internal photons and 2 internal gravitons. 
\noindent
The associated electromagnetic current to this process is 
\begin{multline}
    -i \, \sqrt{4E_1E_2}\, j_\nu^{(2, 0)(\text{a})}(q^2)= 6\times \int \frac{d^{d+1} \ell_1}{(2\pi)^{d+1}}\frac{d^{d+1} \ell_2}{(2\pi)^{d+1}} \frac{-P_{\sigma\rho, \alpha\beta}P_{\gamma\delta, \eta\chi}\left(\tau_{\phi^2A}\right)_{\lambda}\left(\tau_{\phi^2h} \right)^{\eta\chi}\left(\tau_{\phi^2h}\right)^{\sigma\rho}}{\left(\ell_1^2+i\epsilon\right)\left(\ell_2^2+i\epsilon\right)\left((q-\ell_1-\ell_2)^2+i\epsilon\right)} \\
    \times\frac{\left(\tau_{A^2h}\right)^{\gamma\delta,\mu,\lambda}(\ell_1, \ell_1+\ell_2)\left(\tau_{A^2h} \right)^{\alpha\beta}{}_{\mu,\nu}(\ell_1+\ell_2, q)}{\left((\ell_1+\ell_2)^2+i\epsilon\right)((p_1-\ell_1)^2-m^2+i\epsilon)((p_1-\ell_1-\ell_2)^2-m^2+i\epsilon)}\ .
\end{multline}
Exploiting the procedure outlined in section \ref{sec:amplitudes}, one gets
\begin{equation}
\begin{aligned}
    j_0^{(2, 0)(\text{a})}(\vec{q}^2)&=-\frac{\kappa^2 Q m^2}{4}\int \frac{d^d \ell_1}{(2\pi)^d}\frac{d^d \ell_2}{(2\pi)^d}\frac{P^{\alpha \beta, 00}P^{\sigma \rho, 00}}{\vec{\ell_1}^2\vec{\ell_2}^2(\vec{q}-\vec{\ell_1}-\vec{\ell_2})^2(\vec{\ell_1}+\vec{\ell_2})^2}\\
    &\times(\tau_{A^2h})_{0\mu, \alpha \beta}(\ell_1, \ell_1+\ell_2)(\tau_{A^2h})^{0\mu}{}_{\sigma \rho}(\ell_1+\ell_2, q)\bigl|_{\ell_i^0=0}\ ,
\end{aligned}
\end{equation}
where the tensor contraction gives
\begin{equation}
\begin{aligned}
   P^{\alpha \beta, 00}&P^{\sigma \rho, 00}(\tau_{A^2h})_{0,\mu, \alpha \beta}(\ell_1, \ell_1+\ell_2)(\tau_{A^2h})^{0,\mu}{}_{\sigma \rho}(\ell_1+\ell_2q)\bigl|_{\ell_i^0=0}\\&=-\kappa^2\frac{(d-2)^2}{(d-1)^2}(\Vec{\ell_1}+\Vec{\ell_2})\cdot\Vec{q}\,(\Vec{\ell_1}+\Vec{\ell_2})\cdot \Vec{\ell_1}\ .
   \end{aligned}
\end{equation}
%\begin{equation}
 %   -i\, j_0^{(2, 0)}{}^{(\text{a})}=-i\, \frac{\kappa^4Qm^3}{2}\frac{(d-2)^2}{(d-1)^2}\int \frac{d^d \ell_1}{(2\pi)^d}\frac{d^d \ell_2}{(2\pi)^d}\frac{\vec{\ell_1}\cdot(\Vec{\ell_1}+\Vec{\ell_2})\ \Vec{q}\cdot(\vec{\ell_1}+\vec{\ell_2})}{\vec{\ell_1}^2\vec{\ell_2}^2(\vec{q}-\vec{\ell_1}-\vec{\ell_2})^2(\vec{\ell_1}+\vec{\ell_2})^2}\ , 
%\end{equation}
Then using the reduction identities in appendix \ref{App:LoopRed}, one gets the current in term of the master integral as
\begin{equation}
    j_0^{(2, 0)(\text{a})}(\vec{q}^2)= \frac{\kappa^4Qm^2}{12}\frac{(d-2)^2}{(d-1)^2}J_{(2)}(\Vec{q}^2)\ , 
\end{equation}
from which, due to the usual relations of the Fourier transform of the master integral, one recovers 
\begin{equation}
    A_0^{(3, 0)(\text{a})}(r)=\frac{Q}{4\pi}m^2G_N^2\frac{16}{3}\frac{(d-2)^2}{(d-1)^2}\rho^3\ .
\end{equation}

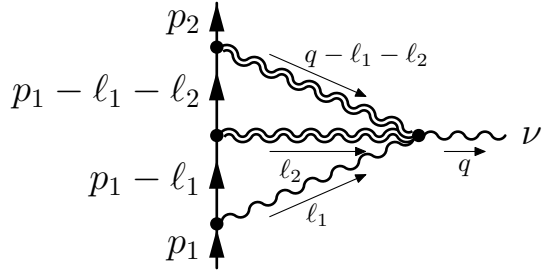
\begin{figure}[h]
    \raggedright
    \Large \encircle{b}
    \begin{minipage}{1\textwidth}
        \centering \begin{equation*}
\begin{fmffile}{221ssss11ad}
\begin{fmfgraph*}(120,100)
\DeclareGraphicsRule{*}{mps}{*}{}
\newcommand{\marrow}[5]{%
    \fmfcmd{style_def marrow#1 
    expr p = drawarrow subpath (1/4, 3/4) of p shifted 6 #2 withpen pencircle scaled 0.4; 
    label.#3 (btex #4 etex, point 0.5 of p shifted 6 #2); 
    enddef;}
    \fmf{marrow#1, tension=0}{#5}}
\unitlength=1mm
 \fmfleft{i1,i2}
 \fmfright{o}
  \fmf{fermion,label=$p_1$,l.s=left,tension=2}{i1,v1}
   \fmf{fermion,label=$p_1-\ell_1$,l.s=left}{v1,a}
  \fmf{fermion,label=$p_1-\ell_1-\ell_2$,l.s=left}{a,v2}
 \fmf{fermion,label=$p_2$,l.s=left,tension=2}{v2,i2}
 \fmffreeze
\fmf{photon,tension=0.1}{v1,v3}
\fmf{dbl_wiggly,tension=0.1}{a,v3}
\fmf{dbl_wiggly,tension=0.1}{v2,v3}
\fmf{photon,tension=0.7}{v3,o}
\marrow{a}{down}{bot}{$q$}{v3,o}
\marrow{b}{down}{bot}{$\ell_1$}{v1,v3}
\marrow{c}{down}{bot}{$\ell_2 \quad \quad$}{a,v3}
\marrow{d}{up}{top}{$\ \ \ \ \ \ \ \ \ \ \ q-\ell_1-\ell_2$}{v2,v3}
\fmfdot{v1,v2,a,v3}
 \fmflabel{${\nu}$}{o}
\end{fmfgraph*}
\end{fmffile}
\end{equation*}
    \end{minipage}
    \caption{2-loop diagram for photon emission with 1 internal photon and 2 internal gravitons.}
    \label{fig:PhotonEMission_2Loop_1Photon2Gravitons}
\end{figure}

The second amplitude we consider is the one in figure \ref{fig:PhotonEMission_2Loop_1Photon2Gravitons}, which has 1 internal photon and 2 internal gravitons.
\noindent
The electromagnetic current associated to the diagram is 
\begin{equation}
\begin{aligned}
    -i \, \sqrt{4E_1E_2}\, &j_\nu^{(2, 0)(\text{b})}(q^2)= 3\times \int \frac{d^{d+1} \ell_1}{(2\pi)^{d+1}}\frac{d^{d+1} \ell_2}{(2\pi)^{d+1}} \frac{-iP_{\sigma \rho, \alpha\beta}P_{\gamma\delta, \eta\chi}\left(\tau_{\phi^2A}\right)^{\mu}\left(\tau_{\phi^2h} \right)^{\eta\chi}\left(\tau_{\phi^2h}\right)^{\sigma \rho}}{\left(\ell_1^2+i\epsilon\right)\left(\ell_2^2+i\epsilon\right)\left((q-\ell_1-\ell_2)^2+i\epsilon\right)}\\
    &\times\frac{\left(\tau_{A^2h^2}\right)^{\alpha\beta, \gamma\delta}{}_{\mu,\nu}(\ell_1, q)}{((p_1-\ell_1)^2-m^2+i\epsilon)((p_1-\ell_1-\ell_2)^2-m^2+i\epsilon)}\ .
    \end{aligned}
\end{equation}
Performing the classical limit, one gets
\begin{equation}
    j_0^{(2, 0)(\text{b})}(\vec{q}^2)= \frac{i}{8}\kappa^2m^2Q \int \frac{d^d\ell_1}{(2\pi)^d} \frac{d^d\ell_2}{(2\pi)^d} \frac{P_{00, \mu\nu}P_{00, \alpha\beta}(\tau_{A^2h^2})^{0,0, \mu\nu, \alpha\beta}(\ell_1, q)\bigl|_{\ell_i^0=0}}{\vec{\ell_1}^2 \vec{\ell_2}^2\left(\vec{q}-\vec{\ell_1}-\vec{\ell_2}\right)^2}\ ,
\end{equation}
where the tensorial contraction of the numerator leads to 
\begin{equation}
    P_{00, \mu\nu}P_{00, \alpha\beta}(\tau_{A^2h^2})^{0,0, \mu\nu, \alpha\beta}(\ell_1, q)\bigl|_{\ell_i^0=0}=i\frac{\kappa^2}{2}\frac{(d-2)(3d-7)}{(d-1)^2}\Vec{\ell_1}\cdot\Vec{q} \ .
\end{equation}
Then using the relation (\ref{LiteRed_2Loop_2}), the current in term of the master integral reads
\begin{equation}
    j_0^{(2, 0)(\text{b})}(\vec{q}^2)= -\frac{m^2Q\kappa^4}{48}\frac{(d-2)(3d-7)}{(d-1)^2} J_{(2)}(\Vec{q}^2)\ ,
\end{equation}
from which considering the usual Fourier transform one gets
\begin{equation}
    A_0^{(3, 0)(\text{b})}(r)=-\frac{Q}{4\pi}m^2G_N^2\frac{4}{3}\frac{(d-2)(3d-7)}{(d-1)^2}\rho^3\ .
\end{equation}

\begin{figure}[h]
    \raggedright
    \Large \encircle{c}
    \begin{minipage}{1\textwidth}
        \raggedright \begin{equation*}
\begin{fmffile}{saasccaaaaascacccaassa}
\begin{fmfgraph*}(120,100)
\DeclareGraphicsRule{*}{mps}{*}{}
\newcommand{\marrow}[5]{%
    \fmfcmd{style_def marrow#1 
    expr p = drawarrow subpath (1/4, 3/4) of p shifted 6 #2 withpen pencircle scaled 0.4; 
    label.#3 (btex #4 etex, point 0.5 of p shifted 6 #2); 
    enddef;}
    \fmf{marrow#1, tension=0}{#5}}
\unitlength=1mm
 \fmfleft{i1,i2}
 \fmfright{o}
 \fmf{fermion,label=$p_1$,l.s=left,tension=3}{i1,v1}
   \fmf{fermion,label=$p_1-\ell_1$,l.s=left}{v1,a}
  \fmf{fermion,label=$p_1-\ell_1-\ell_2$,l.s=left}{a,v2}
 \fmf{fermion,label=$p_2$,l.s=left,tension=3}{v2,i2}
 \fmffreeze
\fmf{photon,tension=0.05}{v2,v3}
\fmf{dbl_wiggly,tension=0.01}{a,b}
\fmf{dbl_wiggly,tension=0.1}{v1,b}
\fmf{dbl_wiggly,tension=0.1}{b,v3}
\fmf{photon,tension=0.4}{v3,o}
\marrow{a}{down}{bot}{$q$}{v3,o}
\marrow{b}{down}{bot}{$\ \ \ \ \ \  \ell_1+\ell_2$}{b,v3}
\marrow{c}{down}{bot}{$\ell_1$}{v1,b}
\marrow{d}{up}{top}{$\ell_2$}{a,b}
\marrow{e}{up}{top}{$\ \ \ \ \quad \ \ \ \ \ \ q-\ell_1-\ell_2$}{v2,v3}
\fmfdot{v1,v2,v3,a,b}
 \fmflabel{${\nu}$}{o}
\end{fmfgraph*}
\end{fmffile}
\end{equation*}
    \end{minipage}
    \caption{2-loop diagram for photon emission with 1 internal photon and 3 internal gravitons.}
    \label{fig:PhotonEMission_2Loop_1Photon3Gravitons}
\end{figure}
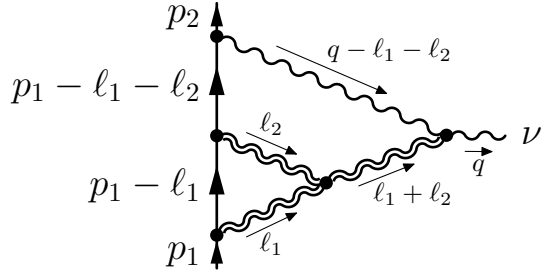

The last diagram that contributes to the process with $l=2, \ j=0$ is the one with an internal 3 graviton vertex, as shown in figure \ref{fig:PhotonEMission_2Loop_1Photon3Gravitons}. 
\noindent
Since the 3 graviton vertex is internal, we must use the expression given in  \eqref{VertDeWitt} as discussed in appendix \ref{App:FeynmanRules}. The current associated to this diagram is
\begin{equation}
\begin{aligned}
    & -i \, \sqrt{4E_1E_2}\, j_\nu^{(2, 0)(\text{c})}(q^2)= 3\times \int \frac{d^{d+1} \ell_1}{(2\pi)^{d+1}}\frac{d^{d+1} \ell_2}{(2\pi)^{d+1}} \frac{P_{\alpha_1\beta_1, \gamma \delta}P_{\alpha_2\beta_2, \eta\chi}P_{\alpha_3\beta_3, \sigma\rho}\left(\tau_{\phi^2A}\right)^{\mu}}{\left(\ell_1^2+i\epsilon\right)\left(\ell_2^2+i\epsilon\right)\left((q-\ell_1-\ell_2)^2+i\epsilon\right)} \\
    & \times\frac{\left(\tau_{\phi^2h} \right)^{\gamma\delta}\left(\tau_{\phi^2h}\right)^{\eta\chi}\left(\tau_{h^3}\right)^{\alpha_1\beta_1,\alpha_2\beta_2, \alpha_3\beta_3}(\ell_1, \ell_2, -\ell_1-\ell_2)\left(\tau_{A^2h} \right)^{\sigma\rho}{}_{\mu, \nu}(q-\ell_1-\ell_2, q)}{\left((\ell_1+\ell_2)^2+i\epsilon\right)((p_1-\ell_1)^2-m^2+i\epsilon)((p_1-\ell_1-\ell_2)^2-m^2+i\epsilon)}\ .
    \end{aligned}
\end{equation}
Referring to eq. \eqref{Electro_Current_Amplitudes}, one gets 
\begin{equation}
\begin{aligned}
    j_0^{(2, 0)(\text{c})}(\vec{q}^2)& = \frac{1}{8}\kappa^2Qm^2\int \frac{d^d \ell_1}{(2\pi)^d}\frac{d^d \ell_2}{(2\pi)^d}\frac{P_{00, \mu\nu}P_{00, \alpha\beta}P_{\gamma \delta, \sigma \rho}}{\vec{\ell_1}^2\vec{\ell_2}^2(\vec{q}-\vec{\ell_1}-\vec{\ell_2})^2(\vec{\ell_1}+\vec{\ell_2})^2}\\
    &\times \left(\tau_{h^3}\right)^{\mu\nu, \alpha\beta, \gamma\delta}(\ell_1, \ell_2, -\ell_1-\ell_2)\left(\tau_{A^2h} \right)^{\sigma \rho}{}_{0,0}(q-\ell_1-\ell_2, q)\bigl|_{\ell_i^0=0}\ ,
    \end{aligned}
\end{equation}
where considering from the beginning the fact that all the expressions that contain internal momenta are symmetric under the exchange of $\ell_1\leftrightarrow \ell_2$, the tensor contraction at the numerator reads
\begin{equation}
    \begin{aligned}
    &P_{\gamma\delta, \sigma\rho}P_{\mu\nu, 00}P_{\alpha\beta, 00}\left(\tau_{h^3}\right)^{\mu\nu, \alpha\beta, \gamma\delta}(\ell_1, \ell_2, -\ell_1-\ell_2)\left(\tau_{A^2h} \right)^{\sigma \rho}{}_{0,0}(q-\ell_1-\ell_2, q)\bigl|_{\ell_i^0=0}\\
    &=-\frac{\kappa^2}{4}\frac{d-2}{(d-1)^2}\Biggl(4(d-1)\Vec{\ell_1}\cdot (\Vec{q}-\vec{\ell_1}-\vec{\ell_2}) \ \vec{\ell_1}\cdot \vec{q}+2(d-1)\Vec{\ell_1}\cdot (\Vec{q}-\vec{\ell_1}-\vec{\ell_2}) \ \vec{\ell_2}\cdot \vec{q} \\
    &+(d-5)\Vec{q}\cdot (\Vec{q}-\vec{\ell_1}-\vec{\ell_2}) \ \vec{\ell_1}\cdot \vec{\ell_2}\Biggl)\ .
    \end{aligned}
\end{equation}
Exploiting the expressions in appendix \ref{App:LoopRed}, the current  finally  reads
\begin{equation}
    j_0^{(2, 0)(\text{c})}(\vec{q}^2) = \frac{\kappa^4m^2Q}{96}\frac{(d-2)(34-29d+7d^2)}{(d-4)(d-1)^2}J_{(2)}(\Vec{q}^2)\ ,
\end{equation}
from which
\begin{equation}
    A_0^{(3, 0)(\text{c})}(r) = \frac{Q}{4\pi} m^2G_N^2 \frac{2}{3}\frac{(d-2)(34-29d+7d^2)}{(d-4)(d-1)^2}\rho^3\ .
\end{equation}
At the end we can sum up all the $j=0$  pieces to obtain 
\begin{equation}
    A_0^{(3, 0)}(r)=\sum_{i=\text{a}, \text{b}, \text{c}}A_0^{(3, 0)(i)}(r)=\frac{Q}{4\pi} m^2G_N^2 \frac{2(d-2)^2(3d-7)}{(d-1)^2(d-4)}\rho^3\ ,
\end{equation}
which is exactly the contribution that appears in (\ref{EM_Potential_deDonder}). In this last expression a divergence in $d=4$ arises. These singularities have to be renormalised, and in section \ref{sec:divergences} the general procedure  to cure them will be shown.

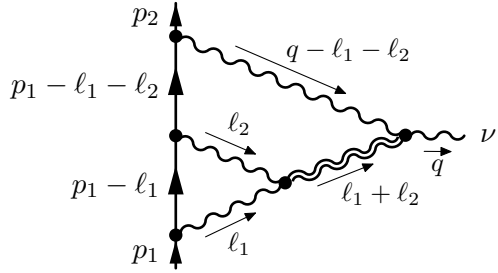
\begin{figure}[h]
    \begin{minipage}{1\textwidth}
        \raggedright \begin{equation*}
\begin{fmffile}{saasccaacccaassa}
\begin{fmfgraph*}(120,100)
\DeclareGraphicsRule{*}{mps}{*}{}
\newcommand{\marrow}[5]{%
    \fmfcmd{style_def marrow#1 
    expr p = drawarrow subpath (1/4, 3/4) of p shifted 6 #2 withpen pencircle scaled 0.4; 
    label.#3 (btex #4 etex, point 0.5 of p shifted 6 #2); 
    enddef;}
    \fmf{marrow#1, tension=0}{#5}}
\unitlength=1mm
 \fmfleft{i1,i2}
 \fmfright{o}
 \fmf{fermion,label=$p_1$,l.s=left,tension=3}{i1,v1}
   \fmf{fermion,label=$p_1-\ell_1$,l.s=left}{v1,a}
  \fmf{fermion,label=$p_1-\ell_1-\ell_2$,l.s=left}{a,v2}
 \fmf{fermion,label=$p_2$,l.s=left,tension=3}{v2,i2}
 \fmffreeze
\fmf{photon,tension=0.05}{v2,v3}
\fmf{photon,tension=0.01}{a,b}
\fmf{photon,tension=0.1}{v1,b}
\fmf{dbl_wiggly,tension=0.1}{b,v3}
\fmf{photon,tension=0.4}{v3,o}
\marrow{a}{down}{bot}{$q$}{v3,o}
\marrow{b}{down}{bot}{$\ \ \ \ \ \ \ \ \ell_1+\ell_2$}{b,v3}
\marrow{c}{down}{bot}{$\ \ \ell_1$}{v1,b}
\marrow{d}{up}{top}{$\ \ \ell_2$}{a,b}
\marrow{e}{up}{top}{$\ \ \ \ \quad \ \ \ \ \ \ q-\ell_1-\ell_2$}{v2,v3}
 \fmflabel{${\nu}$}{o}
\fmfdot{v1,v2,v3,a,b}
\end{fmfgraph*}
\end{fmffile}
\end{equation*}
    \end{minipage}
      \caption{2-loop diagram for photon emission with 3 internal photons and 1 internal graviton.}
    \label{fig:PhotonEMission_2Loop_3Photon1Gravitons}
\end{figure}

Finally, the only process that contributes at $l=2, \ j=2$ is the one which has 3 internal photons and 1 internal graviton, shown in figure \ref{fig:PhotonEMission_2Loop_3Photon1Gravitons}.
\noindent
The electromagentic current associated to this process is then
\begin{equation}
    \begin{aligned}
    -i \, \sqrt{4E_1E_2}\, &j_\nu^{(2, 2)}(q^2)= 3\times \int \frac{d^{d+1} \ell_1}{(2\pi)^{d+1}}\frac{d^{d+1} \ell_2}{(2\pi)^{d+1}} \frac{P_{\gamma\delta, \sigma\alpha}\left(\tau_{\phi^2A}\right)^{\beta}\left(\tau_{\phi^2A}\right)^{\sigma}\left(\tau_{\phi^2A}\right)^{\rho}}{\left(\ell_1^2+i\epsilon\right)\left(\ell_2^2+i\epsilon\right)\left((q-\ell_1-\ell_2)^2+i\epsilon\right)} \\
    &\times\frac{\left(\tau_{A^2h}\right)^{\sigma\alpha}{}_{\beta,\sigma}(\ell_1,-\ell_2)\left(\tau_{A^2h} \right)^{\gamma\delta}{}_{\rho,\nu}(q-\ell_1-\ell_2, q)}{\left((\ell_1+\ell_2)^2+i\epsilon\right)((p_1-\ell_1)^2-m^2+i\epsilon)((p_1-\ell_1-\ell_2)^2-m^2+i\epsilon)}\ ,
    \end{aligned}
\end{equation}
which in the classical regime reads
\begin{equation}
    j_0^{(2, 2)}(\vec{q}^2)=\frac{1}{2}Q^3\int\frac{d^d \ell_1}{(2\pi)^d}\frac{d^d \ell_2}{(2\pi)^d}\frac{P_{\mu\nu, \alpha\beta}\left(\tau_{A^2h} \right)^{\mu\nu}{}_{00}(\ell_1, -\ell_2)\left(\tau_{A^2h} \right)^{\alpha\beta}{}_{00}(q-\ell_1-\ell_2, q)\bigl|_{\ell_i^0=0}}{\vec{\ell_1}^2\vec{\ell_2}^2(\vec{q}-\vec{\ell_1}-\vec{\ell_2})^2(\vec{\ell_1}+\vec{\ell_2})^2}\ .
\end{equation}
Computing the tensor contraction in the numerator as
\begin{equation}
    \begin{aligned}
    P_{\mu\nu, \alpha\beta}&\left(\tau_{A^2h} \right)^{\mu\nu}{}_{0,0}(\ell_1, -\ell_2)\left(\tau_{A^2h} \right)^{\alpha\beta}{}_{0,0}(q-\ell_1-\ell_2, q)\bigl|_{\ell_i^0=0}\\
    &=\frac{\kappa^2}{2}\left(\frac{d-3}{d-1}\Vec{\ell_1}\cdot\Vec{\ell_2}\ \Vec{q}\cdot(\Vec{q}-\Vec{\ell_1}-\Vec{\ell_2})+2\ \Vec{q}\cdot\Vec{\ell_2}\ \Vec{\ell_1}\cdot(\Vec{q}-\Vec{\ell_1}-\Vec{\ell_2})\right)\ , 
    \end{aligned}
\end{equation}
in which again we exploited the symmetry $\ell_1\leftrightarrow \ell_2$, and using the expressions in appendix \ref{App:LoopRed}, we  obtain
\begin{equation}
    j_0^{(2, 2)}(\vec{q}^2)=\frac{Q^3 \kappa^2}{8}\frac{(d-3)(d-2)}{(d-4)(d-1)}J_{(2)}(\Vec{q}^2)\ .
\end{equation}
From this we  get the contribution to the electromagnetic potential
\begin{equation}
    A_0^{(3, 2)}(r)=\frac{Q}{4\pi}\alpha G_N\frac{(d-3)(d-2)}{(d-4)(d-1)}\rho^3\ .
\end{equation}
This last expression matches perfectly the corresponding term of eq. (\ref{EM_Potential_deDonder}). As before,  we notice a divergence for $d=4$ which must be renormalised. 

To summarise, we have shown that from the loop computations in this section one recovers exactly the expression for the electromagnetic potential of the Reissner-Nordstr\"om-Tangherlini solution in de Donder gauge given in eq. \eqref{EM_Potential_deDonder}. 

\section{Divergences and higher-derivative couplings}\label{sec:divergences}
In section \ref{sec:metric} and \ref{sec:gaugepotential} we noticed the appearance of  divergences in  $d=3$ and $d=4$. In this section we show how such divergences are renormalised by non-minimal couplings. After showing 
how to  regularise the divergences, in the first subsection we analyse the structure  of the non-minimal couplings that are needed and the variation of the metric and the potential that they produce, while in the second subsection we use these results to renormalise the divergences.

\subsection*{Regularisation in $d=3$}
In this dimension the only divergences that appear are at 2-loop order for the metric, while the electromagnetic potential is well defined. %The explicit expression of the 2-loop contribution of the metric at $d=3$ is
%\begin{equation}
%    \begin{aligned}
%    h_0^{(d=3)}\bigl|_{2-\text{loop}} =&-\frac{2m\alpha G_N^2}{r^3} -\frac{2m^3G_N^3}{r^3} \\
%    h_1^{(d=3)}\bigl|_{2-\text{loop}}=&\frac{1}{d-3}\left(\frac{\Gamma\left(\frac{d-2}{2}\right)}{\pi^{\frac{d-2}{2}}}\right)^3 \left(\frac{4}{3}m\alpha G_N^2 -\frac{2}{3}m^3G_N^3\right) r^{-3(d-2)} \\
%    h_2^{(d=3)}\bigl|_{2-\text{loop}} = &\frac{1}{d-3}\left(\frac{\Gamma\left(\frac{d-2}{2}\right)}{\pi^{\frac{d-2}{2}}}\right)^3 \left(-4m\alpha G_N^2 +2m^3G_N^3\right) r^{-3(d-2)} \ .
%    \end{aligned}
%\end{equation}
From the explicit expression of the 2-loop contribution of the metric in $d=3$, we notice that $h_0^{(d=3)}\bigl|_{2-\text{loop}}$ is non divergent and can be already compared with the \eqref{hi3} with a perfect match. The other two components are equal up to a multiplicative constant, in particular we have $h_2^{(d=3)}\bigl|_{2-\text{loop}}=-3h_1^{(d=3)}\bigl|_{2-\text{loop}}$. So we will renormalise only the $h_1^{(d=3)}\bigl|_{2-\text{loop}}$ component and then obtain the other one using this relation. We perform a dimensional  regularisation using $d=3+\epsilon$, where $\epsilon$ is a small parameter. Neglecting terms that vanish when $\epsilon \rightarrow 0$, the result is
\begin{equation}
\begin{aligned}
    h_1^{(d=3)}\bigl|_{2-\text{loop}}& =  \frac{1}{\epsilon} \left(\frac{4}{3}\frac{m\alpha G_N^2}{r^3} -\frac{2}{3}\frac{m^3G_N^3}{r^3}\right) +\frac{m^3G_N^3}{r^3}\left(2\log \left(2C_Er\right)-\frac{4}{3}\right) - \\ & -4\frac{m\alpha G_N^2}{r^3}\log \left(2C_Er\right)  \ ,
\end{aligned}
\end{equation}
where following \cite{Mougiakakos:2020laz} we defined the constant $C_E^2\equiv \pi e^{\gamma_{EM}}$, with $\gamma_{EM}$ the Euler-Mascheroni constant. 

\subsection*{Regularisation in $d=4$}
In $d=4$ dimension the metric is divergent at both 1-loop and 2-loop order, while the potential is divergent at 2-loop. 
%\begin{equation}
 %   \begin{aligned}
 %    h_0^{(d=4)}\bigl|_{1-\text{loop}} =&\frac{4}{3} \frac{\alpha G_N}{\pi^2 r^4} + \frac{32}{9}\frac{m^2G_N}{\pi^2r^4}\\
 %    h_1^{(d=4)}\bigl|_{1-\text{loop}}=&\frac{1}{d-4}\left(\frac{\Gamma\left(\frac{d-2}{2}\right)}{\pi^{\frac{d-2}{2}}}\right)^2\left(-\frac{2}{3}\frac{\alpha G_N}{\pi^2}-\frac{40}{9}\frac{G_N^2m^2}{\pi^2}\right) r^{-2(d-2)}\\
  %  h_2^{(d=4)}\bigl|_{1-\text{loop}} = &\frac{1}{d-4}\left(\frac{\Gamma\left(\frac{d-2}{2}\right)}{\pi^{\frac{d-2}{2}}}\right)^2\left(\frac{8}{3}\frac{\alpha G_N}{\pi^2}+\frac{160}{9}\frac{G_N^2m^2}{\pi^2}\right) r^{-2(d-2)} \ .
  %  \end{aligned}
%\end{equation}
Again, from the explicit form of the metric, we can see that at 1-loop order the $h_0^{(d=4)}\bigl|_{1-\text{loop}}$ component is non divergent and no renormalisation procedure has to be carried on. The $h_1^{(d=4)}\bigl|_{1-\text{loop}}$ and $h_2^{(d=4)}\bigl|_{1-\text{loop}}$ components are related by the identity $h_2^{(d=4)}\bigl|_{1-\text{loop}}=-4h_1^{(d=4)}\bigl|_{1-\text{loop}}$ and so we can focus just on one component. Regularizing in $d=4+\epsilon$, and neglecting again terms that vanish when $\epsilon \rightarrow 0$, the result at 1-loop is
\begin{equation}
\begin{aligned}
     h_1^{(d=4)}\bigl|_{1-\text{loop}} & =\frac{1}{\epsilon}\left(-\frac{2}{3}\frac{\alpha G_N}{\pi^2r^4}-\frac{40}{9}\frac{G_N^2m^2}{\pi^2r^4}\right)+\frac{m^2 G_N^2}{\pi^2r^4}\left(-\frac{4}{27}+\frac{40}{9}\log \left(C_E^2 r^2 \right)\right)  \\& +\frac{\alpha G_N}{\pi^2r^4} \left(-\frac{4}{9}+\frac{2}{3}\log \left(C_E^2 r^2 \right)\right)    \ .
\end{aligned}
\end{equation}
% where $C_E$ is the same constant defined before .\\
%\begin{equation}
%    \begin{aligned}
%    h_0^{(d=4)}\bigl|_{2-\text{loop}} (r) =&\frac{1}{d-4}\left(\frac{\Gamma\left(\frac{d-2}{2}\right)}{\pi^{\frac{d-2}{2}}}\right)^3 \left(-\frac{16}{9}\frac{m\alpha G_N^2}{\pi^3} -\frac{320}{27}\frac{m^3G_N^3}{\pi^3}\right) r^{-3(d-2)} \\
%    h_1^{(d=4)}\bigl|_{1-\text{loop}}=&\frac{1}{d-4}\left(\frac{\Gamma\left(\frac{d-2}{2}\right)}{\pi^{\frac{d-2}{2}}}\right)^3 \left(\frac{8}{9}\frac{m\alpha G_N^2}{\pi^3} +\frac{160}{27}\frac{m^3G_N^3}{\pi^3}\right) r^{-3(d-2)}\\
%    h_2^{(d=4)}\bigl|_{1-\text{loop}} = &\frac{1}{d-4}\left(\frac{\Gamma\left(\frac{d-2}{2}\right)}{\pi^{\frac{d-2}{2}}}\right)^3 \left(-\frac{16}{9}\frac{m\alpha G_N^2}{\pi^3} -\frac{320}{27}\frac{m^3G_N^3}{\pi^3}\right) r^{-3(d-2)} \ .
%    \end{aligned}
%\end{equation}
For what concerns the 2-loop order, we can focus just on the $h_1^{(d=4)}\bigl|_{2-\text{loop}}$ component since the relation $h_0^{(d=4)}\bigl|_{2-\text{loop}}=h_2^{(d=4)}\bigl|_{2-\text{loop}}=-2h_1^{(d=4)}\bigl|_{2-\text{loop}}$ holds. We obtain
\begin{equation}
\begin{aligned}
    h_1^{(d=4)}\bigl|_{2-\text{loop}}&=\frac{1}{\epsilon} \left(\frac{8}{9}\frac{m\alpha G_N^2}{\pi^3r^6} +\frac{160}{27}\frac{m^3G_N^3}{\pi^3r^6}\right) + \frac{m ^3G_N^3 }{\pi^3 r^6} \left(\frac{208}{81}-\frac{80}{9}\log \left(C_E^2 r^2 \right)  \right) \\&+ \frac{\alpha m G_N^2}{\pi^3 r^6}\left(\frac{64}{27}-\frac{4}{3}\log \left(C_E^2 r^2\right)  \right) \ ,
\end{aligned}
\end{equation}
Finally, the potential regularised at 2-loop is
\begin{equation}
\begin{aligned}
    A_0^{(d=4)}\bigl|_{2-\text{loop}}&= \frac{1}{\epsilon}\left(\frac{10}{9}\frac{Q(G_Nm)^2}{\pi^4r^6}+\frac{1}{6}\frac{QG_N \alpha}{\pi^4 r^6}\right)
    -\frac{1}{27}\frac{Q(G_Nm)^2}{\pi^4r^6}\left(-28+45\log(C_E^2r^2)\right)\\ & -\frac{1}{36}\frac{QG_N\alpha}{\pi^4r^6}\left(-7+9\log(C_E^2r^2)\right) \ .
    \end{aligned}
\end{equation}

\subsection{Non-minimal couplings}
As shown in \cite{Mougiakakos:2020laz}, the previous divergences can be renormalised considering non-minimal coupling terms linear in the Riemann tensor and quadratic in the scalar field. Since we are considering charged scalars, we generalise their non-minimal couplings to the following form
\begin{equation}\label{NonMinimal_Action}
    \begin{split}
        \delta^{(\mathfrak{n})}S^{ct} =& \sum_{k=0}^{+\infty} (G_Nm)^{\frac{2(\mathfrak{n}-k)}{d-2}}(\alpha G_N)^{\frac{k}{d-2}} \int d^{d+1}x \sqrt{-g}\Bigl(\alpha^{(\mathfrak{n},k)}(d)(D^2)^{\mathfrak{n}-1}RD_{\mu}\phi D^{\mu}\phi \\
        &+\left(\beta_0^{(\mathfrak{n},k)}(d)D_{\mu}D_{\nu}(D^2)^{\mathfrak{n}-2}R + \beta_1^{(\mathfrak{n},k)}(d)(D^2)^{\mathfrak{n}-1}R_{\mu\nu}\right)D^{\mu}\phi D^{\nu}\phi\Bigr)
    \end{split}
\end{equation}
where $D_{\mu}$ is the covariant derivative, and where we consider only positive integer powers of the gravitational and electromagnetic coupling.

Due to properties of Fourier transform, terms with $\mathfrak{n}\geq 2$ and terms proportional to $\beta_0^{(\mathfrak{n},k)}$ and $\beta_1^{(\mathfrak{n},k)}$ do not contribute to the classical limit of the metric \cite{Mougiakakos:2020laz}. We have verified that this statement is valid also for the potential. 
For the sake of simplicity of notation, we define 
\begin{equation}\label{alpha}
    \alpha^{(\mathfrak{n})}(d) = \sum_{k=0}^{+\infty} (G_Nm)^{\frac{-2k}{d-2}}(\alpha G_N)^{\frac{k}{d-2}}\alpha^{(\mathfrak{n},k)}(d)\ ,
\end{equation}
and, in the same way, we can define $\beta_0^{(\mathfrak{n})}$ and $\beta_1^{(\mathfrak{n})}$. Thus, the only contribution to the non-minimal coupling action that will renormalise the metric and the potential is
\begin{equation}
\delta^{(1)}S^{ct} = (G_N m)^{\frac{2}{d-2}}\alpha^{(1)}(d)\int d^{d+1}x  \sqrt{-g}\, R\,  D_{\mu}\phi D^{\mu}\phi \ . \label{theonlynonminimalcoupling}
\end{equation}
In the following, we compute the Feynman rules associated with the counter-terms only in the case $\mathfrak{n}=1$, but for the sake of completeness we give the complete relations considering also the term proportional to $\beta_1^{(1)}$. Then one gets \cite{Mougiakakos:2020laz}
\begin{figure}[H]
%\centering
\begin{minipage}{1.\textwidth}
        \raggedright %\hspace{1cm}
\begin{fmffile}{sseeee33}
 \begin{equation}
 \begin{split}
\begin{gathered}
\begin{fmfgraph*}(80,60)
\fmfleft{i1,i2}
 \fmfright{o}
 \fmflabel{$\mu\nu$}{o}
\fmf{fermion}{i1,v,i2}
\fmf{dbl_wiggly}{v,o}
\fmfv{d.sh=square,d.f=empty,l=$1$,decor.size=6thick,l.d=0cm}{v}
\end{fmfgraph*}
\end{gathered}  \hspace{0.8cm}&= \left(\tau_{\phi^2 h}^{ct}\right)_{\mu\nu}(q) = \\
&=(G_N m)^{\frac{2}{d-2}} i\kappa \left(\alpha^{(1)}(d)\left(-q_{\mu}q_{\nu} + \eta_{\mu\nu}q^2\right) p_1\cdot p_2 + \beta^{(1)}_1(d) \frac{q^2}{2}  p_{1\,\mu}p_{2 \, \nu} \right)
\end{split}
\end{equation}
\end{fmffile}
\end{minipage}%
\end{figure}
\noindent
whose classical limit is
\begin{equation}\label{MassiveCT}
    \left(\tau_{\phi^2 h}^{ct}\right)_{\mu\nu}(q) \simeq -(G_N m)^{\frac{2}{d-2}} i\kappa m^2 \left(\alpha^{(1)}(d)\left(q_{\mu}q_{\nu} + \eta_{\mu\nu}\vec{q}^2\right) + \beta^{(1)}_1(d) \frac{\vec{q}^2}{2}  \delta_{\mu}^{0}\delta_{\nu}^0 \right) \ .
\end{equation}
Another possible vertex is constituted by two photons and one graviton. It arises since we must take into account the electromagnetic gauge symmetry into the covariant derivative. Its contribution is
 \begin{figure}[H]
\begin{minipage}{1\textwidth}
        \raggedright
\begin{fmffile}{sseqwwweee33}
\begin{equation}
\begin{split}
\begin{gathered}
\begin{fmfgraph*}(80,60)
\fmfleft{i1,i2}
\fmfright{o}
 \fmflabel{$\mu$}{i2}
 \fmflabel{$\nu$}{i1}
  \fmflabel{$\alpha\beta$}{o}
\fmf{photon}{i1,v,i2}
\fmf{dbl_wiggly}{v,o}
\fmfv{d.sh=square,d.f=empty,l=$1$,decor.size=6thick,l.d=0cm}{v}
\end{fmfgraph*}
\end{gathered} \hspace{0.8cm} = \left(\tau_{A^2 h}^{ct}\right)_{\alpha\beta,\mu,\nu}(q) = &-(G_N m)^{\frac{2}{d-2}} Q^2  i\kappa \Bigl(\alpha^{(1)}(d)(-q_{\alpha}q_{\beta} + \eta_{\alpha\beta} q^2) 2\eta_{\mu\nu} \\
&+ \beta^{(1)}(d) \frac{q^2}{2}(\eta_{\mu\alpha} \eta_{\nu\beta} + \eta_{\nu\alpha} \eta_{\mu\beta})  \Bigr)
\end{split}\label{2photon1gravitonct}
\end{equation}
\end{fmffile}
\end{minipage}%
\end{figure}
\noindent
However, we find that the 1-loop insertion of this vertex is vanishing in the classical limit, both with an external photon and an external  graviton, and it will be no longer considered in this paper. Now, we can use the counter-term in (\ref{MassiveCT}) to compute the diagrams that will be necessary to the  renormalisation process.

\begin{figure}[h]
%\centering
\begin{minipage}{1.\textwidth}
        \raggedright %\hspace{1cm}
\begin{fmffile}{ssewwweeee33}
 \begin{equation*}
\begin{gathered}
\begin{fmfgraph*}(80,60)
\fmfleft{i1,i2}
 \fmfright{o}
    \fmf{fermion,label=$p_1$}{i1,v}
    \fmf{fermion,label=$p_2$,l.s=right}{v,i2}
    \fmf{dbl_wiggly,label=$q$,l.s=left}{v,o}
    \fmflabel{${\mu\nu}$}{o}
\fmfv{d.sh=square,d.f=empty,l=$1$,decor.size=6thick,l.d=0cm}{v}
\end{fmfgraph*}
\end{gathered}
\end{equation*}
\end{fmffile}
\end{minipage}%
\caption{Insertion of non-minimal coupling at tree level.}
\label{CT1}
\end{figure}
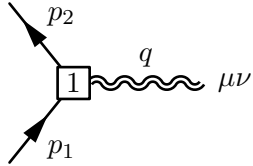

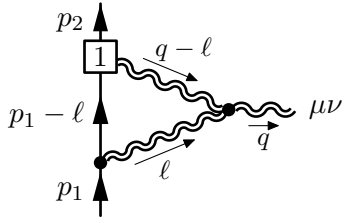
\begin{figure}[h]
\centering
\begin{minipage}{0.4\textwidth}
        \raggedright \begin{equation*}
\begin{fmffile}{dfcsaWQS}
\begin{fmfgraph*}(80,80)
\DeclareGraphicsRule{*}{mps}{*}{}
\newcommand{\marrow}[5]{%
    \fmfcmd{style_def marrow#1 
    expr p = drawarrow subpath (1/4, 3/4) of p shifted 6 #2 withpen pencircle scaled 0.4; 
    label.#3 (btex #4 etex, point 0.5 of p shifted 6 #2); 
    enddef;}
    \fmf{marrow#1, tension=0}{#5}}
\unitlength=1mm
\fmfleft{i1,i2}
 \fmfright{o}
 \fmf{fermion,tension=2,label=$p_1$,l.s=left}{i1,v1}
\fmf{fermion,tension=1,label=$p_1-\ell$,l.s=left}{v1,v2}
 \fmf{fermion,tension=2,label=$p_2$,l.s=left}{v2,i2}
 \fmffreeze
\fmf{dbl_wiggly}{v1,v3}
\fmf{dbl_wiggly}{v2,v3}
\fmf{dbl_wiggly,tension=4}{v3,o}
\fmflabel{${\mu\nu}$}{o}
\fmfv{d.sh=square,d.f=empty,l=$1$,decor.size=6thick,l.d=0cm}{v2}
\fmfdot{v1,v3}
\marrow{a}{down}{bot}{$\ell$}{v1,v3}
\marrow{b}{up}{top}{$\quad q-\ell$}{v2,v3}
\marrow{c}{down}{bot}{$q$}{v3,o}
\end{fmfgraph*}
\end{fmffile}
\end{equation*}
\end{minipage}%
\caption{Insertion of non-minimal coupling at 1-loop with an external graviton.}
\label{CT2}
\end{figure}

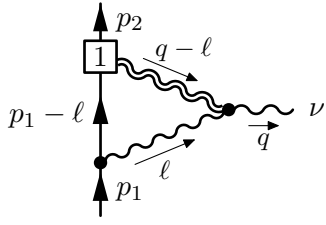
\begin{figure}[h]
\centering
\begin{minipage}{0.4\textwidth}
        \raggedright \begin{equation*}
\begin{fmffile}{dfcdeedsaWQS}
\begin{fmfgraph*}(80,80)
\DeclareGraphicsRule{*}{mps}{*}{}
\newcommand{\marrow}[5]{%
    \fmfcmd{style_def marrow#1 
    expr p = drawarrow subpath (1/4, 3/4) of p shifted 6 #2 withpen pencircle scaled 0.4; 
    label.#3 (btex #4 etex, point 0.5 of p shifted 6 #2); 
    enddef;}
    \fmf{marrow#1, tension=0}{#5}}
\unitlength=1mm
\fmfleft{i1,i2}
 \fmfright{o}
 \fmf{fermion,tension=2,label=$p_1$,l.s=right}{i1,v1}
\fmf{fermion,tension=1,label=$p_1-\ell$,l.s=left}{v1,v2}
 \fmf{fermion,tension=2,label=$p_2$,l.s=right}{v2,i2}
\fmffreeze
\fmf{dbl_wiggly}{v2,v3}
\fmf{photon}{v1,v3}
\fmf{photon,tension=4}{v3,o}
\fmflabel{$\nu$}{o}
\fmfv{d.sh=square,d.f=empty,l=$1$,decor.size=6thick,l.d=0cm}{v2}
\fmfdot{v1,v3}
\marrow{a}{down}{bot}{$\ell$}{v1,v3}
\marrow{b}{up}{top}{$\quad q-\ell$}{v2,v3}
\marrow{c}{down}{bot}{$q$}{v3,o}
\end{fmfgraph*}
\end{fmffile}
\end{equation*}
\end{minipage}
\caption{Insertion of non-minimal coupling at 1-loop with an external photon.}
\label{CT3}
\end{figure}

Following \cite{Mougiakakos:2020laz}, the insertion of the non minimal coupling at tree level gives the process in figure \ref{CT1}. 
\noindent
The associated contribution to the stress-energy tensor is given by
\begin{equation}
    \frac{-i\kappa}{2}\, \sqrt{4E_1E_2}\, \delta^{(1)}T_{\mu\nu}^{(0)}(q^2) = \left(\tau_{\phi^2 h}^{ct}\right)_{\mu\nu}(q) \ .
\end{equation}
Considering the relation between the stress-energy tensor and the metric perturbation, as in \cite{Mougiakakos:2020laz}, we obtain the contributions to the metric components
\begin{equation}\label{CT1_Metric}
\begin{aligned}
    \delta^{(1)}h_0^{(1)}(r) &= 0 \\
    \delta^{(1)}h_1^{(1)}(r) &= \frac{16\alpha^{(1)}(d)\Gamma\left(\frac{d}{2}\right)}{\pi^{\frac{d-2}{2}}} \frac{(G_Nm)^{\frac{d}{d-2}}}{r^d} \\
    \delta^{(1)}h_2^{(1)}(r) &= \frac{-32\alpha^{(1)}(d)\Gamma\left(\frac{d+2}{2}\right)}{\pi^{\frac{d-2}{2}}} \frac{(G_Nm)^{\frac{d}{d-2}}}{r^d}\ ,
\end{aligned}
\end{equation}
where $\alpha^{(1)}(d)$ is defined in (\ref{alpha}). Then, we can consider the insertion in the 1-loop diagram as in figure  \ref{CT2}.
The corresponding contribution to the stress-energy tensor is 
\begin{equation}
\begin{aligned}
    & -\frac{i\, \kappa}{2}\, \sqrt{4E_1E_2}\, \delta^{(1)}T_{\mu\nu}^{(1)}(q^2)= \\& 2\times  \int\frac{d^D\ell}{(2\pi)^D}\frac{-i\, P_{\rho\sigma,\alpha\beta}P_{\eta\xi,\gamma\delta}\left(\tau_{\phi^2h}\right)_{\rho\sigma} \left(\tau_{\phi^2 h}^{ct}\right)_{\eta\xi}(q-\ell)\left(\tau_{h^2h}\right){}^{\mu\nu}{}_{\alpha\beta,\gamma\delta}(l,q) }{((p_1-\ell)-m^2 + i\epsilon)(\ell^2+i\epsilon)((q-\ell)^2+i\epsilon)}
\end{aligned}
\end{equation}
from which using the property in (\ref{Master_CT}), we determine the contributions to the metric \cite{Mougiakakos:2020laz}
\begin{align}
    \delta^{(1)}h_0^{(2)}(r) &= 64 \alpha^{(1)}(d) \frac{(d-2) \Gamma\left(\frac{d}{2}\right)^{2}}{(d-1) \pi^{d-2}}\left(\frac{\left(G_{N} m\right)^{\frac{1}{d-2}}}{r}\right)^{2(d-1)} \\
    \delta^{(1)}h_1^{(2)}(r) &= -64 \alpha^{(1)}(d) \frac{(d-2) \Gamma\left(\frac{d}{2}\right)^{2}}{(d-1) \pi^{d-2}}\left(\frac{\left(G_{N} m\right)^{\frac{1}{d-2}}}{r}\right)^{2(d-1)} \\
    \delta^{(1)}h_2^{(2)} (r)&= 128 \alpha^{(1)}(d) \frac{(d-2) \Gamma\left(\frac{d}{2}\right)^{2}}{(d-1) \pi^{d-2}}\left(\frac{\left(G_{N} m\right)^{\frac{1}{d-2}}}{r}\right)^{2(d-1)} \ .
\end{align}
In addition to the previous diagrams, we compute the insertion in the 1-loop diagram with an external photon, as in figure \ref{CT3}, since it will be necessary in order to renormalise the electromagnetic potential.
\noindent
The contribution to the electromagnetic current is 
\begin{equation}
\begin{aligned}
&-i\, \sqrt{4E_1E_2}\, \delta^{(1)}j_{\nu}^{(1)}(q^2)=\\ &2 \times \int \frac{d^{d+1} \ell}{(2 \pi)^{d+1}} \frac{iP_{\alpha \beta \gamma \delta} \ \left(\tau_{\phi^{2} A}\right)^{\mu}\left(\tau_{\phi^{2} h}^{ct}\right)^{\gamma \delta}\left(q-\ell\right)\left(\tau_{A^{2} h}\right)^{\alpha \beta}{}_{\mu \nu}(\ell, q)}{(\left(p_{1}-\ell\right)^{2}-m^{2}+i \epsilon)(\ell^{2}+i\epsilon)(\left(q-\ell\right)^{2}+i\epsilon)} \ .
\end{aligned}
\end{equation}
%\begin{equation}
%=-i \kappa^2(G_Nm)^{\frac{2}{d-2}} Q\frac{J_{(1)} \, \Vec{q}^2}{2}m^2\alpha^{(1)}(d)\delta^0_{\nu}
%\end{equation}
Considering the classical limit we get
\begin{equation}
    -i\, 2m\, \delta^{(1)}j_{0}^{(1)}(\vec{q}^2)=-iQ \int\frac{d^d\ell}{(2\pi)^d}\frac{P_{\mu\nu, \alpha\beta}\left(\tau_{\phi^{2} h}^{ct}\right)^{\mu\nu}\left(q-\ell\right)\left(\tau_{A^{2} h}\right)^{\alpha \beta}{}_{00}(\ell, q)\bigl|_{\ell^0=0}}{\Vec{\ell}^2(\Vec{q}-\Vec{\ell})^2}\ ,
\end{equation}
from which computing
\begin{equation}
\begin{aligned}
    P_{\mu\nu, \alpha\beta}&\left(\tau_{\phi^{2} h}^{ct}\right)^{\mu\nu}\left(q-\ell\right)\left(\tau_{A^{2} h}\right)^{\alpha \beta}{}_{00}(\ell, q)\bigl|_{\ell^0=0}=\\
    &-\frac{\kappa^2}{2(d-1)}(G_Nm)^{\frac{2}{d-2}}\Bigl(\beta^{(1)}(d)(d-2)(\Vec{\ell}^2\, \Vec{\ell}\cdot\Vec{q}-2\Vec{\ell}\cdot\Vec{q}\, \Vec{\ell}\cdot\Vec{q}+\Vec{\ell}\cdot\Vec{q}\, \Vec{q}^2)-\\
    &2\alpha^{(1)}(d)m^2\Vec{\ell}\cdot\vec{q}(\Vec{q}^2+\Vec{\ell}^2)-2\alpha^{(1)}(d)(d-3)m^2\Vec{\ell}\cdot\Vec{q}\, \Vec{\ell}\cdot\vec{q}\Bigl)\ ,
    \end{aligned}
\end{equation}
and exploiting eqs. \eqref{LR_1Loop_2} and \eqref{LR_1Loop_3}, one finally obtains the contribution to the electromagnetic current
\begin{equation}
    \delta^{(1)}j_{0}^{(1)}(\vec{q}^2)=\frac{1}{4}Q\kappa^2m\alpha^{(1)}(d)(G_Nm)^{\frac{2}{d-2}}\vec{q}^2J_{(1)}(\Vec{q}^2)\ .
\end{equation}
Then using eq. (\ref{Master_CT}), we obtain the contribution to the electromagnetic potential 
\begin{equation}\label{deltaA}
    \delta^{(1)}A_0^{(2)}(r) =  -4 G_N m (G_Nm)^{\frac{2}{d-2}} Q\alpha^{(1)}(d)\frac{\Gamma\left(\frac{d}{2}\right)^{2}}{\pi^{d-1} r^{2(d-1)}} \ . 
\end{equation}
\subsection{Renormalisation}
\subsection*{Renormalisation in d=3}
In four dimensions, the only values of $k$ that respect the constraints described in the previous section are $k=0,1,2$. However, the case $k=2$ does not have a match in the post-Minkowskian expansion and will not be considered.
% , thus it will contribute just with finite terms. 
As a consequence, eq. (\ref{alpha}) becomes
\begin{equation}\label{AlphaCTd3}
    \alpha^{(1)}(3) = \alpha^{(1,0)}(3) + \frac{\alpha G_N}{(G_Nm)^2}\alpha^{(1,1)}(3) \,.
\end{equation}
In order to cure the divergences, $\alpha^{(1)}(3)$ must have the form
\begin{equation}\label{alpha3}
    \alpha^{(1)}(3)=\frac{\omega(3)}{d-3}+\Omega(3)\ ,
\end{equation}
where $\omega(3)$ is the coefficient that we need to fix in order to renormalise the metric, while $\Omega(3)$ is a finite term. In $d=3$, the 2-loop corrections to the metric are given by the diagram in figure \ref{CT1} and, expanding the metric contributions (\ref{CT1_Metric}) in $d=3+\epsilon$, we find the renormalised metric component
\begin{equation}
\begin{aligned}
    h_{1}^{(d=3)}& \bigl|^{renorm}_{2-loop}= h_{1}^{(d=3)}\bigl|_{2-loop} + \delta^{(1)}h_1^{(1)} \\& =\frac{8 G_N^3 m^3}{\epsilon r^3}\left(\omega(3) -\frac{1}{12}+\frac{\alpha}{6G_N m^2}\right)-\frac{4 G_N^3 m^3}{3 r^3}+\frac{8 G_N^3 m^3  }{r^3}\omega(3)+\frac{8 G_N^3 m^3  }{r^3}\Omega(3)\\&+\frac{2 G_N^3 m^3}{r^3} \log \left(2 C_E r\right)-\frac{4 \alpha  G_N^2 m }{r^3}\log \left(2 C_E r\right)-\frac{8 G_N^3 m^3 }{r^3}\omega(3)  \log \left(2 C_E r G_N^2 m^2 \right) \ .
\end{aligned}
\end{equation}
We can fix the $ \omega(3)$ in order to cancel the divergent term in the metric component above. Imposing this condition we obtain for the divergent contribution
\begin{equation}
    \omega(3)=\frac{1}{12}-\frac{\alpha G_N}{6(G_N m)^2}\ .
\end{equation}
For the term independent of $\alpha$ we recognise the coupling already found in \cite{Mougiakakos:2020laz}, while the other is the new piece due to electromagnetic interaction. 
With this choice the metric components become
\begin{equation}
    \begin{aligned}
     h_{1}^{(d=3)}\bigl|^{renorm}_{2-loop}=&\frac{2 G_N^3 m^3}{3 r^3} \left(2 \log \left(\frac{2 C_E r}{G_N m}\right)+ 12\, \Omega(3) -1\right)\\ &-\frac{4 G_N^2 m \alpha}{3 r^3}\left(1  +2   \log \left(\frac{2 C_E r}{G_N m}\right)\right) \ .
    \end{aligned}
\end{equation}
Now it is straightforward to see that this expression matches the metric in (\ref{hi4}) up to finite terms. In fact, although we cannot fix the finite part $\Omega(3)$, since it belongs to the high energy regime, we can keep trace of it, and imposing a strict agreement between the renormalised metric and the one computed classically, the condition
\begin{equation}
    \Omega(3)= 2 \, \omega(3)\log \left(\frac{c}{4 C_E}\right)- \omega(3)+\frac{2}{3} \ 
\end{equation}
must hold, which relates the freedom in the quantum computation with the gauge freedom in the classical one.

% [THIS MEANS THAT THE LOG TERMS ARE JUST A GAUGE FIXING]

\subsection*{Renormalisation in $d=4$}
Referring to (\ref{NonMinimal_Action}), in five dimensions the only allowed values of $k$ are $k=0, 2$. Then the expression in (\ref{alpha}) reads
\begin{equation}
    \alpha^{(1)}(4) = \alpha^{(1,0)}(4)+\frac{\alpha G_N}{(G_Nm)^{2}}\alpha^{(1,2)}(4)\ .
\end{equation}
As we already did in the previous case, in order to cancel the divergences we write
\begin{equation}
    \alpha^{(1)}(d)=\frac{\omega(4)}{d-4}+\Omega(4)\ ,
\end{equation}
where $\omega(4)$ is the coefficient we need to fix in order to renormalise the divergences in $d=4$ and $\Omega(4)$ is a finite contribution. In $d=4+\epsilon$ the metric is renormalised at one loop adding the contribution of the diagram in figure \ref{CT1}. Then we can set the constant $\omega(4)$ in order to vanish the divergence of the metric. The value obtained is
\begin{equation}\label{omega4}
    \omega(4)=\frac{5}{18 \pi}+\frac{G_N \alpha}{(G_N m)^2}\frac{1}{24 \pi}\ ,
\end{equation}
and again we can match the purely gravitational result with \cite{Mougiakakos:2020laz}.
As done for $d=3$ we can obtain a perfect match with the classical expression \ref{hi4} fixing the finite contribution $\Omega(4)$, finding 
\begin{equation}\label{CapOmega4}
    \Omega(4)=\frac{\omega(4)}{2}\log\left(\frac{3 \pi c}{8C_E^2}\right)+\frac{\omega(4)}{6}-\frac{4}{27\pi} \ .
\end{equation}
Finally the renormalised metric component with this choice of coupling constants is
\begin{equation}
\begin{aligned}
    h_{1}^{(d=4)}\bigl|^{renorm}_{1-loop}= h_{1}^{(d=4)}\bigl|_{1-loop} &+ \delta^{(1)}h_1^{(1)} =\frac{56 G_N^2 m^2}{27 \pi ^2 r^4}-\frac{\alpha  G_N}{9 \pi ^2 r^4} +\frac{16 G_N^2 m^2 }{\pi  r^4} \Omega(4)
   \\&+ \frac{20 G_N^2 m^2}{9 \pi ^2 r^4} \log \left(\frac{C_E^2 r^2}{G_N m}\right)+\frac{\alpha  G_N}{3 \pi ^2 r^4}  \log \left(\frac{C_E^2 r^2}{G_N m}\right)\ .
\end{aligned}
\end{equation}
For the 2-loop computation the steps are the same of the case above but using the 1-loop diagram in figure \ref{CT2}. The computations are independent from the previous case and lead to the same constants $\omega(4)$ and $\Omega(4)$ defined above.
At this order the renormalised metric component is found to be
\begin{equation}
\begin{aligned}
    h_{1}^{(d=4)}\bigl|^{renorm}_{2-loop}= h_{1}^{(d=4)}\bigl|_{2-loop} &+ \delta^{(1)}h_1^{(2)} =-\frac{112 G_N^3 m^3}{81 \pi ^3 r^6}+\frac{16 \alpha  G_N^2 m}{9 \pi ^3 r^6} -\frac{64 G_N^3 m^3 \Omega(4) }{3 \pi ^2 r^6}
    \\&-\frac{80 G_N^3 m^3}{27 \pi ^3 r^6} \log \left(\frac{C_E^2 r^2}{G_N m}\right)-\frac{4 \alpha  G_N^2 m}{9 \pi ^3 r^6} \log \left(\frac{C_E^2 r^2}{G_N m}\right) \ .
\end{aligned}
\end{equation}

Then the exact same considerations lead to the computation of the renormalised electromagnetic potential, whose considering (\ref{deltaA}), the 2-loop renormalised contribution is 
\begin{multline}
    A_0^{(d=4)}\bigl|^{renorm}_{2-loop}=A_0^{(d=4)}\bigl|_{2-loop}+\delta^{(1)}A_0^{(2)}=\frac{Q}{4\pi}\Biggl(-\frac{8}{27}\frac{(G_Nm)^2}{\pi^3r^6}+\frac{1}{9}\frac{G_N\alpha}{\pi^3r^6}+\\
    -\frac{16(G_Nm)^2}{\pi^2 r^6}\Omega(4)+\frac{20}{9}\frac{(G_Nm)^2}{\pi^3r^6}\log\left(\frac{G_Nm}{C_E^2r^2}\right)+\frac{1}{3}\frac{G_N\alpha}{\pi^3r^6}\log\left(\frac{G_Nm}{C_E^2r^2}\right)\Biggl)\ ,
\end{multline}
in which the very same counter term in (\ref{omega4}) is used in order to cancel the divergences. Finally using the relation (\ref{CapOmega4}) the renormalised electromagnetic potential perfectly match the classical calculation in section \ref{sec:dedonder}.

\section{Discussion}\label{Sec:Discussion}

In this paper we have shown that the Reissner-Nordstr\"om-Tangherlini solution, describing the metric and the potential generated by a static and  spherically symmetric object of charge $Q$ and mass $m$ in any dimension, can be derived from scattering amplitudes describing  the emission of either a graviton or a photon from a scalar field with the same charge and mass. Our analysis was carried out up to third post-Minkowskian order, that is up to 2 loops. This generalises to the case of charged scalars the work of \cite{Jakobsen:2020ksu} and \cite{Mougiakakos:2020laz}, where the chargeless case was considered at one loop  and up to three loops respectively. All these works extend the original results of \cite{Donoghue:2001qc} and \cite{Bjerrum-Bohr:2002fj}, where amplitude computations were applied to derive the metric of all black hole solutions in four dimensions up to second post-Minkowskian order. 

We have used the techniques outlined in \cite{Bjerrum-Bohr:2018xdl} to extract the classical contributions from the scattering amplitudes. Considering the mass of the scalar to be much larger than the transferred momentum and the loop momenta, one can integrate over the time-like components of the latter in such a way that the propagator of the massive internal scalar drops out of the integral. This projects the quantum scattering amplitude on a tree graph similar to the one considered in  earlier work  \cite{Duff:1973zz}.

The amplitude computations in this paper are performed in de Donder gauge. 
In extracting the metric and the potential from the amplitudes, one finds that in this gauge they both develop singularities. In particular, both the metric and the potential are singular at two loops in five dimensions, and the metric is also singular at one loop in five dimensions and at two loops in four dimensions \cite{Jakobsen:2020ksu,Mougiakakos:2020laz}. In order to cure such divergences, one adds counter-terms from non-minimal couplings. In \cite{Mougiakakos:2020laz} it was shown that a specific higher-derivative coupling, namely the one given in eq. \eqref{theonlynonminimalcoupling}, generates all the counter-terms that are needed to cancel all the divergences coming from graviton loops up to fourth post-Minkowsian order. We show that by simply adding to  the coefficient in front of this term a contribution proportional to the square of the charge, one cancels all the divergences arising from amplitudes in which both photons and gravitons  run in loops up to third post-Minkowskian order. 

After the renormalisation procedure, logarithmic terms are produced both in the metric and the potential. These terms exactly match the logarithms that one generates performing a post-Minkowskian expansion of the Reissner-Nordstr\"om-Tangherlini solution in de Donder gauge, implying that their occurrence is purely an artifact of the gauge choice. In particular, one can perform the same expansion in a gauge in which such terms do not occur, and in \cite{Jakobsen:2020ksu} it was indeed shown that extracting the metric from scattering amplitudes in this gauge there are no divergences at one loop in five dimensions. 
In the context of the world-line formalism, these 
higher-derivative couplings had already been introduced in \cite{Goldberger:2004jt}, and shown not to contribute to any physical observable. The equivalence between the scattering amplitude and world-line approaches was shown in \cite{Mogull:2020sak}.

There are various ways in which this work can be extended.  As mentioned in section \ref{sec:divergences}, the vertex in eq. \eqref{2photon1gravitonct} does not contribute to the renormalisation process up to 2 loops, and extending our analysis at higher loops would reveal  whether this property is general or it is an artifact of the order of the computation. It would also be interesting to perform these higher loop computations in different gauges, generalising the 1-loop analysis of \cite{Jakobsen:2020ksu}. In particular, we expect that in the gauge in which no divergences arise, the non-minimal coupling in eq. \eqref{theonlynonminimalcoupling} does not contribute to the classical limit. Finally, one could extend the analysis to the emission of gravitons and photons from fermionic matter, extending the four-dimensional 1 loop analysis of \cite{Donoghue:2001qc,Bjerrum-Bohr:2002fj}, in order to explore the richer structure of black hole solutions in higher dimensions \cite{Emparan:2008eg}.

The techniques to extract classical contributions from quantum processes have been widely applied in the last few years to determine the dynamics of massive objects from two-body scattering amplitudes \cite{Cheung:2018wkq,Bern:2019crd,Bern:2019nnu,Chung:2019yfs,Cheung:2020gyp,Bjerrum-Bohr:2019kec,Mogull:2020sak,Guevara:2018wpp,Guevara:2019fsj,Guevara:2020xjx,Bautista:2021wfy,Moynihan:2019bor,Emond:2020lwi}, and they have also been extended to higher-dimensions  \cite{Cristofoli:2020uzm,KoemansCollado:2018hss} (for a review see   \cite{Bjerrum-Bohr:2022blt}). This research program can be applied to make high-precision computations for the post-Minkowskian dynamics of binary bound systems, which could lead to predictions  in the context of  gravitational-wave emission \cite{Buonanno:2022pgc}.

\vskip 1cm

\section*{Acknowledgments}
SD would like to thank S. Mougiakakos for discussions at various stages of this project.
The work of FR is partially supported by the MIUR PRIN Grant 2020KR4KN2 ``String Theory as a bridge between Gauge
Theories and Quantum Gravity''.
 
 \vskip 1cm
 
\appendix
\section{Feynman Rules}\label{App:FeynmanRules}
In this section we list  all the Feynman rules used in the paper.  We also write down explicitly some of the expressions for the contraction of the 3-graviton vertex with two propagators.
%At the end of the list we will study in deep the presence of two different expressions of the three graviton vertex and we will obtain a useful analytic expression of this vertex contracted with two propagators.
\begin{itemize}
\item Scalar propagator of mass $m$:
\begin{equation}
\begin{fmffile}{PropScalar}
\begin{fmfgraph*}(50,0)
 \fmfleft{i}
 \fmfright{o}
 \fmf{fermion,label=$q$,l.s=left}{i,o}
\end{fmfgraph*}
\end{fmffile}
      =\frac{i}{q^2 - m^2+ i\epsilon} \ .
\end{equation}

\item Photon propagator in  the Feynman gauge:
       \begin{equation}
\begin{fmffile}{PropPhoton}
\begin{fmfgraph*}(50,0)
 \fmfleft{i}
 \fmfright{o}
 \fmf{photon,label=$q$,l.s=left}{i,o}
 \fmflabel{$\mu$}{i}
 \fmflabel{$\nu$}{o}
\end{fmfgraph*}
\end{fmffile}
    \hspace{0.5cm} =-\frac{i\eta^{\mu\nu}}{q^2 +i\epsilon} \ .
\end{equation}

\item Graviton propagator in de Donder gauge:
\begin{equation}
\begin{fmffile}{PropGraviton}
\begin{fmfgraph*}(50,0)
\DeclareGraphicsRule{*}{mps}{*}{}
\newcommand{\marrow}[5]{%
    \fmfcmd{style_def marrow#1 
    expr p = drawarrow subpath (1/4, 3/4) of p shifted 6 #2 withpen pencircle scaled 0.4; 
    label.#3 (btex #4 etex, point 0.5 of p shifted 6 #2); 
    enddef;}
    \fmf{marrow#1, tension=0}{#5}}
\unitlength=1mm
 \fmfleft{i}
 \fmfright{o}
 \fmf{dbl_wiggly, label=$q$, l.s=left}{i,o}
 \fmflabel{$\mu\nu$}{o}
 \fmflabel{$\alpha\beta$}{i}
 %\marrow{a}{up}{top}{$q$}{i,o}
\end{fmfgraph*}
\end{fmffile}
    \hspace{0.8cm} =i\frac{P_{\alpha\beta,\mu\nu}}{q^2 +i\epsilon} \ ,
        \end{equation}
with $P_{\alpha\beta,\mu\nu}$ defined by
\begin{equation}\label{PropGraviton}
    P_{\alpha\beta,\mu\nu} \equiv  \frac{1}{2} \Big( \eta_{\mu\alpha}\eta_{\nu\beta}+\eta_{\mu\beta}\eta_{\nu\alpha}-\frac{2}{d-1}\eta_{\mu\nu}\eta_{\alpha\beta} \Big) \ .
\end{equation}

\item 2 scalars - 1 photon vertex:
\begin{fmffile}{VertScalPhot}
\begin{equation} \label{2scalars1photon}
\begin{gathered}
\begin{fmfgraph*}(70,50)
 \fmfleft{i1,i2}
 \fmfright{o}
 \fmf{fermion,label=$p$}{i1,v}
 \fmf{fermion,label=$p'$,l.s=right}{v,i2}
 \fmf{photon}{v,o}
 \fmflabel{${\mu}$}{o}
 \fmfdot{v}
\end{fmfgraph*}
       \end{gathered}
            \hspace{0.8cm}
            =\left(\tau_{\phi^2A}\right)^\mu(p,p')=-iQ\left(p+p'\right)^\mu \ . 
        \end{equation}
\end{fmffile}
\item 2 scalars - 1 graviton vertex \cite{Bjerrum:2015}:
\begin{fmffile}{VertScalGrav}
\begin{equation}\label{2scalars1graviton}
\begin{split}
\begin{gathered}   
\begin{fmfgraph*}(70,50)
\DeclareGraphicsRule{*}{mps}{*}{}
\newcommand{\marrow}[5]{%
    \fmfcmd{style_def marrow#1 
    expr p = drawarrow subpath (1/4, 3/4) of p shifted 6 #2 withpen pencircle scaled 0.4; 
    label.#3 (btex #4 etex, point 0.5 of p shifted 6 #2); 
    enddef;}
    \fmf{marrow#1, tension=0}{#5}}
\unitlength=1mm
 \fmfleft{i1,i2}
 \fmfright{o}
 \fmf{fermion,label=$p$}{i1,v}
 \fmf{fermion,label=$p'$,l.s=right}{v,i2}
 \fmf{dbl_wiggly}{v,o}
 \fmflabel{${\mu\nu}$}{o}
 \fmfdot{v}
 %\marrow{a}{up}{top}{$\ q$}{v,o}
\end{fmfgraph*}
\end{gathered}
            \hspace{0.8cm}&=\left(\tau_{\phi^2 h}\right) ^{\mu\nu}(p,p',m)=-\frac{i\kappa}{2}\Big(
            \big( p ^\mu p'{}^\nu + p ^\nu p'{}^\mu\big) - \eta ^{\mu\nu}\big(p\cdot p' - m^2 \big)\Big) \ . 
\end{split}
\end{equation}
\end{fmffile}
\item 2 photons - 1 graviton vertex \cite{Bjerrum:2015}:
\vspace{0.5cm}
\begin{fmffile}{VertPhotonGraviton}
       \begin{equation*}
   \begin{gathered}
\begin{fmfgraph*}(70,50)
\DeclareGraphicsRule{*}{mps}{*}{}
\newcommand{\marrow}[5]{%
    \fmfcmd{style_def marrow#1 
    expr p = drawarrow subpath (1/4, 3/4) of p shifted 8 #2 withpen pencircle scaled 0.4; 
    label.#3 (btex #4 etex, point 0.5 of p shifted 8 #2); 
    enddef;}
    \fmf{marrow#1, tension=0}{#5}}
\unitlength=1mm
 \fmfleft{i1,i2}
 \fmfright{o}
 \fmf{photon}{i1,v}
 \fmf{photon}{v,i2}
 \fmf{dbl_wiggly}{v,o}
 \fmflabel{${\mu\nu}$}{o}
 \fmflabel{$\alpha$}{i1}
 \fmflabel{$\beta$}{i2}
 \fmfdot{v}
 \marrow{a}{down}{bot}{$\ \ \ p$}{i1,v}
 \marrow{b}{up}{top}{$\ \ \ \  p'$}{v,i2}
 \marrow{c}{up}{top}{$q$}{v,o}
\end{fmfgraph*}
  \end{gathered}  
\hspace{0.8cm}
      =\left(\tau_{A^2h}\right) ^{\mu\nu,\alpha,\beta}(p,p')
       \end{equation*}
\end{fmffile}  
\begin{equation}
\begin{split}\label{2photon1graviton}
=i\kappa&\left( P_{(4)}  ^{\mu\nu\alpha\beta} (p\cdot p')+ \frac{1}{2} \left( \frac{1}{2} \eta ^{\mu\nu} (p ^\alpha p'{}^\beta + p ^\beta p' {}^\alpha) + \eta ^{\alpha\beta} (p^ \mu p'{}^ \nu + p^ \nu p'{}^ \mu) \right.\right.\\ &\left.\left. -\frac{1}{2}\left(
    \eta^ {\beta\nu} (p^ \mu p'{}^ \alpha + p^ \alpha p'{}^ \mu) + \eta^ {\alpha\nu} (p^ \mu p'{}^ \beta + p^ \beta p'{}^ \mu)+ \eta^ {\alpha\mu} (p^ \nu p'{}^\beta + p^\beta p'{}^\nu) \right.\right.\right.\\&\left.\left.\left.+ \eta^{\beta\mu} (p^\nu p'{}^\alpha + p^\alpha p'{}^\nu)
    \right)\right)\right) \  ,
\end{split}
\end{equation}
where $P_{(4)}^{\mu\nu, \alpha\beta}$ is \eqref{PropGraviton} in four space-time dimensions, {\it i.e.} for $d=3$.
\item 2 photons - 2 graviton vertex \cite{Bjerrum:2015}:\vspace{0.5cm}
\begin{figure}[H]
\begin{fmffile}{Vert2Photons2Gravitons}
\begin{equation*}
\begin{gathered}
\begin{fmfgraph*}(70,50)
\DeclareGraphicsRule{*}{mps}{*}{}
\newcommand{\marrow}[5]{%
    \fmfcmd{style_def marrow#1 
    expr p = drawarrow subpath (1/4, 3/4) of p shifted 6 #2 withpen pencircle scaled 0.4; 
    label.#3 (btex #4 etex, point 0.5 of p shifted 6 #2); 
    enddef;}
    \fmf{marrow#1, tension=0}{#5}}
\unitlength=1mm
 \fmfleft{i1,i2}
 \fmfright{o1,o2}
 \fmf{photon}{i1,v}
 \fmf{photon}{v,i2}
 \fmf{dbl_wiggly}{v,o1}
  \fmf{dbl_wiggly}{v,o2}
 \fmflabel{${\mu\nu}$}{o2}
 \fmflabel{${\rho\sigma}$}{o1}
 \fmflabel{$\alpha$}{i1}
 \fmflabel{$\beta$}{i2}
 \fmfdot{v}
  \marrow{a}{down}{bot}{$\ p$}{i1,v}
 \marrow{b}{up}{top}{$\ \ \ p'$}{v,i2}
\end{fmfgraph*}
\end{gathered}\hspace{0.8cm}
=\left({\tau}_{A^2h^2}\right){}_{\mu\nu,}{}_{\rho\sigma,\alpha,\beta}(p,p')
\end{equation*}
\end{fmffile}
\begin{equation}\label{VertA2h2}
\begin{aligned} 
\quad=&- \frac{i\kappa^{2}}{4}\left( \left( p_{ \beta} p'_{ \alpha}-\eta_{\alpha \beta}p\cdot p'\right)\left(\eta_{\mu \rho} \eta_{\nu \sigma}+\eta_{\mu \sigma} \eta_{\nu \rho}-\eta_{\mu \nu} \eta_{\rho \sigma}\right)+\eta_{\mu \rho}\left(\eta_{\alpha \beta}\left(p_{ \nu} p'_{ \sigma}+p_{ \sigma} p'_{ \nu}\right) \right. \right. \\
&\left.-\eta_{\alpha \nu} p_{ \beta} p'_{ \sigma}-\eta_{\beta \nu} p_{ \sigma} p'_{ \alpha}
-\eta_{\beta \sigma} p_{ \nu} p'_{ \alpha}-\eta_{\alpha \sigma} p_{ \beta} p'_{ \nu}+p\cdot p'\left(\eta_{\alpha \nu} \eta_{\beta \sigma}+\eta_{\alpha \sigma} \eta_{\beta \nu}\right)\right) \\
&+\eta_{\mu \sigma}\left(\eta_{\alpha \beta}\left(p_{ \nu} p'_{ \rho}+p_{ \rho} p'_{ \nu}\right)-\eta_{\alpha \nu} p_{ \beta} p'_{ \rho}-\eta_{\beta \nu} p_{ \rho} p'_{ \alpha} -\eta_{\beta \rho} p_{ \nu} p'_{ \alpha}-\eta_{\alpha \rho} p_{ \beta} p'_{ \nu}\right.\\
&\left.+p\cdot p' \left(\eta_{\alpha \nu} \eta_{\beta \rho}+\eta_{\alpha \rho} \eta_{\beta \nu}\right)\right) +\eta_{\nu \rho}\left(\eta_{\alpha \beta}\left(p_{ \mu} p'_{ \sigma}+p_{ \sigma} p'_{ \mu}\right)-\eta_{\alpha \mu} p_{ \beta} p'_{ \sigma}-\eta_{\beta \mu} p_{ \sigma} p'_{ \alpha}\right.\\
&\left.-\eta_{\beta \sigma} p_{ \mu} p'_{ \alpha}-\eta_{\alpha \sigma} p_{ \beta} p'_{ \mu}+p\cdot p'\left(\eta_{\alpha \mu} \eta_{\beta \sigma}+\eta_{\alpha \sigma} \eta_{\beta \mu}\right)\right) +\eta_{\nu \sigma}\left(\eta_{\alpha \beta}\left(p_{ \mu} p'_{ \rho}+p_{ \rho} p'_{ \mu}\right)\right.\\
&\left. -\eta_{\alpha \mu} p_{ \beta} p'_{ \rho}-\eta_{\beta \mu} p_{ \rho} p'_{ \alpha}\right.\left.-\eta_{\beta \rho} p_{ \mu} p'_{ \alpha}-\eta_{\alpha \rho} p_{ \beta} p'_{ \mu}+p\cdot p'\left(\eta_{\alpha \mu} \eta_{\beta \rho}+\eta_{\alpha \rho} \eta_{\beta \mu}\right)\right)\\
& -\eta_{\mu \nu}\left(\eta_{\alpha \beta}\left( p_{ \rho} p'_{ \sigma}+p_{ \sigma} p'_{ \rho} \right)-\eta_{\alpha \rho} p_{ \beta} p'_{ \sigma}-\eta_{\beta \rho} p_{ \sigma} p'_{ \alpha}-\eta_{\beta \sigma} p_{ \rho} p'_{ \alpha}-\eta_{\alpha \sigma} p_{ \beta} p'_{ \rho}\right.\\
&\left.+p\cdot p'\left(\eta_{\alpha \rho} \eta_{\beta \sigma}+\eta_{\beta \rho} \eta_{\alpha \sigma}\right)\right) -\eta_{\rho \sigma}\left(\eta_{\alpha \beta}\left(p_{ \mu} p'_{ \nu}+p_{ \nu} p'_{ \mu}\right)-\eta_{\alpha \mu} p_{ \beta} p'_{ \nu}-\eta_{\beta \mu} p_{ \nu} p'_{ \alpha}\right.\\
&\left.-\eta_{\beta \nu} p_{ \mu} p'_{ \alpha}-\eta_{\alpha \nu} p_{ \beta} p'_{ \mu}+p\cdot p'\left(\eta_{\alpha \mu} \eta_{\beta \nu}+\eta_{\beta \mu} \eta_{\alpha \nu}\right)\right) \\
&+\left(\eta_{\alpha \rho} p_{ \mu}-\eta_{\alpha \mu} p_{ \rho}\right)\left(\eta_{\beta \sigma} p'_{ \nu}-\eta_{\beta \nu}p'_{\sigma}\right)
+\left(\eta_{\alpha \sigma} p_{ \nu}-\eta_{\alpha \nu} p_{ \sigma}\right)\left(\eta_{\beta \rho} p'_{ \mu}-\eta_{\beta \mu}p'_{\rho}\right)
\\
&+\left.\left(\eta_{\alpha \sigma} p_{ \mu}-\eta_{\alpha \mu} p_{ \sigma}\right)\left(\eta_{\beta \rho} p'_{ \nu}-\eta_{\beta \nu}p'_{\sigma}\right)
+\left(\eta_{\alpha \rho} p_{ \nu}-\eta_{\alpha \nu} p_{ \rho}\right)\left(\eta_{\beta \sigma} p'_{ \mu}-\eta_{\beta \mu}p'_{\sigma}\right)
\right) \ . 
\end{aligned}
\end{equation}
\end{figure}       
\item 3 graviton vertex \cite{DeWitt:1967ub,Sannan:1986tz}:
\vspace{0.5cm}
\begin{figure}[H]\begin{fmffile}{VertGravitonDeWitt}
       \begin{equation*}
       \begin{gathered}
       \begin{fmfgraph*}(70,50)
\DeclareGraphicsRule{*}{mps}{*}{}
\newcommand{\marrow}[5]{%
    \fmfcmd{style_def marrow#1 
    expr p = drawarrow subpath (1/4, 3/4) of p shifted 8 #2 withpen pencircle scaled 0.4; 
    label.#3 (btex #4 etex, point 0.5 of p shifted 6  #2); 
    enddef;}
    \fmf{marrow#1, tension=0}{#5}}
\unitlength=1mm
 \fmfleft{i1,i2}
 \fmfright{o}
 \fmf{dbl_wiggly}{i1,v}
 \fmf{dbl_wiggly}{v,i2}
 \fmf{dbl_wiggly}{v,o}
 \fmflabel{${\mu_3\nu_3}$}{o}
 \fmflabel{$\mu_1\nu_1$}{i1}
 \fmflabel{$\mu_2\nu_2$}{i2}
 \fmfdot{v}
 \marrow{a}{down}{bot}{$\ \ \  p_1$}{i1,v}
 \marrow{b}{up}{top}{$\ \ \ p_2$}{i2,v}
 \marrow{c}{up}{top}{$\ p_3$}{o,v}
\end{fmfgraph*}       \end{gathered}\hspace{1cm}
=\left(\tau_{h^3}\right)^{\mu_1\nu_1,\mu_2\nu_2,\mu_3\nu_3}(p_1,p_2,p_3)
\end{equation*}
\end{fmffile}
\begin{equation}\label{VertDeWitt}
\begin{aligned}
     = -2i\kappa \text{Sym} \ \Big( &-\frac{1}{4}P_3\big(p_1 \cdot p_2 \eta^{\mu_1\nu_1}\eta^{\mu_2\nu_2}\eta^{\mu_3\nu_3} \big) -\frac{1}{4}P_6\big(p_1^{\mu_2} p_1^{\nu_2} \eta^{\mu_1\nu_1}\eta^{\mu_3\nu_3} \big) \\&+ \frac{1}{4}P_3\big(p_1 \cdot p_2 \eta^{\mu_1\mu_2}\eta^{\nu_1\nu_2}\eta^{\mu_3\nu_3} \big)
      +\frac{1}{2}P_6\big(p_1 \cdot p_2 \eta^{\mu_1\nu_1}\eta^{\mu_2\mu_3}\eta^{\nu_2\nu_3} \big) \\ &+P_3\big(p_1^{\mu_2} p_1^{\nu_3} \eta^{\mu_1\nu_1}\eta^{\nu_2\mu_3} \big)-\frac{1}{2}P_3\big(p_1^{\nu_2} p_2^{\mu_1} \eta^{\nu_1\mu_2}\eta^{\mu_3\nu_3} \big)
     \\ &+\frac{1}{2}P_3\big(p_1^{\mu_3} p_2^{\nu_3}\eta^{\mu_1\mu_2}\eta^{\nu_1\nu_2} \big) +\frac{1}{2}P_6\big(p_1^{\mu_3} p_1^{\nu_3}\eta^{\mu_1\mu_2}\eta^{\nu_1\nu_2} \big) \\ &+
     P_6\big(p_1^{\mu_2} p_2^{\nu_3}\eta^{\nu_2\mu_1}\eta^{\nu_1\mu_3} \big) +P_3\big(p_1^{\mu_2} p_2^{\mu_1}\eta^{\nu_2\mu_3}\eta^{\nu_3\nu_1} \big) \\ &-P_3\big(p_1 \cdot p_2 \eta^{\nu_1\mu_2}\eta^{\nu_2\mu_3}\eta^{\nu_3\mu_1} \big) \Big) \ .
\end{aligned}\end{equation}
\end{figure}
Here "Sym" indicates that a symmetrisation has to be performed on each index pair, while "$P$" indicates that a summation has to be performed on all the distinct permutations of the momentum-index triplets (the subscript gives the number of terms).\footnote{For example: $$\text{Sym}\big[P_3\big(p_1^{\mu_2} p_2^{\mu_1}\eta^{\nu_2\mu_3}\eta^{\nu_3\nu_1} \big)\big] = p_1^{\mu_2} p_2^{\mu_1}\eta^{\nu_2\mu_3}\eta^{\nu_3\nu_1} + p_1^{\mu_3} p_3^{\mu_1}\eta^{\nu_3\mu_2}\eta^{\nu_2\nu_1} + p_2^{\mu_3} p_3^{\mu_2}\eta^{\nu_3\mu_1}\eta^{\nu_1\nu_2} \ .$$}
\item Expanding the Einstein-Hilbert action around a background field, one can derive the vertex for 2 internal and 1 external gravitons \cite{Donoghue:1994dn}:
\vspace{0.5cm}
\begin{figure}[H]\begin{fmffile}{VertGravitonDonoghue}
       \begin{equation*}
       \begin{gathered}
\begin{fmfgraph*}(70,50)
\DeclareGraphicsRule{*}{mps}{*}{}
\newcommand{\marrow}[5]{%
    \fmfcmd{style_def marrow#1 
    expr p = drawarrow subpath (1/4, 3/4) of p shifted 9 #2 withpen pencircle scaled 0.4; 
    label.#3 (btex #4 etex, point 0.5 of p shifted 9 #2); 
    enddef;}
    \fmf{marrow#1, tension=0}{#5}}
\unitlength=1mm
 \fmfleft{i1,i2}
 \fmfright{o}
 \fmf{dbl_wiggly}{i1,v}
 \fmf{dbl_wiggly}{v,i2}
 \fmf{dbl_wiggly}{v,o}
 \fmflabel{${\mu\nu}$}{o}
 \fmflabel{$\alpha\beta$}{i1}
 \fmflabel{$\gamma\delta$}{i2}
 \fmfdot{v}
 \marrow{a}{down}{bot}{$\ \ \ l$}{i1,v}
 \marrow{b}{up}{top}{$\ \ \ \ \ \ \ l-q$}{v,i2}
 \marrow{c}{up}{top}{$q$}{v,o}
\end{fmfgraph*}
\end{gathered}
    \hspace{0.8cm}
=\left(\tau_{h^2h}\right){}^{\mu\nu}{}_{\alpha\beta,\gamma\delta}(l,q) 
\end{equation*}
\end{fmffile}
\end{figure}
\vspace{-1cm}
\begin{equation}\label{VertDonoghue}
\begin{aligned}
=\frac{i \kappa}{2}&\bigg( P^{(4)}_{\alpha \beta,  \gamma \delta}\left(l^{\mu} l^{\nu}+(l-q)^{\mu}(l-q)^{\nu}+q^{\mu} q^{\nu}-\frac{3}{2} \eta^{\mu \nu} q^{2}\right) \\
&+2 q_{\lambda} q_{\sigma}\left({I_{\alpha \beta}}^{\lambda \sigma} {I_{\gamma \delta}}^{\mu \nu}+{I_{\gamma \delta}}^{\lambda \sigma} {I_{\alpha \beta}}^{\mu \nu}-{I_{\alpha \beta}}^{\lambda \mu }{I^{\sigma\nu}}_{\gamma \delta}-{I_{\alpha \beta}}^{\sigma \nu} {I_{\gamma \delta}}^{\lambda \nu}\right) \\
&+q_{\lambda} q^{\mu}\left(\eta_{\alpha \beta} I^{\lambda \nu}{ }_{\gamma \delta}+\eta_{\gamma \delta} {I^{\lambda \nu}}_{\alpha \beta}\right)+q_{\lambda} q^{\nu}\left(\eta_{\alpha \beta} {I_{\gamma \delta}}^{\lambda \mu}+\eta_{\gamma \delta} {I_{\alpha \beta}}^{\lambda \mu}\right)\\
&-q^{2}\left(\eta_{\alpha \beta} {I_{\gamma \delta}}^{\mu \nu}+\eta_{\gamma \delta} {I_{\alpha \beta}}^{\mu \nu}\right)-\eta^{\mu \nu} q^{\lambda} q^{\sigma}\left(\eta_{\alpha \beta} {I_{\gamma \delta \lambda \sigma}}+\eta_{\gamma \delta} I_{\alpha \beta \lambda \sigma}\right) \\
&+2 q ^ { \lambda } \left({I_{\alpha \beta}}^{\sigma \lambda} I_{\gamma \delta \sigma \nu}(l-q)^{\mu}+{I_{\alpha \beta}}^{\lambda\sigma} I_{\gamma \delta \sigma}{}^\mu(l-q)^{\nu} -{I_{\gamma \delta}}^{\lambda \sigma} I_{\alpha \beta \sigma}{}^\nu l^{\mu}-{I_{\gamma \delta}}^{\lambda \sigma} I_{\alpha \beta \sigma}{}^\mu l^{\nu}\right) \\
&+q^{2}\left({I_{\alpha \beta}}^{\sigma \mu} {I_{\gamma \delta ,\sigma}}^{\nu}+I_{\alpha \beta, \sigma}{ }^\nu {I_{\alpha \delta}}^{\sigma \mu}\right)+\eta^{\mu \nu} q^{\lambda} q_{\sigma}\left({I_{\gamma \delta}}^{\rho \sigma} I_{\alpha \beta, \lambda \rho}+{I_{\alpha \beta}}^{\rho \sigma} I_{\gamma \delta, \lambda \rho}\right) \\
&+\left(l^{2}+(l-q)^{2}\right)\left({I_{\alpha \beta}}^{\sigma \mu} {I_{\gamma \delta, \sigma}}^{\nu}+I^{\sigma \nu}{ }_{\alpha \beta} {I_{\gamma \delta, \sigma}}^{\mu}-\frac{1}{2} \eta^{\mu \nu} P^{(4)}_{\alpha \beta, \gamma \delta}\right) \\
&-l^{2} \eta_{\gamma \delta} {I_{\alpha\beta}}^{\mu \nu}-(l-q)^{2} \eta_{\alpha \beta} {I_{\gamma \delta}}^{\mu \nu}\bigg) \ ,
\end{aligned}\end{equation}
 where we have introduced the tensor $ I_{\mu\nu\alpha\beta} \equiv \frac{1}{2}( \eta_{\mu\alpha}\eta_{\nu\beta} + \eta_{\mu\beta}\eta_{\nu\alpha})$. The $\mu\nu$ graviton line corresponds to the external graviton, and the vertex is symmetric under exchange of the two internal lines.
\end{itemize}

It is useful to derive an analytical expression of the vertex in eq. \eqref{VertDonoghue} contracted with two graviton propagators, which can be then used directly to compute the amplitudes in which such vertex occurs. Defining
% We gave two different expressions of the tree gravitons vertex. These are different since they refer to different vertices. In particular the vertex computed by Donoghue (\ref{VertDonoghue}) has been derived with the background field method, taking the background metric expanded as $\Bar{g}_{\mu\nu}(x) = \eta_{\mu\nu} + \kappa\, h^{ext.}_{\mu\nu}(x)$ and then picking only the terms with one external field and two quantum fields \cite{Donoghue:1994dn}. This vertex is not symmetric in the exchange of all the graviton legs (as can be seen in the last line of \ref{VertDonoghue}) and has to be used only with an external graviton. For internal three graviton vertices we have to use the expression in (\ref{VertDeWitt}). The vertex in \eqref{VertDonoghue} is useful because it has an expression in terms of tensors that can more easily be contracted with the propagators. In particular we will give an analytical expression of the vertex contracted with two graviton propagators, which is a contraction that is common to all the diagrams with such a vertex (as can be seen from the amplitudes computations). Therefore we define this contraction as
\begin{equation}\label{VertTildeDef}
    \left(\tilde{\tau}_{h^2h}\right){}^{\mu\nu}{}_{\alpha\beta,\gamma\delta} \equiv {{P}_{\alpha\beta}}^{\rho\sigma}{{P}_{\gamma\delta}}^{\eta\xi}\left(\tau_{h^2h}\right){}^{\mu\nu}{}_{\rho\sigma,\eta\xi} 
\end{equation}
and using the proprieties
\begin{equation}
    \begin{aligned}
    &P_{\alpha\beta,\gamma\delta}P_{(4)}^{\gamma\delta,\mu\nu} = {I^{\mu\nu}}_{\alpha\beta} \quad && \eta^{\alpha\beta}P_{\alpha\beta,\mu\nu} = -\frac{2}{d-1}\eta_{\mu\nu} \\
    &P_{\alpha\beta,\gamma\delta}{I^{\gamma\delta}}_{\mu\nu} = P_{\alpha\beta,\mu\nu} \quad &&
   P_{\alpha\beta,\gamma\delta}P_{\rho\sigma,\mu\nu}P_{(4)}^{\gamma\delta,\rho\sigma}  = P_{\alpha\beta,\mu\nu} \ , \end{aligned}
\end{equation}
we obtain
\begin{equation}\label{VertTilde}
\begin{aligned}
    \left(\tilde{\tau}_{h^2h}\right){}&^{\mu\nu}{}_{\alpha\beta,\gamma\delta}(l,q) 
    =\frac{i \kappa}{2}\left( P_{\alpha \beta ,\gamma \delta}\left(l^{\mu} l^{\nu}+(l-q)^{\mu}(l-q)^{\nu}+q^{\mu} q^{\nu}-\frac{3}{2} \eta^{\mu \nu} q^{2}\right) \right. \\
&+2 q_{\lambda} q_{\sigma}\left({P_{\alpha \beta}}^{\lambda \sigma} {P_{\gamma \delta}}^{\mu \nu}+{P_{\gamma \delta}}^{\lambda \sigma} {P_{\alpha \beta}}^{\mu \nu}-{P_{\alpha \beta}}^{\lambda \mu }{P^{\sigma\nu}}_{\gamma \delta}-{P_{\alpha \beta}}^{\sigma \nu} {P_{\gamma \delta}}^{\lambda \nu}\right)  \\
&-\frac{2}{d-1}q_{\lambda} q^{\mu}\left(\eta_{\alpha \beta} P^{\lambda \nu}{ }_{\gamma \delta}+\eta_{\gamma \delta} {P^{\lambda \nu}}_{\alpha \beta}\right)-\frac{2}{d-1}q_{\lambda} q^{\nu}\left(\eta_{\alpha \beta} {P_{\gamma \delta}}^{\lambda \mu}+\eta_{\gamma \delta} {P_{\alpha \beta}}^{\lambda \mu}\right) \\
&+\frac{2}{d-1}q^{2}\left(\eta_{\alpha \beta} {P_{\gamma \delta}}^{\mu \nu}+\eta_{\gamma \delta} {P_{\alpha \beta}}^{\mu \nu}\right)+\frac{2}{d-1}\eta^{\mu \nu} q^{\lambda} q^{\sigma}\left(\eta_{\alpha \beta} {P_{\gamma \delta ,\lambda \sigma}}+\eta_{\gamma \delta} P_{\alpha \beta, \lambda \sigma}\right)  \\
&+2 q ^ { \lambda } \left({P_{\alpha \beta}}^{\sigma \lambda} P_{\gamma \delta, \sigma}{}^\nu(l-q)^{\mu}+{P_{\alpha \beta}}^{\lambda\sigma} P_{\gamma \delta ,\sigma}{}^\mu(l-q)^{\nu} -{P_{\gamma \delta}}^{\lambda \sigma} P_{\alpha \beta, \sigma}{}^\nu l^{\mu}-{P_{\gamma \delta}}^{\lambda \sigma} P_{\alpha \beta, \sigma}{}^\mu l^{\nu}\right)  \\
&+q^{2}\left({P_{\alpha \beta}}^{\sigma \mu} {P_{\gamma \delta, \sigma}}^{\nu}+P_{\alpha \beta, \sigma}{ }^\nu {P_{\alpha \delta}}^{\sigma \mu}\right)+\eta^{\mu \nu} q^{\lambda} q_{\sigma}\left({P_{\gamma \delta}}^{\rho \sigma} P_{\alpha \beta ,\lambda \rho}+{P_{\alpha \beta}}^{\rho \sigma} P_{\gamma \delta, \lambda \rho}\right)  \\
&+\left(l^{2}+(l-q)^{2}\right)\left({P_{\alpha \beta}}^{\sigma \mu} {P_{\gamma \delta, \sigma}}^{\nu}+P^{\sigma \nu}{ }_{\alpha \beta} {P_{\gamma \delta, \sigma}}^{\mu}-\frac{1}{2} \eta^{\mu \nu} P_{\alpha \beta ,\gamma \delta}\right)  \\
&\left.+\frac{2}{d-1}\left(l^{2} \eta_{\gamma \delta} {P_{\alpha\beta}}^{\mu \nu}+(l-q)^{2} \eta_{\alpha \beta} {P_{\gamma \delta}}^{\mu \nu}\right)\right) \ .
\end{aligned}
\end{equation}
For instance, to compute the 1-loop amplitude in figure \ref{1loop_grav} one needs $\left(\tilde{\tau}_{h^2h}\right){}^{\mu \nu}{}_{00,00}$, and from eq. \eqref{VertTilde} one gets \cite{Mougiakakos:2020laz}
 \begin{equation}
 \begin{aligned}
\left(\tilde{\tau}_{h^2h}\right){}^{\mu \nu}{}_{00,00}& (l, q)= \frac{i\kappa}{2}\frac{1}{d-1}\left((d-2)\left(l^{\mu} l^{\nu}+(l-q)^{\mu}(l-q)^{\nu}+q^{\mu} q^{\nu}+\frac{3}{2} \eta^{\mu \nu} \vec{q}^{2}\right) \right.\\
&-
\left.2(d-2)\left(\vec{l}^{2}+\left(\vec{l}+\vec{q}\right)^{2}\right)\left(\delta_{0}^{\mu} \delta_{0}^{\mu}-\frac{1}{4}\eta^{\mu \nu}\right)-2(d-3) \vec{q}^{2} \delta_{0}^{\mu} \delta_{0}^{\mu}\right) \ .
\end{aligned}
\end{equation}

\section{Master integrals}\label{App:masterintegrals}

We derive here all the identities needed for the evaluation of \eqref{MetricPerturb_loop_Expansion} and \eqref{ClassicalEM_potential_withCurrent}. These computations involve the Fourier transform with respect to the transferred momentum $\vec{q}$ of the master integral $J_{(l)}(\vec{q}^2)$. The relevant integrals are 
\begin{equation}\label{antitransform1}
    \int \frac{d^d\vec{q}}{(2\pi)^d}  \frac{J_{(l)}(\vec{q}^2)}{\vec{q}^2}e^{i\vec{q}\cdot\vec{x}}
\end{equation}
and
\begin{equation}\label{antitransform2}
    \begin{split}
        \int \frac{d^d\vec{q}}{(2\pi)^d} \frac{q_iq_j}{\vec{q}^2} \frac{J_{(l)}(\vec{q}^2)}{\vec{q}^2}e^{i\vec{q}\cdot\vec{x}} \ .
    \end{split}
 \end{equation}
The sunset master integral with $l$-loop has been defined in \eqref{EMT_propto_MasterIntegral} and it can be rewritten  as (\cite{Mougiakakos:2020laz}, Eq. (2.31))
\begin{equation}\label{masterintegral_solved}
        J_{(l)}(\vec{q}^2)=\frac{\Gamma\left( l+1-\frac{ld}{2}\right)\Gamma\left(\frac{d-2}{2}\right)^{l+1}}{(4\pi)^{\frac{ld}{2}}\Gamma\left(\frac{(l+1)(d-2)}{2}\right)} (\vec{q}^2)^{\frac{l(d-2)}{2}}\ ,
\end{equation}
and substituting it inside \eqref{antitransform1} we have 
\begin{equation}\label{antitransform_mid}
      \int \frac{d^d\vec{q}}{(2\pi)^d}  \frac{J_{(l)}(\vec{q}^2)}{\vec{q}^2}e^{i\vec{q}\cdot\vec{x}}=\frac{\Gamma\left( l+1-\frac{ld}{2}\right)\Gamma\left(\frac{d-2}{2}\right)^{l+1}}{(4\pi)^{\frac{ld}{2}}\Gamma\left(\frac{(l+1)(d-2)}{2}\right)} \int \frac{d^d\vec{q}}{(2\pi)^d} |\vec{q}|^{l(d-2)-2} e^{i\vec{q}\cdot\vec{x}} \ .
\end{equation}

To solve \eqref{antitransform1} we can compute the inverse Fourier transform $F(r)$ for a generic spherically symmetric function $f(q)$, defined as
\begin{equation}\label{antitransform_general}
    F(r) \equiv \int\frac{d^dq}{(2\pi)^d}f(q)e^{i\vec{x}\cdot\vec{q}} \ .
\end{equation}
Rewriting both $\vec{q}$ and $\vec{x}$ in hyper-spherical coordinates, we can decompose the $d$-dimensional plane wave in Bessel functions $J_\nu(qr)$ \cite{Avery:2017hh}, and assuming spherical symmetry one gets 
\begin{equation}\label{antitranform_F_Bessel}
     F(r) = (2\pi)^{-\frac{d}{2}}r^{1-\frac{d}{2}} \int_0^\infty q dq \ \ q^{\frac{d}{2}-1} f(q) J_{\frac{d}{2}-1}(qr) \ .
\end{equation}
Using the definition of the $\nu$-th order inverse Hankel transform of a generic function $g(q)$ (defined for $q>0$) 
\begin{equation}
    \mathscr{H}_\nu \{g(q)\} \equiv \int_0^\infty q dq \ g(q) J_\nu(qr) \ ,
\end{equation}
eq. \eqref{antitranform_F_Bessel} becomes
\begin{equation}\label{antitranform_F_Hankel}
     F(r) = (2\pi)^{-\frac{d}{2}}r^{1-\frac{d}{2}}  \mathscr{H}_{\frac{d}{2}-1} \{q^{\frac{d}{2}-1}f(q) \} \ ,
\end{equation}
which relates a generic spherically symmetric function $F(r)$ with his Fourier transform $f(q)$. The $(\frac{d}{2}-1)$-th order inverse Hankel transform of a generic power $\alpha$ of the momentum is
\begin{equation}\label{hankelTransform}
    \mathscr{H}_{\frac{d}{2}-1}\{q^\alpha\} = \frac{2^{\alpha+1}}{r^{\alpha+2}}\frac{\Gamma \left(\frac{d+2\alpha+2}{4}\right)}{\Gamma \left(\frac{d-2\alpha-2}{4}\right)} \ ,
\end{equation}
obtained inverting Item (2) of the Table 9.2 of \cite{Poularikas:2018tr}. In our case $f(q)=q^{l(d-2)-2}$ as can be seen from \eqref{antitransform_mid} and therefore $\alpha = \frac{d}{2}+l(d-2)-3$.
From \eqref{hankelTransform} and  \eqref{antitranform_F_Hankel} we find 
\begin{equation}\label{AppIdent1}
    \int \frac{d^d\vec{q}}{(2\pi)^d}  \frac{J_{(l)}(\vec{q}^2)}{\vec{q}^2}e^{i\vec{q}\cdot\vec{x}} = \left(\frac{\Gamma\left(\frac{d-2}{2}\right)}{4\pi^{\frac{d}{2}}}\frac{1}{r^{d-2}}\right)^{l+1}= \left(\frac{\rho}{4\pi}\right)^{l+1} \ .
\end{equation}

Analogously, in order to solve \eqref{antitransform2} we make the same computation for $\alpha = \frac{d}{2}+l(d-2)-5$, obtaining the identity
\begin{equation}
    \begin{split}
        \int \frac{d^d\vec{q}}{(2\pi)^d}  \frac{J_{(l)}(\vec{q}^2)}{\vec{q}^4}e^{i\vec{q}\cdot\vec{x}} =  \left(\frac{\Gamma\left(\frac{d-2}{2}\right)}{4\pi^{\frac{d}{2}}}\right)^{l+1}\frac{r^{d-4+l(d-2)}}{(2 - l(d-2)) (d-4+ l(d-2))} \ .
\end{split}
\end{equation}
Applying the operator $\partial_i\partial_j$ on both sides we finally obtain 
\begin{equation}\label{AppIdent2}
    \begin{split}
        \int \frac{d^d\vec{q}}{(2\pi)^d} \frac{q_iq_j}{\vec{q}^2} \frac{J_{(l)}(\vec{q}^2)}{\vec{q}^2}e^{i\vec{q}\cdot\vec{x}} =&  \frac{1}{2-l(d-2)} \left(\delta_{ij} + (l+1)(2-d)\frac{x_ix_j}{r}\right) \left(\frac{\rho}{4\pi}\right)^{l+1}\ .
    \end{split}
 \end{equation}
In both the identities \eqref{AppIdent1} and \eqref{AppIdent2} the relation between the post-Minkowskian and the loop expansions is explicit.

From the analysis above one can also derive the identity \cite{Mougiakakos:2020laz}
\begin{equation}\label{Master_CT}
    \int \frac{d^{d} \vec{q}}{(2 \pi)^{d}}J_{(1)}\left(\vec{q}^{2}\right) e^{i \vec{q} \cdot \vec{x}} =-\frac{\Gamma\left(\frac{d}{2}\right)^{2}}{2 \pi^{d} r^{2(d-1)}} \ ,
\end{equation}
  which is needed for the evaluation of the momentum integrals in section \ref{sec:divergences}.
  
\section{Loop integral reduction}\label{App:LoopRed}

The classical limit of both the stress-energy tensor and the electromagnetic current arise extracting from the loop integrals the contributions proportional to the master integral. In this appendix we list such contributions for all the loop integrals that occur in the amplitude computations in sections \ref{sec:metric} and \ref{sec:gaugepotential}. For some of  these relations we made use 
of  the \texttt{LiteRed} package of Mathematica \cite{LiteRed}.
% Notice that  at 2-loop order all the expressions are meant to be symmetric under the exchange of $\ell_1\leftrightarrow\ell_2$. This is true because the momenta that appear in the loop integrals are internal, and every amplitude is symmetric under the exchange of internal lines. 
\subsubsection*{1-loop}
\begin{equation}
    \int \frac{d^d \ell}{(2\pi)^d} \frac{\vec{\ell}^2}{\vec{\ell}^2(\vec{q}-\vec{\ell})^2} = 0\ , 
\end{equation}
\begin{equation}\label{LR_1Loop_1}
    \int \frac{d^d \ell}{(2\pi)^d} \frac{\vec{\ell}\cdot \vec{q}}{\vec{\ell}^2(\vec{q}-\vec{\ell})^2} = \frac{1}{2}J_{(1)}(\vec{q}^2)\ , 
\end{equation}
\begin{equation}\label{LR_1Loop_2}
    \int \frac{d^d \ell}{(2\pi)^d} \frac{\vec{\ell}\cdot \vec{q}\, \Vec{\ell}^2}{\vec{\ell}^2(\vec{q}-\vec{\ell})^2} = 0 \ , 
\end{equation}
\begin{equation}\label{LR_1Loop_3}
    \int \frac{d^d \ell}{(2\pi)^d} \frac{\vec{\ell}\cdot \vec{q}\, \vec{\ell}\cdot \vec{q}}{\vec{\ell}^2(\vec{q}-\vec{\ell})^2} = \frac{1}{4}\Vec{q}^2\, J_{(1)}(\vec{q}^2)\ , 
\end{equation}

\subsubsection*{2-loop}

\begin{equation}\label{LiteRed_2Loop_2}
      \int \frac{d^d\ell_1}{(2\pi)^d} \frac{d^d\ell_2}{(2\pi)^d} \frac{\vec{\ell_1}\cdot \vec{q}}{\vec{\ell_1}^2 \vec{\ell_2}^2\left(\vec{q}-\vec{\ell_1}-\vec{\ell_2}\right)^2}=\frac{1}{3}J_{(2)}\left(\vec{q}^{2}\right)
\end{equation}

\begin{equation}
      \int \frac{d^d\ell_1}{(2\pi)^d} \frac{d^d\ell_2}{(2\pi)^d} \frac{\vec{\ell_1}\cdot \vec{\ell_2}}{\vec{\ell_1}^2 \vec{\ell_2}^2\left(\vec{q}-\vec{\ell_1}-\vec{\ell_2}\right)^2}=\frac{1}{6}J_{(2)}\left(\vec{q}^{2}\right)
\end{equation}

\begin{equation}
      \int \frac{d^d\ell_1}{(2\pi)^d} \frac{d^d\ell_2}{(2\pi)^d} \frac{\vec{\ell_1}\cdot\Vec{q}\ \Vec{\ell_1}^2}{\vec{\ell_1}^2 \vec{\ell_2}^2\left(\vec{q}-\vec{\ell_1}-\vec{\ell_2}\right)^2\left(\vec{\ell_1}+\vec{\ell_2}\right)^{2}}=0
\end{equation}

\begin{equation}
      \int \frac{d^d\ell_1}{(2\pi)^d} \frac{d^d\ell_2}{(2\pi)^d} \frac{\vec{\ell_2}\cdot\Vec{q}\ \Vec{\ell_1}^2}{\vec{\ell_1}^2 \vec{\ell_2}^2\left(\vec{q}-\vec{\ell_1}-\vec{\ell_2}\right)^2\left(\vec{\ell_1}+\vec{\ell_2}\right)^{2}}=0
\end{equation}

\begin{equation}
      \int \frac{d^d\ell_1}{(2\pi)^d} \frac{d^d\ell_2}{(2\pi)^d} \frac{\vec{\ell_1}\cdot\Vec{\ell_2}\ \Vec{q}\cdot\Vec{\ell_1}}{\vec{\ell_1}^2 \vec{\ell_2}^2\left(\vec{q}-\vec{\ell_1}-\vec{\ell_2}\right)^2\left(\vec{\ell_1}+\vec{\ell_2}\right)^{2}}=\frac{1}{6}J_{(2)}\left(\vec{q}^{2}\right)
\end{equation}

\begin{equation}
      \int \frac{d^d\ell_1}{(2\pi)^d} \frac{d^d\ell_2}{(2\pi)^d} \frac{\vec{\ell_1}^2\ \Vec{q}^2}{\vec{\ell_1}^2 \vec{\ell_2}^2\left(\vec{q}-\vec{\ell_1}-\vec{\ell_2}\right)^2\left(\vec{\ell_1}+\vec{\ell_2}\right)^{2}}=0
\end{equation}

\begin{equation}
      \int \frac{d^d\ell_1}{(2\pi)^d} \frac{d^d\ell_2}{(2\pi)^d} \frac{\vec{\ell_1}\cdot\Vec{\ell_2}\ \Vec{q}^2}{\vec{\ell_1}^2 \vec{\ell_2}^2\left(\vec{q}-\vec{\ell_1}-\vec{\ell_2}\right)^2\left(\vec{\ell_1}+\vec{\ell_2}\right)^{2}}=\frac{1}{2}J_{(2)}\left(\vec{q}^{2}\right)
\end{equation}

\begin{equation}\label{LR_Loop2_l1q_l2q}
      \int \frac{d^d\ell_1}{(2\pi)^d} \frac{d^d\ell_2}{(2\pi)^d} \frac{\vec{\ell_1}\cdot\Vec{q}\ \vec{\ell_2}\cdot\Vec{q}}{\vec{\ell_1}^2 \vec{\ell_2}^2\left(\vec{q}-\vec{\ell_1}-\vec{\ell_2}\right)^2\left(\vec{\ell_1}+\vec{\ell_2}\right)^{2}}=\frac{2d-7}{6(d-4)}J_{(2)}\left(\vec{q}^{2}\right)
\end{equation}

\begin{equation}
      \int \frac{d^d\ell_1}{(2\pi)^d} \frac{d^d\ell_2}{(2\pi)^d} \frac{\vec{\ell_1}\cdot\Vec{q}\ \vec{\ell_1}\cdot\Vec{q}}{\vec{\ell_1}^2 \vec{\ell_2}^2\left(\vec{q}-\vec{\ell_1}-\vec{\ell_2}\right)^2\left(\vec{\ell_1}+\vec{\ell_2}\right)^{2}}=\frac{d-3}{3(d-4)}J_{(2)}\left(\vec{q}^{2}\right) \ .
\end{equation}

\vskip 1cm

% \newpage
\bibliographystyle{JHEP}
\bibliography{biblio}

\end{document}